\documentclass{article}

\usepackage{graphicx}  % standard LaTeX graphics tool;
\usepackage{amsmath}   % For getting proper math eqns
\usepackage[compress]{cite}
\usepackage{amssymb}   % Used for getting various symbols eg. gtrsim, lesssim etc.
\usepackage{bm} % For bold math (esp of lower greek letters)

\usepackage{mathrsfs}

\def\eq#1{{Eq.~(\ref{#1})}}

\def\frab#1#2{\left(\frac{#1}{#2}\right)}

\def\b#1{{\bm#1}}

\def\ket#1{|#1\rangle}                    %%%%   ket 
\def\bra#1{\langle #1|}                   %%%%   bra
\def\bk#1#2#3{{\langle #1|#2|#3\rangle}}  %%%%   bracket
\def\amp#1#2{\langle #1 | #2\rangle}      %%%%   amplitude

\title{Obtaining the Non-relativistic Quantum Mechanics from Quantum Field Theory: Issues, Folklores and Facts\\
\vskip 1.1cm
{\textit{\Large What happens to the anti-particles when you take the non-relativistic limit of QFT?\\
\vskip 2cm
}}}

\author{T. Padmanabhan\\
IUCAA, Post Bag 4, Ganeshkhind,
 Pune - 411 007, India.\\
email: paddy@iucaa.in}

\begin{document}

\maketitle

\begin{abstract}
 Given the classical dynamics of a non-relativistic particle in terms of a Hamiltonian or an action, it is relatively straightforward to obtain the non-relativistic quantum mechanics (NRQM) of the system. These standard procedures, based on either the Hamiltonian or the path integral, however, do not work in the case of a relativistic particle. As a result we do not have a single particle description of relativistic quantum mechanics (RQM). Instead, the correct approach requires a transmutation of dynamical variables from the position and momentum of a  single particle to a field and its canonical momentum. Particles, \textit{along with antiparticles},  reappear in a very non-trivial manner as the excitations of the field. The fact that one needs to adopt  completely different languages to describe  relativistic and non-relativistic free particle implies that obtaining the NRQM limit of QFT is conceptually non-trivial. I examine this limit in several approaches (like, for e.g., Hamiltonian dynamics, Lagrangian and Hamiltonian path integrals, field theoretic description etc.) and identify the precise issues which arise when one attempts to obtain the NRQM from QFT in each of these approaches. The dichotomy of description between NRQM and QFT  does not originate just  from the square root in the Hamiltonian or from the demand of Lorentz invariance, as it is sometimes claimed. The real difficulty has its origin in the necessary existence of antiparticles to ensure a particular notion of relativistic causality. Because of these conceptual issues, it turns out that one \textit{cannot}, in fact,  obtain  some of the popular descriptions of NRQM by \textit{any} sensible limiting procedure applied to QFT. To obtain NRQM from QFT in a  seamless manner, it is necessary to 
 work with NRQM expressed in a language  closer to that of QFT. This fact has several implications, especially for the operational notion of space coordinates in quantum theory. A close examination of these issues, which arise when quantum theory is combined with  special relativity,  could offer insights in the context of attempts to combine quantum theory with general relativity.  
\end{abstract}

\tableofcontents

\section{Motivation and Summary}

Given the classical theory of a non-relativistic particle, there is  a systematic way of obtaining its quantum version (NRQM), using either  a Hamiltonian approach or one based on  path integrals. For a system with, say, $H(\bm{x},\bm{p})=(\bm{p}^2/2m) + V(\bm{x})$,  these approaches lead to the the same quantum theory.
This success, however, turns out to be more of an exception than a rule in the description of Nature.   There is  no guarantee that the standard (Hamiltonian or path integral) procedures of quantization  will allow you to construct a quantum theory --- \textit{in terms of the same dynamical variables} --- if you try to impose  some   extra  constraints, like for e.g Lorentz invariance,\footnote{Unless otherwise specified, I use the expression `Lorentz invariance' to mean intrinsic Lorentz invariance and not manifest Lorentz invariance.} general covariance, or the notion of relativistic causality, which exist in the classical theory. 

An important example of a well-defined physical system, 
which has a simple classical description, but does not have a corresponding quantum description in terms of the \textit{same} dynamical variables
is provided by a relativistic free particle.
The usual procedures which work for NRQM do not work  in this case.  
Bringing together the principles of special relativity and quantum mechanics leads to \textit{change in the dynamical variables}, existence of antiparticles and several other complications leading, eventually, to --- what is called --- Quantum Field Theory (QFT). The formalism and the language are completely different in QFT and in NRQM. 
 
Though we have all learnt to live at peace with this development for decades, it is downright surprising  when you think about it. 

We do know that  both QFT and NRQM work quite well in their  respective domains. In the classical limit, the \textit{equation of motion} describing a relativistic particle \textit{does} go over to those describing a non-relativistic particle\footnote{Aside: But the \textit{Lagrangian} for the relativistic particle does \textit{not} go over to that of a non-relativistic particle in this limit, contrary to what some text books would like us to believe; instead, the Lagrangian  blows up. If you subtract  the constant $mc^2$ --- which blows up --- and redefine the Lagrangian, you lose Lorentz invariance of the action. In fact, it is not possible to construct any Lorentz invariant action for a relativistic free particle which will give the $L_{NR}=(1/2)mv^2$ in the $c\to \infty$ limit; see chapter 15 of Ref. \cite{tpsbtp}.} when you take the limit $c\to \infty$. This suggests that, in the corresponding quantum avatars,  one should be able to get NRQM from QFT by taking the $c\to \infty$ limit. 
But if the language and even the dynamical variables used in QFT and NRQM are completely different, how can you get NRQM from QFT seamlessly?
 Several text books and published literature deal with these issues rather too glibly (and inadequately).   A large part of this paper will be devoted to pointing out that the transition from QFT to NRQM is \textit{not} possible if your aim is to reproduce many of the conventional descriptions of NRQM. Towards the end of the paper, I will describe how this can be achieved using one specific formulation of NRQM.
 
\textit{ Notation:} Latin indices range over $0,1,2,....n=D-1$ where, usually, $D=4$. The Greek indices range spatial coordinates $1,2,...n=D-1$. I will set $\hbar=1, c=1$ when it will not lead to any confusion. The signature is mostly negative. I denote by $p.x$ the on-shell dot product in which $p_0$ is a given function of $\bm{p}$, like e.g., $p_0=(\bm{p}^2+m^2)^{1/2}$ while $p_ax^a$ will denote the off-shell dot product. I will omit the superscripts in $x^i,p^i$ etc when it is clear from the context, like e.g., use the notation $\psi(x)$ for $\psi(x^i)$. 
The symbol $\equiv$ in an equation tells you that the equation is used to define some quantity.
 
 \subsection{Does the Emperor have clothes?}\label{sec:emperor}
 
Let me briefly describe a series of issues which arise when you try to think of NRQM as the $c\to\infty$ limit of QFT.\footnote{Throughout the paper, I will only deal with a non-interacting, massive, scalar field because it is enough to illustrate the issues I am interested in. Spin and interactions add extra complications which I want to avoid so that I can highlight these issues in the simplest possible context.} These should alert you that the situation is not as straightforward as the folklore might suggest. 
 
 (1) In NRQM, a description based on  the  Schroedinger wave function $\psi(x)$, (which is a \textit{c-number complex function }  in the coordinate representation) has a distinct technical advantage over the one based on Heisenberg picture. In QFT, however, the Heisenberg picture is better suited for the description and one  uses, say,  a, \textit{real}, scalar field \textit{operator} $\hat \phi(x)$ which satisfies the Klein-Gordon equation. Of course, operators remain operators and real functions remain real when you take $c\to\infty$ limit; so to get $\psi(x)$ from $\hat \phi(x)$ one has to do something more than just taking the $c\to\infty$ limit. A favourite procedure adopted in the textbooks
  is the identification of $e^{-i mt}\bk{{\bm k}}{\hat \phi(x)}{0}$ (where $\ket{{\bm{k}}}$ is a one-particle state with momentum $\bm{k}$) with the Schroedinger wave function. While it is trivial to show that this function, in the appropriate limit, satisfies the Schroedinger equation, this construction is rather ad hoc. More importantly, it leads to another serious issue: 
 
 \textit{What happens to anti-particles when you take the $c\to \infty$ of QFT?} After all, a massive anti-particle has every right to remain at rest (or  in a low energy state) such that it could be described by NRQM! So when you take the appropriate limit of QFT you should be able to get the NRQM of \textit{both} particles and anti-particles in a seamless manner. Many of the conventional procedures (including the one mentioned above) will not do this.
 At best you will get the Schroedinger equation for the particle and will have to forget about the anti-particle which, of course, is unsatisfactory.\footnote{For example, there are (wrong) claims in the literature that the \textit{real} scalar field in QFT has no NRQM limit because, for a real scalar field, the particle is the same as the anti-particle. If anti-particles vanish mysteriously when you take the NRQM limit, then, of course, real scalar fields cannot have an NRQM limit. As we shall see, this is  incorrect.} 
 
 (2) Another issue of interpretation has to do with the very different roles played by the spatial coordinate $\bm{x}$ in QFT and NRQM. In QFT we will deal with $\hat \phi (t,\bm{x})$ which is an operator with both $t$ and $\bm{x}$ acting as labels. This is necessary since Lorentz transformations will mix space and time; so if $t$ is a label so should be $\bm{x}$. But in NRQM the spatial coordinate itself will acquire an operator status $\hat{\bm{x}}(t)$ labeled by $t$. Stated in another way, the dynamical variables
 in NRQM
 are $\hat{x}^\alpha(t)$ and $\hat{p}_\beta(t)$ obeying the equal time commutation rule (ETCR), $[\hat{x}^\alpha(t),\hat{p}_\beta(t)]= i \delta^\alpha_\beta$. On the other hand, in QFT the dynamical variables are $\hat{\phi}(x)$ and $\hat{\pi}(x)$ which obey the ETCR given by $[\hat{\phi}(t,\bm{x}), \hat{\pi}(t,\bm{y})] = i\delta (\bm{x-y})$. But there is no way of obtaining the position operator of NRQM from the basic field operators of QFT. Text books do pay homage to this fact by mumbling something about the inability to localize a particle in QFT but that does not answer the technical question of how the appropriate limit has to be taken so that you get the dynamical variables and the ETCR of NRQM from the dynamical variables and ETCR of the QFT. This, in fact, turns out to be impossible; \textit{you can't go there from here.} As we shall see, to make the seamless transition you need to describe  NRQM in a language which is closer to that of QFT; not the other way around.
 
 (3) Similar --- and sometimes worse --- difficulties arise when you approach the problem in the language of path integrals.\footnote{Most of the intuition you develop about path integrals is based on the quadratic momentum dependence of the Hamiltonian, making this intuition pretty much useless in the study of even a relativistic free  particle.} 
 Whenever we have a well-defined classical action,  we could try to  quantize the system in terms of the path integral by performing the sum over all paths, connecting two events $x_1$ and $x_2$, in the expression:
\begin{equation}
 G(x_2,x_1)=\sum_{\bm{x}(t)}\exp iA[\bm{x}(t)]
 \label{gassum}
\end{equation}
This works like a charm in NRQM. What is more, the resulting expression $G_{NR}(x_2,x_1)$  has an equivalent interpretation  as the matrix element of the time evolution operator:
\begin{equation}
 G_{NR}(x_2,x_1)=\bk{\bm{x}_2}{\exp[-i(t_2-t_1)H]}{\bm{x}_1}=\amp{t_2,\bm{x}_2}{t_1,\bm{x}_1}
 \label{gasme}
\end{equation}
The interpretation  relies on the fact that $\ket{\bm{x}_1}$ and $\ket{\bm{x}_2}$ are the eigenkets of a position operator $\hat{\bm{x}}(0)$ with eigenvalues $\bm{x}_1$ and $\bm{x}_2$; so we can use $G_{NR}(x_2,x_1)$ to propagate the wave function $\amp{t_1,\bm{x}_1}{\psi}$ to give
$\amp{t_2,\bm{x}_2}{\psi}$. We run into several issues when we try to do any of these in an attempt to obtain a RQM.

To begin with, there are some technical issues in performing the sum in \eq{gassum}; most of the procedures which work well in NRQM do not work in this case.
(This is because  these procedures in NRQM work only if the Hamiltonian is quadratic in momentum.)
There is one procedure, based on Euclidean lattice regularization, which \textit{does} give the sensible result leading to what is usually called the Feynman propagator $G_R(x_2,x_1)$ in QFT. But the interpretation of this propagator is  nontrivial because, roughly speaking, it contains information about both the particle and the antiparticle. Hence, it cannot be expressed in the form
$G_R(x_2,x_1)=\amp{t_2,\bm{x}_2}{t_1,\bm{x}_1}$; in fact,
we do not have an analogue of position operator $\hat{\bm{x}}(t)$ or its eigenstates, $\ket{t,\bm{x}}$ in a Lorentz invariant QFT;
so one does not have an analogue of \eq{gasme} with the same interpretation in RQM.\footnote{There is a folklore belief that, when you take the $c\to \infty $ limit of the Feynman propagator $G_R(x_2,x_1)$, you will get the non-relativistic propagator $G_{\rm NR}(x_2, x_1)$. As I will show in Sec.\ref{sec:nroff}, this is again \textit{not} true without extra, ad-hoc, assumptions.} 
 
 Thus there are serious issues in obtaining the NRQM based on position eigenstates $\ket{t,\bm{x}}$ and a wave function $\amp{t,\bm{x}}{\psi}$
 as a sensible limiting case of QFT. This conclusion remains valid irrespective of the procedure --- Hamiltonian or path integral --- adopted to construct the quantum theory of a relativistic particle. 
 
 \subsection{Preview and Summary}\label{sec:presum}
 
  Let me next summarize the structure of the rest of the paper and the key results. In Sec.\ref{sec:hofp}, I begin by constructing the quantum theory of a ``free particle''\footnote{I will define a ``free particle'' to be one for which neither the Lagrangian nor the Hamiltonian depend on the spacetime coordinates explicitly and $H(\bm{p}) = H(|\bm{p}|)$. Throughout this paper we will be concerned only with a free particle. As you will see, such a system itself creates enough problems!} described by the Hamiltonian $ H= H(|\bm{p}|)$. Since this form covers both non-relativistic and relativistic free particles, it is possible to compare the two situations at one go by studying such a system and probe why we cannot extend the standard ideas of NRQM to construct a RQM.   Since a well-defined momentum operator and its eigenstates $\ket{\bm{p}}$ exist, it is possible to develop the quantum theory in momentum representation in a straightforward manner. Neither the square root structure of the Hamiltonian for a relativistic particle nor the requirement of Lorentz invariance introduces any serious difficulties in the momentum representation. But Lorentz invariance  requires using a relativistically invariant normalization for momentum eigenkets (viz., $\amp{\bm{p}'}{\bm{p}} = 2\omega_{\bm{p}}\ (2\pi)^n \delta (\bm{p}- \bm{p}')$ with $\omega_{\bm{p}}=(\bm{p}^2+m^2)^{1/2}$; see \eq{four}).
  
  The first real difficulty arises when we try to introduce a (conjugate) position representation. In the relativistic theory we cannot introduce localized particle position states $\ket{\bm{x}}$ as eigenstates of a position operator because no sensible position operator can be constructed. We can still attempt to define states $\ket{\bm{x}}$, \textit{labeled by} spatial coordinates $\bm{x}$, as Fourier (like) transforms of the momentum eigenstates  $\ket{\bm{p}}$, but with a relativistically invariant integration measure. This leads to a Lorentz invariant propagator for the system, given by:\footnote{Recall that I denote by $p.x$ the on-shell dot product in which $p_0=(\bm{p}^2+m^2)^{1/2}$ while $p_ax^a$ will denote the off-shell dot product.} 
 \begin{equation}
G_+(x_2,x_1)\equiv \int  \frac{d^n\bm{p}}{(2\pi)^n}\,\frac{1}{2\omega_{\bm{p}}} \, \exp(- i p.x); \qquad x\equiv x_2-x_1  
\end{equation} 
with the D-dimensional (spacetime) momentum space representation:\footnote{This is built from the so called ``positive frequency'' modes of the Klein-Gordan (KG) equation and has to be distinguished from the Feynman propagator we will come across later on. We shall develop all these results in detail in the coming sections.}
\begin{equation}
 G_+(p)\equiv \int d^Dx G(x)e^{ip_ax^a}=\delta(p^2-m^2)\theta(p^0); \qquad D=n+1
\end{equation} 
 But the trouble is that the  states $\ket{\bm{x}}$ we have defined (and used to construct $G_+(x_2,x_1)$), do \textit{not} represent localized particles. The amplitude $\amp{\bm{x}}
  {\bm{y}}$ will not be a Dirac delta function $\delta(\bm{x}-\bm{y})$. So, even though defining $\ket{\bm{x}}$ as Fourier-like transform of  $\ket{\bm{p}}$, allows us to define a Lorentz invariant propagator for the system $G_+(x_2,x_1)$,  there is no way of introducing a relativistic wave function 
  in the coordinate representation,
  $\psi(x)=\amp{t,\bm{x}}{\psi}$, in the absence of position eigenstates $\ket{t,\bm{x}}$. In fact, the propagator  $G_+(x_2,x_1)$ does not satisfy the correct composition law or the limiting behaviour which are necessary for it  to ``propagate'' a wave function. 
 
So the straightforward Hamiltonian approach does not lead to an RQM such that we can obtain the NRQM as a limiting case. The utility of this discussion, for our purpose, is different. In
 Section \ref{sec:ffrpr}, I show how the above description leads to a natural notion of (non-Hermitian) field operators \textit{both in NRQM and RQM}. Here we see the first glimpse of an approach in which a  natural transition  from QFT to NRQM could be possible entirely in terms of field operators. We do not use position \textit{operator} $\hat x^\alpha$ at all and both $t$ and $\bm{x}$ remain c-number labels, even after we have obtained the NRQM. The propagator obtained in Sec.\ref{sec:hofp} can be expressed in terms of the field operators, again,  both in NRQM and in QFT. In the relativistic case, the field operators are Lorentz invariant but they do not commute on spacelike surfaces. Hence they cannot be used to construct physical observables directly. (This requires some more work  and leads to  the notion of antiparticle \textit{both in QFT and NRQM};  see Sec.\ref{sec:seamless}.) 
 
 The discussion in these two Secs. \ref{sec:hofp} and \ref{sec:ffrpr} tells us that: (i) Lorentz invariance or the square root in the Hamiltonian do not introduce any serious conceptual difficulties in developing RQM. (ii) The fact that particles are non-localizable in RQM leads to difficulties in defining the position eigenkets but these difficulties can be handled by working in momentum representation and introducing necessary Fourier transforms. (iii) But when we do that, the resulting propagator $G_+$ does not satisfy the composition law necessary for it to propagate a wave function. In fact, we cannot even properly define $\psi(x)=\amp{t,\bm{x}}{\psi}$ in the absence of position eigenkets $\ket{t,\bm{x}}$. (iii) The formalism leads to the concept of a field operator  both in NRQM and QFT but we run into trouble with the notion of causality in QFT. This is related to the particle states not being localizable but, as we shall see later, the issue is deeper and is linked to the existence of antiparticles. 
 
 In Sec.\ref{sec:prfrpi} we look at the  same (free particle) system, described by a Hamiltonian $H(|\bm{p}|)$, from the path integral perspective. In Sec.\ref{sec:hpi}, I show how the \textit{Hamiltonian} path integral is indeed straightforward to evaluate for such systems --- even for the relativistic case with a square root Hamiltonian.  If you use the standard measure $d^n\bm{x} d^n\bm{p}$ in the Hamiltonian path integral, you get the correct answer in NRQM; but, in the case of RQM, you get a propagator --- called Newton-Wigner propagator --- which is \textit{not} Lorentz invariant.  It is possible to tinker with the path integral measure --- taking a cue from our discussion in Sec.\ref{sec:hofp} --- and arrange matters so that the resulting propagator \textit{is} Lorentz invariant. This procedure again leads to the same propagator $G_+(x_2,x_1)$ obtained earlier. This  also means that we inherit all the difficulties encountered earlier. 
 
 In Sec.\ref{sec:lpi}, I study the same system using a \textit{Lagrangian} path integral. Again, there is a natural way of defining the measure for this path integral which leads to the correct result in NRQM. The same procedure, when applied to the relativistic Lagrangian, leads to nonsense --- that is, the path integral does not exist for any choice of the measure. The fact that, for the relativistic particle,  the Hamiltonian path integral exists while the Lagrangian path integral does not can be traced to the structure of the Hamiltonian. One can write down a general condition which must be satisfied by the Hamiltonian if the Lagrangian and Hamiltonian approaches have to lead to the same result. The square root Hamiltonian of the relativistic particle violates this condition. This is probably the only occasion in which the square root in the Hamiltonian leads to a serious technical difficulty. 
 
 There is, however, another --- rather elegant --- procedure for defining the Lagrangian path integral for a relativistic particle. This makes use of the geometric interpretation of the relativistic action as the path length in the Euclidean space. You can then define the path integral in an Euclidean lattice and obtain a continuum limit using a natural regularization.  I do this in Sec.\ref{sec:lattice} and show that the resulting propagator $G_R(x_2,x_1)$ is the standard Feynman propagator in QFT with the Fourier space representation:
 \begin{equation}
G_R(p)=\int d^Dx G_R(x)\exp(i p_ax^a)=\frac{i}{(p^2-m^2+i\epsilon)}                                                                 
\end{equation} 
In Sec.\ref{sec:jacobi}, I show that this particular path integral approach for the relativistic case is very similar to the path integral based on the Jacobi action for  a non-relativistic free particle. This mathematical identification clarifies several peculiar features of the Feynman propagator.
 I also discuss briefly some aspects of reparametrisation invariance and its connection with the Jacobi action. 
 
 Obtaining the Feynman propagator from a path integral prescription is gratifying but this does not again help in our task of obtaining NRQM from QFT.
 In Sec.\ref{sec:nroff}, I discuss the non-relativistic limit of $G_R(x_2,x_1)$ and show that it does \textit{not} reduce to the propagator $G_{\rm NR}(x_2,x_1)$ of NRQM. So while the lattice regularization provides a natural way of obtaining $G_R(x_2,x_1)$, it does not help us in obtaining the NRQM limit in a seamless manner.
 Once again, we cannot use $G_{\rm R}(x_2,x_1)$ to propagate a relativistic wave function because $G_{\rm R}(x_2,x_1)$ does not obey the correct composition law and does not have the appropriate limit.
 In Sec.\ref{sec:compo}, I provide a brief discussion of the different composition laws obeyed by relativistic and non-relativistic propagators and how the relativistic composition law goes over to a non-relativistic one in the $c\to\infty$ limit. This discussion clarifies several issues discussed in the literature. 
 
 In Sec.\ref{sec:hofp},  we  have obtained $G_+(x_2,x_1)$ as a matrix element of a time evolution operator provided the states $\ket{\bm{x}}$ are defined via Fourier transform from the eigenkets $\ket{\bm{p}}$ of the momentum operator. On the other hand, $G_R(x_2,x_1)$ was obtained in Sec.\ref{sec:lattice}  from a lattice regularization procedure, applied to the path integral,  and it is not clear whether it is also a matrix element of the time evolution operator. Strictly speaking, it is not. However, it is possible to express it as such a matrix element using a particular integral representation of the time evolution operator. I do this in Sec.\ref{sec:fmatrix} and show how this approach connects up with the  discussion in Sec.\ref{sec:jacobi}. 
 
 These results show how difficult it is to obtain the NRQM from QFT in a  straightforward manner. We run into difficulties  both in the Hamiltonian approach and in the path integral approach. The lattice regularization of the relativistic path integral does lead to the QFT propagator $G_R(x_2,x_1)$. But this propagator does not have a single particle NRQM limit. This is to be expected because $G_R(x_2,x_1)$ contains information about \textit{both} particles and anti-particles. In the NRQM limit, it should therefore represent the dynamics of \textit{both} the particle and the anti-particle rather than just a single particle. I show how this result arises --- thereby answering the question raised in the subtitle of this paper!--- in the last two sections. 
 
 Sections \ref{sec:seamless} and \ref{sec:prcor} identify the necessary ingredients for the NRQM to arise in the appropriate limit of QFT. This is done by using a \textit{pair} of field operators rather than a single relativistically invariant operator. Such a pair restores microscopic causality in QFT and collectively describes a particle-antiparticle system. This behaviour survives in the NRQM limit and we obtain the Schroedinger equation for \textit{two} field operators, one describing the particle and the other describing the anti-particle. They co-exist at equal footing in the NRQM limit.

So, I have good news and bad news. Good news is that one \textit{can} obtain NRQM, as a limiting case of QFT, if --- but only if --- we interpret NRQM in terms of a field operator satisfying the Schroedinger equation \textit{\`{a} la} (what is usually called, quite misleadingly, as) the ``second quantized'' approach. The bad news is that you \textit{cannot} get the standard formalism (viz. the stuff we teach kids in QM101, in which $x^\alpha$ and $p_\beta$ are treated as operators and $\psi(t,\bm{x})=\amp{t,\bm{x}}{\psi}$ is a ``wave function'' etc.) as a natural limiting case of QFT. Section \ref{sec:impli} discusses some of the broader implications of this result.

While the main focus of this paper is on the conceptual issues (and it does clarify and highlight several of them), there are also many interesting results of technical nature which either do not exist in the previous literature or not adequately discussed. I mention below some of them:

(a) The Hamiltonians for both the relativistic and the non-relativistic (free) particle depends only on their momentum.  Section \ref{sec:hofp} discusses such systems, for which $H(\bm{p},\bm{x})=H(|\bm{p}|)$, in a \textit{unified} manner and identifies the reasons why, in spite of this simplicity, we do not have an RQM but we have an NRQM. The unified, focused, discussion should have found a place in textbooks but it had not.

(b) The most natural way of defining a path integral, either from a Hamiltonian $H(\bm p)$ or from a Lagrangian $L(\dot{\bm x})$, is by time-slicing. (We look for other ``sophisticated''  methods only when this approach fails but alas, often, without investigating \textit{why} exactly it failed!). Section \ref{sec:hpi} explains what happens (or goes wrong) when you attempt time slicing with the Hamiltonian for a relativistic particle; I have not seen such an explicit discussion, e.g., about the issues regarding choice of measure, see \eq{measure1}, in published literature. Section \ref{sec:lpi} takes up the corresponding question in the case of the Lagrangian path integral. I show that there is a natural way of defining the time-sliced path integral leading to \eq{gengofp} and use this to clearly contrast the NR case with the relativistic case. I have not seen such a discussion --- leading to e.g., \eq{disaster1} and the discussion in the  two paragraphs following \eq{disaster1} --- in the literature.

(c) One consequence of the above analysis is the following: It clearly shows that the Hamiltonian and the Lagrangian time-slicing procedures are not equivalent --- another fact which is inadequately stressed in literature. I also identify the formal condition, \eq{hlequiv}, for their equivalence which I have not seen in the literature, at least not in this context (though it night exist buried somewhere in the literature on formal path integral techniques).

(d) Much of the discussion in different subsections of Sec. \ref{sec:lattice} is new. In particular, the discussion in Sec. \ref{sec:comlattice} leading to e.g.,  to the interpretation in \eq{tpnotefour}, the NR limit of lattice regularization in \eq{14oct9} to \eq{14oct17}, comments in the last paragraph of Sec. \ref{sec:jacobi} leading to \eq{notaction} are either entirely new or highlights aspects inadequately discussed in the literature.

(e) Section \ref{sec:nroff} shows that you cannot get the NR propagator from the $c\to\infty$ limit of the Feynman propagator. Again, I have not seen an explicit discussion of this (correct) result in the textbook literature. The result in Sec. \ref{sec:fmatrix} is new and clarifies the structure of the $G_F$ from an alternative point of view.

(f) Section \ref{sec:seamless} emphasizes the fact that the standard KG field is built from \textit{two} fields which in the NR limit represent the particle and the antiparticle. This, by itself, is not new and exists in several textbooks including my own \cite{tpqft}. But it assumes importance in the context of \eq{mystery1} which \textit{I claim nobody understands}, in spite of it being  the key equation in QFT, allowing the formalism to work. The fact that the path integral for, ostensibly, a single relativistic particle actually describes the propagation of \textit{two} particles is the key issue here and the discussion in Sec. \ref{sec:seamless} provides the backdrop for it.

Of course, these technical results are just the trees in the wood of conceptual discussion and, hopefully, the reader will not miss the latter for the former.

\section{Quantum theory of a  system with the Hamiltonian $H(\bm{p},\bm{x})=H(|\bm{p}|)$}\label{sec:hofp}

The classical dynamics of a free  particle is completely described by the action which has no explicit dependence on the space or time coordinates:
 \begin{equation}
  A=\int dt L(\dot {\bm{x}})=\int dt[\bm{p}\cdot \dot{\bm{x}} -H(\bm{p})]
  \end{equation}
 in terms of  a well-defined Lagrangian $L(\dot {\bm{x}}) = L(|\dot{\bm{x}}|)$ or a Hamiltonian $H(\bm{p}) =H(|\bm{p}|) $. In the case of a \textit{non-relativistic} free particle we take:
  \begin{equation}
   L_{NR}(\dot{\bm{x}})= \frac{1}{2} m \dot{\bm{x}}^2; \qquad
  H_{NR}(\bm{p})=\frac{\bm{p}^2}{2m}
 \end{equation} 
 while, for the \textit{relativistic} free particle, we have:\footnote{The Hamiltonian for a classical relativistic particle is positive definite and  the square root is taken with positive sign in $H(\bm{p})=+(\bm{p}^2+m^2)^{1/2}$. This is the classical system we want to quantize --- not a strange one with $H(\bm{p})=\pm(\bm{p}^2+m^2)^{1/2}$ which has no classical meaning.}
  \begin{equation}
   L_{R}(\dot{\bm{x}})= - m (1-\dot{\bm{x}}^2)^{1/2}; \qquad
  H_{R}(\bm{p})=(\bm{p}^2+m^2)^{1/2}
 \end{equation} 
  In either case, the Lagrangian and the Hamiltonian  are independent of $\bm{x}$ and we can deal with both of them at one go. The classical equations of motion are easy to solve leading to $\bm{p}=\bm{p}_0=$ constant, $x^\alpha(t)={F}^\alpha t
  +x^\alpha(0)$,  where ${F}^\alpha\equiv (\partial {H}/\partial {p}_\alpha) = $ constant. That is the end of the story. 

What about the quantum theory? If one does not bring in any extra symmetry considerations, then the quantum theory of any system with $H=H(|\bm{p}|)$ is also trivial in the Heisenberg picture. We upgrade the position and momentum to operators satisfying the commutation rule
 $[x^\alpha, p_\beta]=i\delta^\alpha_\beta$ which can be concretely implemented --- in the space of normalizable complex functions --- in the \textit{momentum} representation with $\hat{x}^\alpha = i \partial/\partial p_\alpha$. Since the Hamiltonian commutes with momentum, $\hat{\bm{p}}(t) = \hat{\bm{p}}(0)$. It is trivial to integrate the operator equation for $x^\alpha$  and obtain $\hat{x}^\alpha(t) = \hat{F}^\alpha t + \hat{x}^\alpha(0)$ where $\hat{F}^\alpha\equiv (\partial \hat{H}/\partial \hat{p}_\alpha) = $ constant. Since we have solved  the operator equations, we can answer any question about the quantum dynamics. Obviously, this procedure should work for $ H_{NR}(\bm{p})=\bm{p}^2/2m$ as well as for $H_{R}(\bm{p})=(\bm{p}^2+m^2)^{1/2}$. 
 
 So it is not the form of the Hamiltonian which creates problems when we try to construct relativistic quantum mechanics (RQM) of a free, single, particle. But we do know that combining principles of special relativity and quantum theory does require more drastic modifications of the description and, in fact, we cannot have a viable, single particle quantum theory based on, say, a relativistically invariant wave function. The question arises as to why this is the case.

When you move from NRQM to RQM, there are \textit{two}  new ingredients which come in. First, the Hamiltonian for a free particle changes from $H_{\rm NR}(\bm{p}) = \bm{p}^2/2m$ to $H_R=+(\bm{p}^2 + m^2)^{1/2}$ with corresponding changes in the dynamical equations. Second, we want the physics to respect Lorentz invariance rather than Galilean invariance. 
As we have seen above, the square root structure of the Hamiltonian does not create any new conceptual issues when we use the momentum representation and Heisenberg picture.\footnote{If you attempt to write the Schroedinger equation with $H_R$ in the coordinate representation, the square root in the Hamiltonian will make the equation non-local. But then, if you write the Schroedinger equation for a non-relativistic particle moving under the action of  a non-polynomial potential $V(\bm{x})$ in the \textit{momentum} representation, you will again get a non-local Schroedinger equation. So this by itself is not a conceptual difficulty; but merely a technical nuisance.}
The next suspect, of course, is the requirement of Lorentz invariance. As we shall see,
 the issue of maintaining Lorentz invariance requires having the correct, relativistically invariant, integration measure in the momentum space when we describe, say, the momentum eigenstates of particles.  Roughly speaking, you can ensure that a classical theory is relativistically invariant, if you ensure that the dynamical equations are relativistically invariant. But in quantum theory, you need to ensure that \textit{both} the dynamical equations (for the operators in Heisenberg picture, say) as well as the description of quantum states in the Hilbert space are relativistically invariant.\footnote{It is possible to address some of these issues, very formally, in terms of the structure of the Lorentz group and Galilean group and their interrelationship. We will, however, adopt a more transparent and down-to-earth approach in this paper.} The first requirement --- viz. relativistic invariance of dynamical equations --- can be ensured by using a relativistically invariant action or Hamiltonian; but the second requirement  does not have a direct analogue in \textit{classical} relativistic mechanics. We will see that this requirement is the root cause of several nontrivial features in QFT.
  We will now see in some detail the mathematical consequences of these requirements.
  
  \subsection{Propagators in momentum and coordinate spaces}
 
 Since a Hermitian momentum operator has  to exist for the proper definition of $ H(\bm{p})$, we start by introducing a complete set of orthonormal momentum eigenkets, $\ket{\bm{p}}$ which must exist for any system described by a Hamiltonian of the form $ H(\bm{p})$, including NRQM and RQM. We would then like $\amp{\bm{p}'}{\bm{p}}$ to be proportional to $\delta (\bm{p}- \bm{p}')$. This works in NRQM but the integration over $d^n\bm{p}\delta (\bm{p}- \bm{p}')$ is not Lorentz invariant. The relativistically invariant measure for momentum integration is $d\Omega_{\bm{p}} \equiv d^n \bm{p}/(2\pi)^n (1/\Omega_{\bm{p}})$ with $\Omega_{\bm{p}} = 2\omega_{\bm{p}}$. So we need to  postulate:
 \begin{equation}
 \amp{\bm{p}'}{\bm{p}} = (2\pi)^n \Omega_{\bm{p}}\ \delta (\bm{p}- \bm{p}'); \qquad d\Omega_{\bm{p}} \equiv \frac{d^n \bm{p}}{(2\pi)^n} \frac{1}{\Omega_{\bm{p}}}
 \label{four}
\end{equation} 
so that $\amp{\bm{p}'}{\bm{p}} d\Omega_{\bm{p}} = \delta (\bm{p}'- \bm{p}) d^n \bm{p}$.
 In NRQM we can take $\Omega_{\bm{p}}$ to be a constant, or even unity; but in RQM the Lorentz invariance of the measure for momentum integration $d\Omega_{\bm{p}}$ requires the factor
 $\Omega_{\bm{p}} = 2\omega_{\bm{p}}$. By keeping the choice of $\Omega_{\bm{p}}$ unspecified in the algebraic expressions we take care of both the cases at one go; further, in the non-relativistic limit, $\omega_{\bm{p}}$ can be approximated by the constant $m$ allowing us to take the limit seamlessly. 
 With this definition, the resolution of unity and the consistency condition on the momentum eigenkets become:
 \begin{equation}
1 \equiv \int d\Omega_{\bm{p}'} \ket{\bm{p}'}\bra{\bm{p}'};\qquad 
\ket{ \bm{p}} \equiv \int d\Omega_{\bm{p}'} \ket{\bm{p}'}\amp{\bm{p}'}{\bm{p}}
\label{fortynine}
\end{equation} 
 These relations can be taken care of by the choices in \eq{four}.
 In the integration measure as well as in the Dirac delta function, we have introduced a factor $\Omega_{\bm{p}}$ which, of course, cancels out in the right hand side of th second relation in \eq{fortynine}. 
 
 Given these momentum eigenstates, we can  define a natural \textit{momentum space} propagator by the rule:
 \begin{equation}
 G(t_b,\bm{p}_b; t_a,\bm{p}_a) \equiv \bk{\bm{p}_b}{e^{-it \hat{H}(\bm{p})}}{\bm{p}_a}
 =(2\pi)^n \Omega_{\bm{p}_b}\ \delta (\bm{p}_a- \bm{p}_b)\exp-itH(\bm{p}_b);
 \label{propdef2}
\end{equation}
where $t\equiv t_b-t_a$. Given any arbitrary state $\ket{\phi}$ in the Hilbert space we can ``propagate'' the complex function $\phi(t_a,\bm{p}_a)\equiv \amp{\bm{p}}{\phi}$ by this propagator:
\begin{equation}
 \phi(t_b,\bm{p}_b)=\int d\Omega_{\bm{p}_a} G(t_b,\bm{p}_b; t_a,\bm{p}_a)\phi(t_a,\bm{p}_a)=\phi(t_a,\bm{p}_b)\exp-itH(\bm{p}_b)
 \label{momevl}
\end{equation} 
So the momentum space evolution is just a change in phase.
Since momentum operator generates translation in space, it seems natural to introduce
a \textit{position} space propagator by the definition:
 \begin{equation}
 G(t_b,\bm{x}_b; t_a,\bm{x}_a)\equiv \int d\Omega_{p_a}d\Omega_{p_b}G(t_b,\bm{p}_b; t_a,\bm{p}_a)\exp i(\bm{p}_b \cdot \bm{x}_b -\bm{p}_a \cdot \bm{x}_a)
 \label{gdef1}
\end{equation}
Using \eq{propdef2} in \eq{gdef1} and
performing the integrations, we get the  propagator, $G(x)\equiv G(t_b,\bm{x}_b; t_a,\bm{x}_a)$ where $x=x_b-x_a$, for both NRQM and RQM at one go,
 in the form:
 \begin{equation}
G(x) = \int  d\Omega_{\bm{p}} \, \exp(- i p\cdot x) = \int \frac{d^n \bm{p}}{(2\pi)^n}\ \frac{1}{\Omega_{\bm{p}}} \, \exp(-ip\cdot x)
\label{ggen}
\end{equation} 
where we have introduced the 4-component object (in both NRQM and RQM) by
$
 p^a = (H(\bm{p}), \bm{p})
$ which, of course, is a genuine four-vector in RQM and just a convenient notation in NRQM.
For later reference, note that the standard spatial Fourier transform (defined with the measures $d^n\bm{x}$ and $d^n\bm{p}/(2\pi)^n)$ of this propagator is given by:
\begin{equation}
 G_{\bm{p}} (t)\equiv  \int d^n\bm{x} G(t, \bm{x}) \, e^{-i\bm{p\cdot x}} = \frac{1}{\Omega_{\bm{p}}}\, e^{ - i tH(\bm{p})}
 \label{spaceft}
\end{equation}
Let us now consider the two cases, NRQM and RQM. 
In NRQM we get:
 \begin{equation}
 G_{\rm NR}(x) = \int \frac{d^n \bm{p}}{(2\pi)^n} \, \exp\left[i\left(\bm{p\cdot x} - \frac{p^2}{2m} t\right)\right]
 = \frab{m}{2\pi it }^{n/2} \exp\frab{im|\bm{x}|^2}{2t} 
 \label{nrqmg}
\end{equation} 
and in RQM we have,\footnote{While discussing the general expressions for the propagator, valid in both NRQM and RQM, we will denote it by $G(x)$ with no subscripts. The propagator in NRQM is unique and will be noted by $G_{NR}$. In RQM and QFT, we will encounter different types of propagators denoted with different subscripts. This particular one carries the subscript $+$, since it is made of positive frequency solutions of the KG equation.} with $x^2 \equiv x_a x^a$:
\begin{equation}
 G_+(x)\equiv 
\int  \frac{d^n\bm{p}}{(2\pi)^n}\,\frac{1}{2\omega_{\bm{p}}} \, \exp(- i p.x)=F(x^2)
 \label{rqmg}
\end{equation}
which is clearly Lorentz invariant. For spacelike separations, $F$ can be expressed in terms of a Bessel function and decays exponentially; for timelike separations, it can be expressed in terms of a Hankel function and oscillates; it has a singular behaviour on the light cone. (See e.g., \cite{tpqft}). So obtaining a Lorentz invariant propagator is not an issue at all. If we take the $c\to\infty$ limit of 
$ G_+(x)$, we get:
\begin{equation}
 \lim_{c\to\infty}G_+(x)=\frac{e^{-i(mc^2)t}}{2m}\left[G_{NR}-\frac{i\hbar}{mc^2}\frac{\partial G_{NR}}{\partial t}+ ....\right]
 \label{nrlim1}
\end{equation} 
In this expression, the overall factor $(1/2m)$ is irrelevant; the factor $e^{-i(mc^2)t}$ is unavoidable because the rest energy $mc^2$ will always contribute  to the phase. The second and higher order terms within the  square bracket in \eq{nrlim1} vanish in the $c\to\infty$ limit. So one can think of the non-relativistic propagator being recovered  in the limit:
\begin{equation}
 \lim_{c\to\infty} [(2m) e^{i(mc^2)t}]G_+(x)=G_{NR}
 \label{plustonr}
\end{equation} 
which seems reasonable.
So far, so good.

\subsection{The problems in defining localized particle states}

We would, however, like to think of this real space propagator, \textit{defined} though the Fourier transform in \eq{gdef1} to be the same as the matrix element of the time evolution operator:
\begin{equation}
 G(x_b,x_a)=G(t, \bm{x}) = \bk{\bm{x}_b}{e^{-it \hat{H}(\bm{p})}}{\bm{x}_a} 
 \label{propdef1}
\end{equation}
for some suitable states $\ket{\bm{x}}$. 
  To do this we need to introduce the states $\ket{\bm{x}}$ labeled by the spatial coordinates. In NRQM they could be thought of as the eigenkets of the operator $\hat{\bm{x}}(0)$. For a more general system described by an arbitrary $H(\bm{p})$ like, for e.g. in RQM, we do not have the natural notion of such a position operator. But we can take a cue from the previous results and use
 the property that the momentum operator is the generator of spatial translations (which holds both in NRQM and RQM) to define $\ket{\bm{x}}$ along the following lines:\footnote{Here, as well as in most of the discussions which follow, we are interested in expressions at a given time $t$, taken to be $t=0$, for convenience. The notion of a state $\ket{x}=\ket{t,\bm{x}}$ such that $\ket{\bm{x}}\equiv\ket{0,\bm{x}}$ will be introduced later in
 \eq{nintynine}.}
 \begin{equation}
 \ket{ \bm{x}} \equiv e^{-i\bm{x}\cdot \hat{\bm{p}}}|\bm{0}\rangle = \int d\Omega_{ \bm{p}} e^{-i\bm{p\cdot x}}C_{ \bm{p}} |\bm{p}\rangle ; \qquad C_{ \bm{p}} \equiv \amp{\bm{p}}{\bm{0}};\qquad \amp{\bm{p}}{\bm{x}} = C_{\bm{p}} e^{-i\bm{x\cdot p}}
 \label{fifty}
\end{equation} 
 This defines $\ket{\bm{x}}$ in terms of a single function $C_{\bm{p}}$. Inserting a complete set of momentum eigenstates in the matrix element in \eq{propdef1}, and using the last relation in \eq{fifty}, we can evaluate the propagator explicitly in terms of $C_{\bm{p}}$. We get:
 \begin{equation}
 G(x) = \int d\Omega_{\bm{p}} | C_{\bm{p}}|^2 \exp(- i p.x)
 \label{fiftysix}
\end{equation} 
where we have again defined the 4-component object
$
 p^a = (H(\bm{p}), \bm{p})
$ taking care of both NRQM and RQM. 

In NRQM, it is natural to take the measure in the momentum space integration with $\Omega_{\bm{p}} =$ constant; similarly, we can also set $C_{\bm{p}}=1$. With these choices and using $H_{\rm NR} = \bm{p}^2/2m$ in \eq{fiftysix}, we immediately obtain the NRQM propagator given by \eq{nrqmg}.
In RQM, we want to obtain a Lorentz invariant propagator. In \eq{fiftysix}, the measure $d\Omega_{\bm{p}}$ as well as the function $\exp(-ip\cdot x)$ are Lorentz invariant. Therefore, the propagator will be Lorentz invariant if we take $C_{\bm{p}}=$ constant. It is conventional to scale things so that  $C_{\bm{p}}=1$. Then the propagator is given by the  expression in \eq{rqmg}.
 We have thus arrived at a Lorentz invariant propagator for RQM which can also be interpreted as the matrix element of the time evolution operator through \eq{propdef1}. Unfortunately, the situation is not so simple when we study it more closely.

To begin with, note that the only difference between the relativistic and non-relativistic propagators is in the  $(1/\Omega_{\bm{p}})$ factor which we can take to be a constant (or even unity) in NRQM but is $(1/2\omega_{\bm{p}})$ in QFT. As we shall see, this makes all the difference.
 From the definition of $\ket{\bm{x}}$ in \eq{fifty}, it follows that:
 \begin{equation}
  \amp{\bm{y}}{\bm{x}} = \int d\Omega_{\bm{p}}  \, e^{-i\bm{p \cdot (x-y)}} |C_{\bm {p}}|^2
  \label{fiftytwo1}
\end{equation} 
If you want localized particle positions, this expression should be proportional to a Dirac delta function. This in turn requires $|C_{\bm {p}}|^2=2\omega_p$ to give $d\Omega_{\bm{p}}|C_{\bm {p}}|^2=[d^n\bm{p}/(2\pi)^n]$. But we get a Lorentz invariant propagator from \eq{fiftysix} only if $|C_{\bm {p}}|^2=$ constant in \eq{fiftysix}! So, while the propagator defined through \eq{propdef1} can be made Lorentz invariant, we do not know \textit{what} it propagates because $\ket{\bm{x}}$ do not represent localized particle states! (The difficulty in localizing particles states in RQM is discussed extensively in the literature; see, e.g., \cite{F,F12,F13,F14,F15}.)

Furthermore, with this  Lorentz invariant choice $C_{\bm{p}}=1$ we also have the result
\begin{eqnarray}
 \int d^n \bm{x} \, \ket{\bm{x}}\amp{\bm{x}}{\bm{p}} &=& \int d^n \bm{x} \, e^{i\bm{p \cdot x}} \ket{\bm{x}}
 =\int d^n \bm{x} \,e^{i\bm{p \cdot x}}\int d\Omega_{q'} \, e^{-i\bm{q \cdot x}} \ket{\bm{q}}\nonumber\\
 &=& \int d^n {q} \,\frac{1}{\Omega_q}\, \delta(\bm{p}-\bm{q})\ket{\bm{q}} = \frac{1}{\Omega_{\bm{p}}} \ket{\bm{p}}
 \label{twofour}
\end{eqnarray} 
So we cannot use the states $\ket{\bm{x}}$ for the resolution of identity.  Equation (\ref{twofour}) also shows that it is the combination $\Omega_{\bm{p}}d^n \bm{x}$ rather than 
$d^n \bm{x}$ which behaves better. For example, while 
 the measure of integration $d^n \bm{x}$ is not Lorentz invariant, the combination  $\Omega_{\bm{p}}d^n \bm{x}$ is. (We will discuss this aspect in greater detail later on.) In the case of $d^n \bm{p}$, we could work from the Lorentz invariant combination $d^4p \delta(p^2-m^2) \theta(p^0)\propto d^n \bm{p}/2\omega_p$ but there is no natural analogue\footnote{Taking a cue from momentum space, one can redefine the integration measure as $d^4x \delta(x^2-\mu^2)$ where $\mu$ specifies the spacelike hypersurface $t^2-\mathbf{x}^2=\mu^2$. I have explored this possibility \cite{tpun}  but it leads to problems.} for that in the case of $d^n \bm{x}$.  The best one can do is to write, for any state $\ket{\psi}$ the  relation:
\begin{equation}
 \ket{\psi}=\int d\Omega_{\bm{p}}\ket{\bm{p}}\amp{\bm{p}}{\psi}=\int d\Omega_{\bm{p}}\int [d^n \bm{x}\ \Omega_{\bm{p}}]\ket{\bm{x}}\amp{\bm{x}}{\bm{p}}\amp{\bm{p}}{\psi}
 =\int \frac{d^n \bm{p} d^n \bm{x}}{(2\pi)^n}\ket{\bm{x}}\amp{\bm{x}}{\bm{p}}\amp{\bm{p}}{\psi}
 \label{measure2}
 \end{equation} 
 which is Lorentz invariant if the left-hand-side is. 
So there is some kind of resolution of identity in \textit{phase} space:
\begin{equation}
1=\int \frac{d^n \bm{p} d^n \bm{x}}{(2\pi)^n}\ket{\bm{x}}\amp{\bm{x}}{\bm{p}}\bra{\bm{p}}
 \end{equation}
 but not in normal space.
We will come across this combination again later, while computing phase space path integrals.
Note, for future reference, that the natural extension of $\ket{\bm{x}} = \ket{0,\bm{x}}$ for $t\neq 0$  is defined as the state $\ket{x} = \ket{t,\bm{x}}$ through  
the relation 
\begin{equation}
\ket{x} \equiv  \ket{t,\bm{x}}\equiv e^{i Ht} \,\ket{\bm{x}} = \int d\Omega_p \ e^{i{p . x}} \ket{\bm{p}}
\label{nintynine}
\end{equation} 
  where $H(\bm{p})$ is the Hamiltonian.\footnote{The sign in the exponential is correct and gives $\bra{t,\bm{x}}=\bra{0,\bm{x}}\exp(-iHt)$ which is the correct relation; see e.g., 1.2.2 of Ref. \cite{tpqft}.}

The propagator we have obtained also has another nice property which arises directly from the definition in \eq{propdef2}. It satisfies the \textit{first order} differential equation 
 \begin{equation}
 (i \partial_t - H(\bm {p}) )\, G =0
 \label{itg}
\end{equation} 
for any $H(\bm {p})$. In the specific case of the relativistic free particle, the structure of \eq{rqmg} tells you that it \textit{also} satisfies the equation
\begin{equation}
[\partial_a\partial^a + m^2] G_+(x)=0
\label{boxg}
\end{equation} 
The zeros in the right hand sides of \eq{itg} and \eq{boxg} are closely related to the fact that the definition in \eq{propdef1} --- as well as the form of the final propagator --- is valid for both $t>0$ and $t<0$ . \textit{Nowhere did we assume that $t>0$ to obtain the form of the propagator.} 
The time evolution operator in quantum theory 
 $U(t_2,t_1) \equiv \exp[-iH(t_2-t_1)]$ evolves a state from $t=t_1 $ to $t=t_2$ irrespective of the time ordering of $t_2$ and $t_1$; that is, this is a valid evolution operator for \textit{both} $t_2>t_1$ and $t_2<t_1$. For example, in NRQM, given a wave function $\psi(t,\bm{x})$ we can determine the wave function at all the earlier times and later times.\footnote{The same results hold even in a relativistic theory where $H(t_2-t_1)$ will be replaced by an integral of $d\Sigma^a P_a$ over a spacelike hypersurface $\Sigma$ of the four-momentum $P_a$ and the evolution proceeds from one spacelike hypersurface to another.}
 Therefore the expression for the  propagator in NRQM, defined as the matrix element $\bk{\bm{x}_2}{U(t_2,t_1)}{\bm{x}_1}$, is valid for both $t_2>t_1$ and $t_2<t_1$.

Sometimes it is convenient to \textit{define} another propagator by multiplying $G$ by a theta function in time, getting $U(x_2,x_1)=\theta(t)G(x_2,x_1)$ which will satisfy the differential equation
 \begin{equation}
 (i \partial_{t_2} - H)\, U =i\delta(t_2-t_1)\amp{\bm{x}_2}{\bm{x}_1}
\end{equation}
The right hand side will reduce to $i\delta(x_2-x_1)$ in NRQM but not in the relativistic theory.
When we bring in Lorentz invariance, we run into trouble regarding the time ordering. The notion of, say, $t_2>t_1$ is well-defined only if the events $x_2$ and $x_1$ are separated by a time-like interval. When the events are separated by a spacelike interval, we can always choose a Lorentz frame such that $t_2 = t_1$ and hence $G(x_2,x_1) = \amp{\bm{x}_2}{\bm{x}_1}$. If  $G(x_2,x_1) $ does not vanish for spacelike intervals, then multiplying $G(x_2,x_1)$ by $\theta (t_2-t_1)$ will not lead to a Lorentz invariant construct. 

\subsection{Propagator does not propagate the wave functions}

The  reason why this propagator $G_+(x)$, (in spite of (i) being  defined as a time evolution operator for the relativistic Hamiltonian through \eq{propdef1} and (ii)
 Lorentz invariant), cannot be used to define a  single particle RQM is the following: We cannot use it to propagate a wave function with standard probabilistic interpretation in real space. To see this let us recall how this becomes feasible in NRQM.
The dynamics of a free particle in NRQM  can be described using the propagator $G_{\rm NR}(x_b, x_a)$ which relates the Schroedinger wave function at two different times through the relation:
\begin{equation}
 \psi(x_b) = \int d^n \bm{x}_a\ G_{\rm NR}(x_b, x_a) \psi(x_a)
 \label{nrprop} 
\end{equation} 
This provides the physical interpretation for $G_{\rm NR}(x_b, x_a)$ as the amplitude for the particle to propagate from the event $\mathcal{A}$ to the event $\mathcal{B}$. One can immediately draw two key conclusions from the existence of  a relation like \eq{nrprop}.

(1) Consistency of \eq{nrprop} in the limit $t_b\to t_a$ tells you that $G_{\rm NR}(x_b, x_a)$ must satisfy the boundary condition 
 \begin{equation}
  \lim_{t_b\to t_a} G_{\rm NR}(x_b, x_a)=\delta(\bm{x}_b - \bm{x}_a)
  \label{coin1}
 \end{equation}

(2) The propagator must satisfy the transitivity condition (also called the composition law) given by 
 \begin{equation}
  G_{\rm NR}(x_b, x_a) = \int d^n\bm{x}_1\ G_{\rm NR}(x_b, x_1)\,G_{\rm NR}(x_1, x_a)
  \label{complaw}
 \end{equation} 
 This is an extremely stringent condition on the form of the propagator $G_{\rm NR}(x_b, x_a)$. In the case of a free particle,  $G_{\rm NR}(x_b, x_a)$ must be a function of $x_b-x_a$ alone. It is then straightforward to show (see page 5 of \cite{tpqft}) that the spatial Fourier transform $G_{\rm NR}(t, \bm{p})$ must have the form 
 \begin{equation}
  G_{\rm NR}(t, \bm{p}) \equiv \int d^n \bm{x}\, G_{\rm NR}(t, \bm{x})\, \exp(-i\bm{p\cdot x}) = \exp[-itF({\bm{p}})]
  \label{spaceft1}
 \end{equation} 
 That is, $ G_{\rm NR}(t, \bm{p})$, the propagator in momentum space, is a unit norm complex function with a phase that is linear in time.

 Neither of these conditions, in \eq{coin1}, \eq{complaw}, hold for $G(x)$. The  condition in \eq{coin1} is violated because $G(0,\bm{x}_b-\bm{x}_a)=\amp{\bm{x}_b}{\bm{x}_a}$ is not a Dirac delta function; this is the same issue of $\ket{\bm{x}_b}$ not representing a localized particle state. 
 The  condition in \eq{complaw} is violated because the spatial Fourier transform of $G$, given by \eq{spaceft} is not of the form in \eq{spaceft1}. So the idea of ``propagation of a wave function'' in \eq{nrprop} does not work in RQM.
 
 It is interesting to ask how \eq{coin1} is reproduced in the non-relativistic limit. 
 Using \eq{itg}, we can rewrite \eq{nrlim1}, in the limit of $c\to\infty$, as
 \begin{equation}
 G_+(x)\approx\frac{e^{-i(mc^2)t}}{2m}\left[G_{NR}+\frac{\lambda_C^2}{2}\nabla^2 G_{NR}+ ....\right];
 \qquad \lambda_C\equiv\frac{\hbar}{mc}
 \label{nrlim2}
\end{equation} 
Taking the limit of $t_2\to t_1$ we find
\begin{equation}
G_+(\bm{x}_2-\bm{x}_1)\approx\frac{1}{2m}\left[\delta(\bm{x}_2-\bm{x}_1)+\frac{\lambda_C^2}{2}\nabla^2\delta(\bm{x}_2-\bm{x}_1) + ....\right];
 \label{nrlim3}
\end{equation} 
with a highly singular second term. This implies that
\begin{equation}
 (2m)\int d\bm{x}_1G_+(\bm{x}_2-\bm{x}_1)\psi(\bm{x}_1)\approx\psi(\bm{x}_2)-\frac{\lambda_C^2}{2}\nabla^2\psi(\bm{x}_2)
\end{equation} 
The second term is nonlocal and probes the wave function over a region of the size of the Compton wavelength $\lambda_C$. Clearly this non-localisability of the particle state is the cause for the trouble which vanishes in the $c\to\infty$ limit. So the propagator $G_+(x)$ cannot be used to propagate anything consistently in RQM.
 
 One might think that the propagation equation \eq{momevl} in momentum space should lead to similar equation in real space in terms of the Fourier transform $\psi(t,\bm{x})$ of  $\phi(t,\bm{p})$.  This is indeed true but the propagator which will appear in that expression  is \textit{not} the Lorentz invariant one, defined by \eq{gdef1}. We could define the Fourier transform $\psi(t,\bm{x})$ of  $\phi(t,\bm{p})$ with either the measure $d^n\bm{p}$ or with $d\Omega_{\bm{p}}$ and both approaches lead to similar difficulty. The $\Omega_p$ factors will come in the way when you try to translate \eq{momevl} into something like \eq{nrprop} with $G(t_b,\bm{p}_b; t_a,\bm{p}_a)$ replaced exactly by $G(t_b,\bm{x}_b; t_a,\bm{x}_a)$.
 For example, if you define $\psi(t,\bm{x})$ with the  Lorentz invariant measure as:
 \begin{equation}
  \psi(x_b)\equiv \int d\Omega_b \phi(t_b,\bm{p}_b) \exp (i\bm{p}_b\cdot\bm{x}_b)
 \end{equation} 
 and use \eq{momevl} you will find that:
 \begin{equation}
\psi(x_b) = \int d^n \bm{x}_a\ K(x_b, x_a) \psi(x_a) 
\label{psievl}
\end{equation} 
 with
 \begin{eqnarray}
  K({x}_b; {x}_a)&\equiv& \int d\Omega_{p_a}d\Omega_{p_b}\ [\Omega_{p_b}
  G(t_b,\bm{p}_b; t_a,\bm{p}_a)]\exp i(\bm{p}_b \cdot \bm{x}_b -\bm{p}_a \cdot \bm{x}_a)\nonumber\\
  &=&\int  \frac{d^n\bm{p}}{(2\pi)^n}\,  \, \exp(- i p.x) =2i\frac{\partial G_+}{\partial t}
  \label{kxaxb}
 \end{eqnarray} 
This $K({x}_b; {x}_a)$ does propagate $\psi$ but it is not  Lorentz invariant. As you can see, the extra factor of $\Omega_{p_b}$ in the integrand ensures that $K({x}_b; {x}_a)$ reduces to a Dirac delta function when $t\to0$, ensuring the consistency with \eq{psievl}. The combination $K({x}_b; {x}_a)d^n \bm{x}_a$ behaves as a Lorentz scalar though neither $K({x}_b; {x}_a)$ nor $d^n \bm{x}_a$ individually is, thereby allowing us to define $\psi$  as a Lorentz scalar. Thus we can define a propagation relation only with a propagator which is not Lorentz invariant.\footnote{In NRQM, we can treat both momentum eigenstates $\ket{\bm p}$ and position eigenstates $\ket{\bm x}$ at an equal footing while in the RQM momentum eigenstates $\ket{\bm p}$ acquires a preferred status. Notice, however, that even in textbook NRQM, there is one key difference between these descriptions. The probability density $\rho\equiv|\amp{\bm x}{\psi}|^2$ in position space satisfies a continuity equation $\partial_t\rho+\nabla\cdot\bm{j}=0$ while we do not have a corresponding continuity equation for the probability density $\bar\rho\equiv|\amp{\bm p}{\psi}|^2$ in momentum space. This is hardly emphasized in the text books.}
The propagator $K(x_b,x_a)$ is sometimes called the Newton-Wigner propagator. (For a small sample of literature dealing with Newton-Wigner states and related topics, see  
\cite{A1,B,B4,C5,D,F1,F2,F3,F4,F5,F6,F7,F8,F11}.)

This is the propagator you get if you forget all about Lorentz invariance and study a system with the Hamiltonian $H=(\bm{p}^2 + m^2)^{1/2}$ as though you are doing NRQM with this Hamiltonian. In this case,  we will be working with   $\Omega_{\bm p}=1$ in \eq{four} and will take $C_{\bm p}=1$ in \eq{fifty}. Equation~(\ref{propdef1}) will then lead to $K(x_b,x_a)$. We will also recover standard resolution of identity for the states $\ket{\bm{x}}$ in \eq{twofour} because we have set $\Omega_{\bm{p}} =1$. Everything will proceed exactly as in NRQM except for the fact that $p_0=(\bm{p}^2+m^2)^{1/2}$ in \eq{kxaxb}. This propagator will satisfy the standard composition law and the boundary condition in \eq{coin1} and \eq{complaw} which is, of course, necessary for a propagation law of the form \eq{psievl} to hold. Finally, if you take the $c\to \infty$ limit $K(x_b,x_a)$ will reduce to $G_{\rm NR}(x_b,x_a)$ (except for the understandable factor $\exp(-imt)$). So the square root in the Hamiltonian is of no real consequence in developing a quantum theory, \textit{if} you are willing to sacrifice Lorentz invariance. Needless to say, this is too high a price to pay. 

The fact that spatial integration with the measure $ d^n \bm{x}$
 is not Lorentz invariant also means that a relation like \eq{complaw} has no hope of surviving in a Lorentz invariant theory if the propagators are Lorentz invariant. The standard procedure to define invariant spatial integration is to use a (variant of the) combination like $d\Sigma^a F_1\partial_a F_2 =  d^n \bm{x} F_1\partial_0 F_2$ for two scalar functions $F_1,F_2$. This, however, does not help us to define a wave function for a relativistic particle. But it again raise the question as to how the correct composition law in \eq{complaw} is recovered in the non-relativistic limit; we will discuss this issue in Sec.\ref{sec:compo}.
 
Some of these ideas involving the states $\ket{\bm{k}}$ and $\ket{x}$ are usually expressed by introducing a one particle ``wave function''
which, as we know, is not an useful notion. Nevertheless, to connect up with previous literature, let me briefly mention how this comes about. Consider a state $\ket{\Psi}$ defined in terms of a function $F(\bm{k})$ by
 \begin{equation}
 \ket{\Psi} \equiv \int d\Omega_{\bm{k}} F(\bm{k}) \ket{\bm{k}} 
  \label{14oct18}
 \end{equation} 
 We clearly have $F(\bm{p}) = \amp{\bm{p}}{\Psi}$. Given the definition of $\ket{x}$ in \eq{nintynine}, we see that 
 \begin{equation}
 \amp{x}{\Psi} = \int d\Omega_{\bm{k}}\ e^{-ikx}\, F(\bm{k}) = \bar F(x)
 \end{equation} 
 It is  easy to show that this function $\bar F(x)$ satisfies the relativistic Schroedinger equation 
 \begin{equation}
 i\partial_t \bar F(x) = (-\nabla^2+ m^2)^{1/2} \, \bar F(x) = \hat H(\hat{\bm{p}}) \bar F(x)
 \end{equation}
 By acting on the left hand side with $i\partial_t$ again, we see that $\bar F(x)$ \textit{also} satisfies the Klein-Gordon equation $(\Box +m^2)\bar (x) =0$. The fact that $\bar F(x)$, which is analogous to single particle wave function, and the operator $A(x)$ both satisfy the Klein-Gordon equation sometimes create (avoidable) confusion in the literature.
 
 Because of the $2\omega_{\bm{k}}$ factor in the measure $d\Omega_{\bm{k}}$, the  $\bar F(x)$ is not a straightforward Fourier transform of $F(\bm{k})e^{-i\omega_k t}$ in RQM. This is also reflected in the fact that while $\ket{\Psi}$ has a straightforward expansion in terms of $\ket{\bm{k}}$, the corresponding expansion is non-local when we attempt it\footnote{Recall our notation 
 $
 \ket{\bm{x}} \equiv \ket{x}_{t=0} = \ket{0,\bm{x}}
 $.  We will use the same notation for all physical quantities.}
 in terms of $\ket{\bm{x}}$.
 The norm of the state $\ket{\Psi}$ can be expressed in two equivalent ways: 
 \begin{equation}
 \int d\Omega_{\bm{k}} \ F^*(t,\bm{k})\, F(t, \bm{k}) = i \int d \Sigma^a \bar F^*(x)\, \overleftrightarrow{\partial_a} \, F(x)
 \end{equation} 
which shows that it is fairly and natural in the momentum space but involves what is called the Klein-Gordon inner product in real space.

 \section{Fields from propagators in NRQM and RQM}\label{sec:ffrpr}
 
 The fact that the relativistic propagator does not propagate a wave function while the non-relativistic propagator does, leads to the first point of departure between the two. Even though a useful notion of wave function fails to exist in the relativistic case, the propagator does lead to a natural notion of field \textit{operators} (not c-number wave functions) in both NRQM and RQM. They can be introduced in a unified way, and as we shall see later, actually facilitate a seamless transition from QFT to NRQM. This section introduces this idea which we will explore further in Sec.\ref{sec:seamless}. 
 
 To do this, recall that the $\ket{\bm{p}}$ represents the state with a single particle having a momentum $\bm{p}$ and energy $H(\bm{p})$ both in NRQM and RQM. When  a particle is in an external field or when its interacts with other particles, it could evolve from, say, a state 
 $\ket{\bm{p}_1}$ to $\ket{\bm{p}_2}$.  Such a process can be equivalently thought of as annihilating a particle in state $\ket{\bm{p}_1}$, leading to a no-particle state, which we will denote by $\ket{0}$, followed by a creation of a particle in  $\ket{\bm{p}_2}$ from $\ket{0}$. To specify these processes, we can
 introduce a pair of operators $A_{\bm{p}}$ and $A^\dagger_{\bm{p}}$ (``creation'' and ``annihilation'' operators) which obey the following relations: 
\begin{equation}
\left[  A_{\bm{p}}, A^\dagger_{\bm{q}}\right] = (2\pi)^n \, \Omega_p \delta (\bm{p}- \bm{q});
 \qquad A_{\bm{p}}\ket{0}=0;  \qquad  \ket{\bm{p}} \equiv A^\dagger_{\bm{p}}\ket{0} 
 \label{hundred}
\end{equation} 
  The first relation defines the commutator structure of the creation and annihilation operators in the momentum space with the Dirac delta function in the right hand side defined with the invariant measure containing the factor $\Omega_{\bm{p}}$. The second relation defines the unique no-particle state $\ket{0}$ as the one annihilated by $A_{\bm{p}}$ for all $\bm{p}$. The third relation constructs the momentum eigenstate from $\ket{0} $ by the action of the creation operator. All these work \textit{both} in NRQM and RQM.
   Combining \eq{nintynine} and \eq{hundred} we find that $\ket{x}$ can be expressed in the form:
\begin{equation}
 \ket{x} =  \int d\Omega_p  A^\dagger_p \, e^{ip.x}\ket{0} \equiv A^\dagger(x)\ket{0} 
\end{equation}
where we have defined the operator:
\begin{equation}
 A(x) \equiv \int d\Omega_p \, A_p e^{-ip.x};\qquad A^\dagger(x) \equiv \int d\Omega_p \, A_p^\dagger e^{ip.x}
 \label{fop1}
\end{equation} 
  So we find that the state $\ket{x}$ can be obtained from the state $\ket{0}$ by the action of a 
  non-Hermitian ``field operator'' $A^\dagger(x)$ both in NRQM and in RQM. The propagator we obtained earlier can now be expressed in the form:
  \begin{equation}
   \amp{x_2}{x_1}=\amp{t_2,\bm{x}_2}{t_1,\bm{x}_1}=\bk{0}{A(x_2)A^\dagger(x_1)}{0}
   =\int d\,\Omega_{\bm{p}}e^{-ip.x}= G_+(x_2;x_1)
   \label{fourfour}
  \end{equation} 
 with the four component object $(\bm{p},H(\bm{p}))$. Again this relation is valid both in NRQM and RQM allowing seamless limiting process. 
 
 The  difference between NRQM and QFT is in the interpretation of the amplitude in the left hand side in \eq{fourfour}. In NRQM, the state $\ket{t_1,\bm{x}_1}$ can be defined as the eigenstate of the position operator $\bm{\hat{x}}(t_1)$ at time $t_1$ with eigenvalue $\bm{x}_1$; that is, $\bm{\hat{x}}(t_1)\ket{t_1,\bm{x}_1}=
 \bm{x}_1 \ket{t_1,\bm{x}_1}$. Such an interpretation is not possible in RQM since we do not have a suitable position operator and the states like $\ket{x}$ has to be built from $\ket{\bm{p}}$ by Fourier transform tricks. We also have the equal time result:
 \begin{equation}
  \amp{t_2,\bm{x}_2}{t_2,\bm{x}_1}=\int d\,\Omega_{\bm{p}}e^{i\bm{p}\cdot(\bm{x}_2-\bm{x}_1)}
  \end{equation} 
 which is a Dirac delta function in NRQM but not in RQM, because of the $2\omega_p$ factor in the measure, leading to issue of
 non-localisability of particle position. 
 
 It is trivial to see that the field operator defined in \eq{fop1} \textit{always} obeys the \textit{first order} differential equation:
 \begin{equation}
  [i\partial_t-H(\bm{p})]A=0;\qquad  [-i\partial_t-H(\bm{p})]A^\dagger=0
 \end{equation} 
  including \textit{both} in NRQM and in RQM.  In NRQM, it is just the Schroedinger equation. If $H=(\bm{p}^2+m^2)^{1/2}$ the field operator \textit{also} obeys the Klein-Gordon equation $\Box A(x) = 0 = \Box A^\dagger(x)$.

  A straightforward computation, using \eq{fop1} and \eq{hundred}, shows that the field obeys the commutation rule 
\begin{eqnarray}
 [A(x_2), A^\dagger(x_1)] &=& \int d\,\Omega_{\bm{p}}\int d\,\Omega_{\bm{q}} e^{-ipx_2}e^{iqx_1}[A_{\bm{p}},A^\dagger_{\bm{q}}]\nonumber\\
&=& \int d\,\Omega_{\bm{p}}e^{-ipx} \equiv G_+(x_2;x_1) =\amp{x_2}{x_1}
\label{hundredthree}
\end{eqnarray}
 On a $t_2=t_1$ spacelike hypersurface, 
$[A(t_2,\bm{x}_2), A^\dagger(t_2,\bm{x}_1)]$ is Dirac delta function in NRQM but a finite non-vanishing function in RQM. So the non-localisability of particle position has a counterpart in the field commutator as well. This, in turn, implies that if you try to construct bilinear operators from the field
and treat them as observables, they do not commute on a spacelike hypersurface. The measurement of one observable will affect the other thereby violating the relativistic notion of causality. We will see later on what it implies for RQM and --- more importantly --- for the NRQM as well.

Some of the  unnaturalness in the above expressions can be taken care of by sacrificing manifest Lorentz invariance.
 For the sake of completeness we will briefly describe these constructs and their relationship to Newton-Wigner position operator.
This is usually done by introducing a different set of creation and annihilation operators $a_{\bm{k}}, a^\dagger_{\bm{k}}$ through the relation 
 $
 [(2\pi)^n 2\omega_{\bm{k}}]^{1/2} a_{\bm{k}} \equiv A_{\bm{k}}
 $ 
 etc. A comparison with \eq{hundred} shows that these operators obey the simpler commutation rule 
 \begin{equation}
 \left[ a_{\bm{k}}, a^\dagger_{\bm{p}}\right] = \delta (\bm{k} - \bm{p})
 \end{equation} 
 which is \textit{not} Lorentz invariant. If we also define $f_{\bm{k}}$ by the corresponding rule, 
 $
 [(2\pi)^n 2\omega_{\bm{k}}]^{1/2} f_{\bm{k}} \equiv F_{\bm{k}}
 $, 
 we can write the state $\ket{\Psi}$ in \eq{14oct18} in the form 
 \begin{equation}
 \ket{\Psi} = \int d^n \bm{k}\ f(\bm{k})\ a^\dagger_{\bm{k}} \ket{0}
 \end{equation} 
 We can also define the fields $a(x), a^\dagger(x)$ in terms of $A(x), A^\dagger(x)$ in an analogous fashion. 
 While the relationship between $F(t, \bm{k})$ and $f(t, \bm{k})$ is a simple scaling in momentum space, the corresponding relationship between $\bar F(t, \bm{x})$ and $\bar f(t, \bm{x})$ is much more complicated in real space and is given by
 \begin{equation}
 \bar f(t,\bm{x}) = \int d^n \bm{x}' \, Q(\bm{x}, \bm{x}')\, \bar F(t, \bm{x}')
 \end{equation} 
 where 
 \begin{equation}
 Q(\bm{x}, \bm{x}') = \int d\Omega_{\bm{k}} \, (2\omega_{\bm{k}})^{3/2}\, e^{i\bm{k}\cdot (\bm{x}-\bm{x}')}
 \end{equation} 
 
 One reason people like to work with $a(x)$ and $a^\dagger(x)$ is that it allows defining a set of states $\ket{\bm{x}}_{\rm NW}$ as  eigenstates of a position operator called the Newton-Wigner position operator.  We define  $\ket{\bm{x}}_{\rm NW}$ through the relation 
 $
 \ket{\bm{x}}_{\rm NW} \equiv a^\dagger (\bm{x})\ket{0}
 $. It is then straightforward to verify that these states are eigenstates of an operator $\hat{\bm{x}}_{\rm NW}$, that is 
 $
 \hat{\bm{x}}_{\rm NW} \ket{\bm{x}}_{\rm NW} = \bm{x}\ket{\bm{x}}_{\rm NW}
 $
 where the Newton-Wigner position operator $\hat{\bm{x}}_{\rm NW}$ is defined as 
 \begin{equation}
 \hat{\bm{x}}_{\rm NW} \equiv \int d^n \bm{x}\, a^\dagger(\bm{x})\, \bm{x}\, a(\bm{x}) = \int d^n \bm{p}\, a^\dagger(\bm{p}) \left(i\frac{\partial}{\partial \bm{p}}\right) a(\bm{p})
 \end{equation} 
 This appears to be a natural definition both in position space and in momentum space (where $\bm{x}$ is replaced by $i\partial/\partial \bm{p}$) --- but as we have stressed several times --- it is not Lorentz invariant. If we try to re-express it in terms of Lorentz invariant operators $A_{\bm{p}}, A^\dagger_{\bm{p}}$ and the Lorentz invariant integration measure $d\Omega_{\bm{p}}$, then we get fairly complicated expressions given by
 \begin{equation}
 \hat{\bm{x}}_{\rm NW} = \int d\Omega_{\bm{p}} \, A^\dagger_{\bm{p}} \left[ i \left(\frac{\partial}{\partial \bm{p}} - \frac{\bm{p}}{2\omega_{\bm{p}}^2}\right)\right]\, A_{\bm{p}} = \int d^n \bm{x}\, A^\dagger(\bm{x}) \left[ \bm{x} + \frac{\nabla}{2(m^2 - \nabla^2)^{1/2}}\right] \, A(\bm{x})
 \end{equation} 
 which are obviously not Lorentz invariant. These features once again stress the fact that a single particle description of RQM is not easy to obtain. 

 \subsection{Aside: Some general comments}

I have taken a particular approach to demonstrate the problems which arise when one attempts to introduce a Lorentz invariant, single-particle description in RQM with a natural definition of probability. Given the importance of this issue, it is not surprising that many people have attempted to do it from many other perspectives in the past. All of them requires making some compromise and it is only fair to say that now of them appear natural. This is in fact the major reason, people adhere to the standard interpretation of QFT, in which one no longer attempts an interpretation in terms of a ``relativistic wave function''. Further discussion in this paper will confirm this point of view.

But before I proceed further, it is probably worthwhile to make some general comments about these attempts, which will further clarify the situation. The basic point is strikingly simple: In NRQM you can treat (i) the momentum operator in the position basis $\hat p_\alpha=-i\partial/\partial x^\alpha$ and (ii) the position operator in momentum basis $\hat x^\alpha=i\partial/\partial p_\alpha$ at equal footing. This is because both are unconstrained variables (in a sense which will become clear in a moment) and the corresponding measures of integration are identical in form, being proportional to $d^D\bm{x}$ and $d^D\bm{p}$. A natural generalization to RQM will be\footnote{For example, we will see later that, in the Schwinger's propertime approach one can work with the worldline $x^a(s)$. This provides a natural backdrop for introducing  the operators $\hat x^a$ etc.} to use the momentum operator in the position basis to be $\hat p_a=i\partial/\partial x^a$ and  the position operator in momentum basis to be $\hat x^a=-i\partial/\partial p^a$ (in our mostly negative signature). The essential problem is that the four-momentum is a constrained variable satisfying the condition $p^ap_a=m^2$ while the four-coordinate $x^a$ has no such constraint. This also implies a key difference between the measures of integration in coordinate and momentum spaces. As long as the mass $m$ is treated as a Lorentz invariant, scalar constant this asymmetry will always surface somewhere in the formalism.
As soon as we do this, we also have to treat the coordinate time $\hat x^0$ also as an operator with all sorts of interpretational issues. 
One invariably pays a price for such attempts, for example, in the form of having to make $m$ a variable, dynamical entity rather than retaining it as a parameter, which happens, e.g., in approaches like \cite{horwitz}.

Other attempts to handle this issue demands working with an ensemble of particles (see e.g., \cite{many1,many2} for a sample) --- rather than a single particle theory ---with several peculiar interpretational issues. In addition it it being a many-particle description, one runs into difficulties in defining center-of-mass with expected properties. What is more, the entire formalism lacks the naturalness and one wonders whether this is a remedy worse than disease. We again see that strictly single particle description with a constant mass parameter is not easy to obtain.

There is actually a fundamental reason why such issues arise and one is forced away from a constant mass description (see e.g., \cite{fb}), which I will describe very briefly. Let us assume there exist an operator $X^a$ and quantum states $\ket{\psi}$ etc such that $\bk{\psi}{X^a}{\psi}=x^a$ is the coordinates of a localized event. (I temporarily use capital letters to denote operators to avoid the clutter of adding `hats'). Then, using the facts that (a) Lorentz-Poincare transformation is to be implemented in the Hilbert space by a unitary operator and (b) knowing the transformation rule for the coordinates $x^a$, one can determine the commutation rules of the position and momentum operators. We will then find that the position operator $X^a$ does \textit{not} commute with the operator corresponding to the Casimir invariant $P^2\equiv P^aP_a\equiv M^2$. In fact, you get $[X^a,M^2]=-2iP^a$ which can lead to all sorts of trouble. For example, working in the subspace which excludes zero-mass states, we can re-write this relation as $[X^a,M]=-2iP^a/M$ which will lead to the uncertainty relation (with c-factors reintroduced)
$\Delta X^a \Delta (Mc)\geq (\hbar/2)|\langle P^a/Mc\rangle|$. In a single particle description, we necessarily have $\Delta (Mc)=0$ violating this bound. We now see why single particle description cannot coexist with an operator $X^a$ with standard Lorentz transformation properties. This is the fundamental reason why many previous attempts have to tinker with the mass parameter and either make it a dynamical variable or introduce many-particle description.

Another possible ``way-out'' is to tinker with the notion of localization itself, one possibility being to work with hyperplane-dependent states \cite{fb}. It is difficult to think of these as localized states around an event and the description is definitely not the most natural one. I merely quote this to show that you need to pay a price one way or another; either mass becomes a dynamical variable or one needs a more liberal view of what localization means. These attempts also run into trouble \cite{malament} with the natural notion of causality based on the idea that 
the association of an
operator with a spacetime region implies that one can measure it by
performing operations confined to that region. In fact, as we shall see later, it is the consistency with micro-causality and Lorentz invariance which makes the single particle description extremely difficult to come by.

 \section{Propagators from path integrals}\label{sec:prfrpi}
 
 Let us now consider the above results from the path integral perspective, which is expected to provide an intuitive connection between the classical and quantum mechanics. Path integral formalism also has the advantage that we can work with c-number functions rather than with operators, state vectors etc.  If the classical physics of the system is described by an action $A$, specified as a functional of the relevant paths, then $G(x_b, x_a)$ is expected to arise  from a sum over all paths connecting the events $\mathcal{A}$ and $\mathcal{B}$ with the amplitude for each path being $\exp(iA)$. (The relativistic path integral has been studied in several previous papers in the literature; see, e.g., \cite{A,A1,A2,B,B2,B9,E,H,I}.)
 
 There are \textit{three}  forms for the action functional which we will concentrate on. The first one is the Hamiltonian form of the action:
 \begin{equation}
  A_{\bm{p}}[\bm{p}(t), \bm{x}(t)] \equiv \int_a^b dt [\bm{p \cdot \dot x} - H (\bm{p})]
 \end{equation} 
 where the action $A_{\bm{p}}$ is a functional of  $\bm{p}(t)$ and $\bm{x}(t)$ which are treated as independent. The second is the (more familiar) Lagrangian form of the action:
 \begin{equation}
A_{\bm{x}}[\bm{x}(t)] = \int_a^b dt \, L(\dot{\bm{x}})
\end{equation} 
 in which the action $A_{\bm{x}}$ is a functional of just $\bm{x}(t)$. Finally we can also define a Jacobi action for our system, which is quite different from either of these. It requires a separate treatment which we will take up in Sec.\ref{sec:jacobi}.
  
  In terms of either $A_{\bm{p}}[\bm{p}(t), \bm{x}(t)]$ or $A_{\bm{x}}[\bm{x}(t)]$,  the path integral propagator is formally defined as: 
 \begin{equation} 
G(x_b, x_a) = \sum_{\bm{x}(t)} \exp (iA_{\bm{x}});\qquad  G(x_b, x_a)= \sum_{\bm{x}(t), \bm{p}(t) }\exp(i A_{\bm{p}} )
 \end{equation}
 Of the two, the Lagrangian path integral has an obvious intuitive appeal.  In contrast, the ``sum over paths'' in phase space lacks a simple interpretation  because, classically, a single point in phase space determines the trajectory. Also note that, in the Lagrangian path integral, the paths are continuous but not the momenta while in the Hamiltonian path integral the paths are also discontinuous making the physical picture harder to interpret. So the meaning of the Hamiltonian path integral is not as straightforward as that of the Lagrangian path integral.
 
 If we are assured that both these path integrals lead to the same propagator (as they do in NRQM) one would have  preferred the Lagrangian path integral, at least as a formal expression.\footnote{The issue of measure in defining the path integral is somewhat easier to handle in the Hamiltonian approach than in the Lagrangian approach. For example, when you use time slicing, one needs to add an extra integration measure in the Lagrangian approach which has a natural origin in the Hamiltonian approach. But as a formal expression, Lagrangian path integral makes better intuitive sense.} Unfortunately the Hamiltonian and Lagrangian path integrals are not guaranteed to lead to the same result. In fact, we will see that the most natural definition for Lagrangian path integral does \textit{not}  work in the case of a relativistic particle, while the Hamiltonian path integral can be \textit{made to} work with some extra tinkering of the measure. We will now examine both, stating from the Hamiltonian path integral. 
 
\subsection{Propagator from Hamiltonian path integral}\label{sec:hpi} 
 
Let us work out the Hamiltonian path integral for the ``free particle'' with $H=H(\bm{p})$ taking care of both NRQM and RQM at one go. The standard procedure which we will adopt involves the following steps. 

(i) We discretize the time interval $t_b-t_a$ into $N$ intervals of size $\epsilon$ such that $N\epsilon=t_b-t_a$. At the end of the computation we take the limit of $N\to\infty, \epsilon\to0$ keeping the product $N\epsilon=t_b-t_a$ a constant. 

(ii) We discretize  the action and treat it as a function of $(\bm{p}_j,\bm{x}_j)$ where $j=0,1,2,...N$, with the identifications $\bm{x}_0=\bm{x}_a,\bm{x}_N=\bm{x}_b$ defining the end points. This discretized action is given by:
\begin{eqnarray}
 A_{\bm{p}}&=&\sum_{j=1}^N \left[ \bm{p}_j\cdot (\bm{x}_j-\bm{x}_{j-1}) -\epsilon H(\bm{p}_j)\right]\nonumber\\
 &=& \sum_{j=1}^{N-1} \left( \bm{p}_j - \bm{p}_{j+1} \right)\bm{\cdot x}_j+ \bm{p}_N \bm{\cdot x}_N - \bm{p}_1\bm{\cdot x}_a - \epsilon \sum_{j=1}^{N} H(\bm{p}_j)
 \label{discaction}
\end{eqnarray} 
As we will see, the second form of  $A_{\bm{p}}$ is more convenient for the computation.

(iii) The sum over paths is treated as integrations over $(\bm{p}_j,\bm{x}_j)$. The $\bm{x}_j$ integrations are over $j=1,2,...N-1$ keeping the end points fixed, so that there are $N-1$ integrals to do. The $\bm{p}_j$ integrations are over $j=1,2,...N$ so that there is one extra momentum integration. 

The crucial question, of course, is the choice of measure for the integration. The natural choice is to use just $d\bar\Gamma=d^n \bm{x}d^n \bm{p}/(2\pi)^n$. In this case, the propagator is defined by the integrals over the discretized action, given by the second equation in \eq{discaction}:
\begin{equation}
 G=\int d \bar\Gamma_1 \cdots d \bar\Gamma_{N-1} \, \int \frac{d^n\bm{p}_N}{(2\pi)^n} \ \exp i A_{\bm p}
\end{equation} 
Note that  this choice will lead to the surviving momentum integration (because there are $N$ momentum integrations but only $N-1$ position integrations) to appear with the measure $d^n \bm{p}/(2\pi)^n$. 
At each intermediate step, the  integration over $d\bm{x}_n$ leads to a Dirac delta function on the momentum. (This is the advantage of using the second expression in \eq{discaction}.) On integrating over the momenta, only the contribution from one end point  survives (since there is no corresponding $\bm{x}$ integration) leading to the  propagator:
\begin{equation}
 G(x)=\int \frac{d^n\bm{p}}{(2\pi)^n}\, e^{-ip\cdot x} 
\end{equation} 
defined again using the four-component object $p_a=(H,\bm{p})$. This leads to the standard propagator $G_{NR}$ in \eq{nrqmg} in NRQM. 
But in the case of RQM, the surviving integration over $d^n \bm{p}_N/(2\pi)^n$ will break the Lorentz invariance. 
leading to the Newton-Wigner propagator encountered earlier in \eq{kxaxb}:
\begin{equation}
 K(x)=\int \frac{d^n\bm{p}}{(2\pi)^n}\, e^{-ip\cdot x} =  2i \frac{\partial}{\partial t_b} G_+ (x_b, x_a)
\end{equation} 
This  Newton-Wigner propagator is obviously not Lorentz invariant and is  built from positive frequency solutions of Klein-Gordon equation. This situation
is completely analogous to NRQM; the price we have paid is the lack of Lorentz invariance which, unfortunately, is  too high. 

If we want a Lorentz invariant propagator the final momentum integration measure has to be $d\Omega_p=d^n \bm{p}/(2\pi)^n(1/\Omega_p)$. But this will lead to a wrong result  in the intermediate integrals, if it is used with $d^n \bm{x}$.  To solve this problem, we are forced to  tinker with the choice of measure and choose it to be:
\begin{equation}
 d\Gamma = \left[ d^n \bm{x}\ \Omega_n\right]\, \left[ d^n \bm{p} \ \Omega_n^{-1}\right](2\pi)^{-n}
 \label{measure1}
\end{equation} 
At each intermediate step, this is same as the original choice $d\bar\Gamma=d\bm{x}_n \, d\bm{p}_n/(2\pi)^n$ (since the $\Omega_n$ factors cancel) but the surviving momentum integration will come with an invariant measure.
With this choice, the propagator is now defined by the integrals over the discretized action, with:
\begin{equation}
 G=\int d \Gamma_1 \cdots d \Gamma_{N-1} \, \int d\Omega_N \ \exp i A_{\bm p}
\end{equation} 
At each intermediate step, the  integration over $d\bm{x}_n$ again leads to a Dirac delta function on the momentum.  On integrating over the momenta, only the contribution from one end point  survives (since there is no corresponding $\bm{x}$ integration) leading to the final result: 
\begin{equation}
 G=\sum_{\bm{p}} \sum_{\bm{x}} e^{iA_{\bm{p}}} = \int d\Omega_{\bm{p}_a} \ e^{-i{p}_a. {x}}=G_+
 \label{pi1}
\end{equation}
which matches with the result in \eq{ggen} obtained from the Hamiltonian procedure. 

A somewhat more intuitive way of obtaining these results is as follows: Re-write the Hamiltonian form of the action by eliminating $\dot{\bm{x}}$:
\begin{equation}
 A_p = \bm{p\cdot x}\Big|^b_a - \int_a^b dt \, \left[ \bm{x\cdot \dot{p}} + H(\bm{p})\right]
\end{equation} 
We then define the measure for the sum over $\bm{x}(t)$ such that it  gives a Dirac delta function of $\dot{\bm{p}}$. Then,  the 
 path integral becomes
\begin{equation}
 G=\sum_{\bm{p}} \sum_{\bm{x}} e^{iA_p} =   \sum_{\bm{p}} \delta(\dot{\bm{p}}) e^{i(\bm{p}_b \cdot \bm{x}_b - \bm{p}_a \cdot \bm{x}_a)} \ e^{-i \int dt\, H}
\end{equation} 
The existence of delta function tells you that in the sum $\bm{p}$ (and thus $H(\bm{p})$) remains constant, which immediately leads to the result in \eq{pi1}.

Clearly, a nontrivial choice of measure --- which is not easy to justify from the first principles --- was needed to get the correct result. The final, surviving momentum integral has to come with the measure $d\Omega_{\bm{p}}$ to give a Lorentz invariant result but the intermediate integrations have to be over $d\bm{x}_n \, d\bm{p}_n$ to give the Dirac delta functions. This requires defining the phase space measure by
\eq{measure1} which is the structure we were led to earlier in \eq{measure2}. This is the first instance of our running into a measure problem and of course, it does not arise in NRQM when $\Omega_{\bm{p}}=1$. Since the final answer is $G_+$ we will inherit all the issues discussed in Sec.\ref{sec:hofp}.

\subsection{Propagator from  the Lagrangian path integral}\label{sec:lpi}

There is a fairly general and natural procedure for defining the Lagrangian path integral by time slicing which works very well for the non-relativistic particle but fails for the relativistic particle. To see how this disaster comes about, we will next consider the discretized version of the Lagrangian path integral for both these cases.

To compute the propagator $G(x_b,x_a)$ it is again convenient to divide the time interval $(t_b - t_a)$ into $N$ equal parts of interval $\epsilon$ such that $N\epsilon = t_b - t_a$. In the interval $(t_{n-1},t_n)$ we will approximate the action by $A=\epsilon L(\dot{\bm{x}}) = \epsilon L\left((\bm{x}_n - \bm{x}_{n-1})/\epsilon\right)$. The full propagator is obtained by multiplying the amplitudes for each of the infinitesimal intervals with the intermediate spatial coordinates integrated out. 
 This would lead to an expression for the path integral of the form:
\begin{equation}
\sum_{\bm{x}} \exp i\int L dt = \int \prod_{k=1}^{(N-1)} \, d\bm{x}_k\, M(N,\epsilon) \, \exp i\epsilon L(\bm{\ell}/\epsilon);
\qquad\bm{\ell}\equiv(\bm{x}_n - \bm{x}_{n-1})
\label{disc1}
\end{equation}
where $M(N, \epsilon)$ is a measure which we hope to choose such that the continuum limit exists.

To evaluate this expression, it is convenient to work in the Euclidean sector. (We assume that we can obtain the Lorentzian result by analytic continuation at the end of the calculation). Let us introduce the spatial Fourier transform of the discretized Euclidean amplitude $e^{-\epsilon L(\bm{\ell}/\epsilon)}$ by:
\begin{equation}
 e^{-\epsilon L(\bm{\ell}/\epsilon)} = \int d^n \bm{p}\ F(\bm{p},\epsilon)\, e^{i{\bm{p\cdot \ell}}}
 \label{ftofl}
\end{equation}
The intermediate integrations in \eq{disc1} now lead to a series of Dirac delta functions allowing us to determine the spatial Fourier transform of the propagator in the form:
\begin{equation}
 G(\bm{p}) = C(N,\epsilon)\ \left[ F(\bm{p},\epsilon)\right]^{N}
 \label{gengofp}
\end{equation} 
where $C(N,\epsilon)$ takes care of the integration measure and other numerical constants.
  We now have to take the limit $\epsilon\to 0$, $N\to \infty$ with $N\epsilon =t$. If such a limit exists for a suitable choice of $C(N,\epsilon)$, then we have succeeded in defining the path integral. As we will see, this works for a non-relativistic particle but not for a relativistic particle.

 Let us first consider the non-relativistic case, for which the relevant Fourier transform in \eq{ftofl} is given by:
\begin{equation}
 F(\bm{p})= \int d^n\bm{\ell}\ \exp\left(-i\bm{p\cdot \ell} - \frac{m}{2\epsilon} \bm{\ell}^2\right) = \frab{2\pi \epsilon}{m}^{n/2} \ \exp\left(-\frac{\epsilon \bm{p}^2}{2m}\right)
 \label{fofpnr}
\end{equation} 
Therefore, the Fourier transform of the discretized path integral is given by:
\begin{equation}
  G(\bm{p}) =C(N,\epsilon) (F)^{N} = C(N,\epsilon)\frab{2\pi \epsilon}{m}^{nN/2} \ \exp\left[-\frac{\bm{p}^2}{2m} (N \epsilon)\right]
\end{equation} 
We now see that the exponential factor has a finite limit when $N \epsilon = t_b-t_a$. The prefactor can be made unity by choosing $C(N,\epsilon)=(2\pi \epsilon/m)^{-nN/2} $. We will then get the continuum limit of the propagator to be the one in \eq{nrqmg}. No surprises at all.

Let us next consider the relativistic case. The  conventional action functional for a relativistic particle, analytically continued to Euclidean sector, is given by:
 \begin{equation}
 A_E = - m \int \sqrt{\delta^a_b dx_a dx^b}=- m \int_{t_1}^{t_2} dt \sqrt{1+\bm{v}^2}
\label{sqrtact}
\end{equation} 
The relevant Fourier transform in \eq{ftofl} becomes
\begin{eqnarray}
 F(\bm{p}) &=& \int d^n\bm{\ell}\ \exp[-m(\epsilon^2+\bm{\ell}^2)^{1/2} - i\bm{p\cdot \ell}]\nonumber\\
 &=& \frab{m}{2\pi}^{1/2}\frab{2\pi}{m}^{n/2} \int_0^\infty \frac{d\mu }{\sqrt{\mu }} \, \mu ^{n/2} \exp\left( - \frac{\mu }{2m} \omega_p^2 - \frac{m}{2\mu } \epsilon^2\right)
 \label{fofprel}
\end{eqnarray}
where $\omega_p^2 \equiv \bm{p}^2 + m^2$.
The integral can be expressed in terms of McDonald functions leading to
\begin{equation}
 F(\bm{p}) = \frab{2\pi}{m}^{(n-1)/2} 2 \left(- \frac{m^2 \epsilon^2}{\omega_p^2}\right)^{(n+1)/4} \ e^{-(i\pi/4)(n+1)} \
 K_{-(n+1)/2} (\omega_p\epsilon)
\end{equation}
We, however, only need its form for small $\epsilon$; in this limit, 
 this expression becomes:
\begin{equation}
 F(\bm{p}) = 2m (4\pi)^{(n-1)/2} \, \Gamma\frab{n+1}{2} \frab{1}{\omega_p^2}^{(n+1)/2}
 \label{fofprel1}
\end{equation}
which can also be obtained directly from \eq{fofprel}. 
Therefore, the Fourier transform of the discretized path integral for the relativistic case is given by
\begin{equation}
 G(\bm{p}) = C(N,\epsilon)\left[ F(\bm{p})\right]^{N} \propto \frac{C(N,\epsilon)}{(\bm{p}^2+m^2)^{(n+1)N/2}}
 \label{disaster1}
\end{equation} 
We again need to take the limit of $N\to \infty$, $\epsilon \to 0$ with $N\epsilon = t$  in this expression and obtain a finite result. It is clear that one cannot obtain a finite result for any choice of the measure $C(N,\epsilon)$. Therefore the straightforward approach to obtain the propagator fails. 

The algebraic reason for the different results in the case on non-relativistic and relativistic cases can be traced to the structure of the integrands in \eq{fofpnr} and \eq{fofprel}. Reintroducing the $c$-factors, which occurs in the combination $c\Delta t=c\epsilon$, we note that the discretized action in the relativistic case has the combination $mc(c^2\epsilon^2 +\bm{\ell}^2)^{1/2}$. If we first take the $c\to\infty$ limit in this expression, keeping  $\epsilon$ finite --- which is what we do to get the non-relativistic result --- this gives $mc^2\epsilon+(1/2)m(\bm{\ell}^2/\epsilon)$ and the Fourier transform leads to the result in \eq{fofpnr} except for a finite, irrelevant, phase $-imc^2t$, in the Lorentzian sector. But if you take the $\epsilon\to0$ limit first, keeping $c$ finite ---  which is what we do in the exact relativistic case ---  the action $mc(c^2\epsilon^2 +\bm{\ell}^2)^{1/2}$ becomes 
$mc|\bm{\ell}|$ leading to the result in \eq{fofprel1}. So the fact the $c\epsilon$ goes to either infinity or zero, depending on whether you take the $c\to\infty$ limit first or the $\epsilon\to0$ limit first, makes all the difference.

There is another crucial feature which is worth mentioning. If you take the propagator in NRQM, given by \eq{nrqmg},  and consider its limit when the time interval $t = \epsilon \to 0$, you find that the argument of the  exponential factor is precisely equal to the non-relativistic action; that is, in this limit the propagator has the factor $\exp[i\epsilon L(|\bm{x}_2 - \bm{x}_1|/\epsilon)]  $. So, the propagator for a finite interval can indeed be thought of\footnote{This idea works even in the presence of a potential for a Lagrangian of the form $L = (1/2)m\dot{\bm{x}}^2 - V(\bm{x})$ as first noted by Dirac thereby paving the way for the path integral description of QM.} as arising from a product of infinitesimal propagators. But this result does not generalize to the relativistic propagator. The infinitesimal form of the relativistic propagator is \textit{not} related in any simple manner to the exponential of the action for infinitesimally separated events. This is again closely related to the composition laws obeyed by the two propagators. The composition law in \eq{complaw} can be iterated repeatedly allowing the $G_{\rm NR}$, for a finite interval of time, to be expressed as an integral over the products of the propagators for infinitesimal time separations.  Since the relativistic propagator does not obey this composition law, you cannot do this in a straightforward manner. 

Thus, while the Lagrangian and Hamiltonian path integrals lead to the same result in the NRQM, they differ widely for a relativistic action. The standard approach leads to a nonsensical result in the case of Lagrangian path integral while the Hamiltonian path integral measure has to be chosen carefully to lead to a Lorentz invariant result. 

Why do the two approaches lead to different results?
The Lagrangian and Hamiltonian path integrals  will lead to the same result  only if --- in the discretized version --- the integrals over $\bm{p}$ in the Hamiltonian path integral lead to the corresponding (discretized) Lagrangian form of the action.\footnote{As it turns out, this does happens in the case of NRQM , creating the myth that somehow this should always happen. This is far from true and does not, in general, hold even for an arbitrary `free particle' Hamiltonian $H=H(\bm{p})$, let alone in the presence of interactions.} So this  equivalence will hold only if the following condition holds: 
\begin{equation}
 \int d^n \bm{p} \ M(\bm{p}) \, e^{i\bm{p\cdot \ell} - i \epsilon H(\bm{p})} = f(\epsilon) \, e^{i\epsilon L(\bm{\ell}/\epsilon)}
\label{hlequiv} 
\end{equation}  
where $M(\bm{p})$ is some measure in momentum space and $f(\epsilon)$ is a measure for the Lagrangian path integral. So if the functions $M(\bm{p})$ and 
$f(\epsilon)$ exist, then the two procedures will give the same final result. 
This happens  for the non-relativistic action but not for the relativistic action. 

The time slicing procedure to define the (Hamiltonian or Lagrangian) path integral automatically selects a class of paths which satisfy the following condition: any path which is included in the sum cuts the intermediate time slices at only one point. That is, you only sum over paths which are always going forward (or always going backwards) in time. In either case, it seems reasonable to  interpret the expression in \eq{pi1} with a $\theta(t)$ [or a $\theta(-t)$] factor. But, as we mentioned earlier, $\theta(t)G_+(x)$ is not Lorentz invariant. In fact the whole idea of choosing paths which go only forward in time is not a Lorentz invariant criterion when the events $x_2$ and $x_1$ are separated by a spacelike interval. We will see in the next section that, using the lattice regularization procedure to give meaning to the path integral, bypasses these issues.

\section{Lattice regularization of the path integral}\label{sec:lattice}

So far we have seen that: (a) The Hamiltonian path integral can be made to give the propagator $G_+(x)$ with a specific choice of measure while (b) the straightforward way of computing the Lagrangian path integral does not work.
Interestingly enough, there is another way to define the Lagrangian path integral for the relativistic particle based on a geometric interpretation of the relativistic  action functional. This is based on  a lattice regularization procedure and leads to the Feynman propagator (with $x^2=x_ax^a$):
 \begin{equation}
G_R(x)=\int \frac{d^Dp}{(2\pi)^D}\frac{ie^{-ip_ax^a}}{(p^2-m^2+i\epsilon)}
 =\frac{m}{4\pi^2 i \sqrt{x^2}} \, K_1(im\sqrt{x^2})
\label{gfinft}
\end{equation} 
which is more relevant to standard QFT than  $G_+(x)$. I will briefly describe how this result comes about. (More details of this approach are available in e.g. Ref. \cite{tpqft}, Section 1.6.2)

 We will again work  in the Euclidean space of $D$-dimensions, evaluate the path integral and analytically continue to the Lorentzian space at the end.
The Euclidean action in \eq{sqrtact} can be expressed in the form 
\begin{equation}
 A_E = - m \int_a^b (dt^2+ d\bm{x}^2)^{1/2} = - m \int_a^b d\ell \equiv - m\  \ell 
\end{equation} 
where $\ell(x_b,x_a)$ is the length of a path connecting the events $\mathcal{A}$ and $\mathcal{B}$.  
 Our aim is  to give meaning to the sum over paths
\begin{equation} 
G_R(\mathbf{ x_2, x_1};m)=
\sum_{\mathrm{all}\,{\mathbf x}(s)}\exp
-m\,\ell[\mathbf{ x}(s)] 
\label{qlat}
\end{equation} 
in the Euclidean sector, where
$
 \ell(\mathbf{ x_2,x_1})
$ 
is just the Euclidean length of a path, connecting
${\mathbf x}_1$ and ${\mathbf x}_2$. (We will use $\mathbf{x}$ to denote the position in D-dimensional Euclidean space, in contrast to $\bm{x}$ which was used earlier for position in the $n=D-1$ dimensional space in Lorentzian spacetime. We will also label the $D=n+1$ axes as $(x^1, x^2, \ldots x^j, \ldots x^D)$ with no $x^0$ axis.)
This sum can be given a meaning through the following limiting procedure:

Consider a lattice of points in a $D$-dimensional cubic lattice with a uniform
lattice spacing of $\epsilon$. We will work out $G$ in the 
lattice and will then take the limit of $\epsilon\to 0$
with a suitable measure. To obtain a finite answer, we have to use an overall normalization factor $M(\epsilon)$ in \eq{qlat} as well as 
treat $m$ (which is the only parameter in the problem)
as varying with $\epsilon$  in a specific manner; i.e. we will use a function $\mu(\epsilon)$ in place of $m$ on the lattice
and will  reserve the symbol $m$  for the parameter in the continuum limit.\footnote{Purely from dimensional analysis, we would expect the mass parameter $\mu(\epsilon)$ to scale inversely as lattice spacing;  we will see that this is what happens.}  Thus the sum over paths in the continuum limit is \textit{defined} by the limiting procedure
\begin{equation} 
G_R(\mathbf{ x_2,x_1};m)=
\lim_{\epsilon\to 0}
\left[M(\epsilon){\cal G}_E(\mathbf{ x_2, x_1};
\mu(\epsilon))\right]  
\end{equation} 
where ${\cal G}_E(\mathbf{ x_2, x_1};\mu(\epsilon))$ is the sum defined on a finite lattice with spacing $\epsilon$.

In a lattice the sum can be evaluated in a straightforward manner.
Because of the translation invariance of the problem, ${\cal G}_E$
can only depend on $\mathbf{ x_2-x_1}$; so we can set ${\mathbf x_1}=0$
and call ${\mathbf x_2}=\epsilon {\mathbf R}$ where ${\mathbf R}$ is a
$D$-dimensional vector with integral components:
${\mathbf R}=(n_1,n_2,n_3\cdots n_D)$. Let $C(N,{\mathbf R})$ be the number of
paths of length $N\epsilon$ connecting the origin to the lattice
point $\epsilon{\mathbf R}$. Since all such paths contribute a term
$[\exp-\mu(\epsilon)(N\epsilon)]$ to \eq{qlat}, we get:
\begin{equation} 
{\cal G}_E({\mathbf R};\epsilon)=
\sum^{\infty}_{N=0}C(N;{\mathbf R})\exp\left(-\mu(\epsilon)N\epsilon \right)
\label{latticesum1}
\end{equation} 
It can be shown from elementary combinatorics (see e.g., Sec. 1.6.2 of Ref. \cite{tpqft}) that   the  
$C(N;{\mathbf R})$ satisfies the condition
\begin{equation} 
F^N\equiv\left[\sum_{j=1}^D 2\cos k_j\right]^N
=
\sum_{{\mathbf R}} C(N;{\mathbf R})e^{i\mathbf{ k.R}} 
\label{sevenfive}
\end{equation} 
Therefore,
\begin{eqnarray} 
\sum_{{\mathbf R}}e^{i\mathbf{ k.R}}
{\cal G}_E({\mathbf R};\epsilon)&=&
\sum^{\infty}_{N=0}\sum_{{\mathbf R}}C(N;{\mathbf R})
e^{i\mathbf{ k.R}}\exp\left( -\mu(\epsilon)N\epsilon\right)\nonumber\\
&=&\sum^{\infty}_{N=0}e^{-\mu(\epsilon)\epsilon  N}
F^N
=\left[1-Fe^{-\mu(\epsilon)\epsilon}\right]^{-1} 
\end{eqnarray}
Inverting the Fourier transform, we get
\begin{eqnarray}  
{\cal G}_E({\mathbf R};\epsilon)=\int
{d^D{\mathbf k}\over (2\pi)^D}
{e^{-i\mathbf{ k.R}}\over (1-e^{-\mu(\epsilon)\epsilon}F)}
=\int{d^D{\mathbf k}\over (2\pi)^D}
{e^{-i\mathbf{ k.R}}\over (1-2e^{-\mu(\epsilon)\epsilon}
\sum^D_{j=1}\cos k_j)} 
\end{eqnarray}
Converting to the physical length scales
${\mathbf x}=\epsilon{\mathbf R}$ and
${\mathbf p}=\epsilon^{-1}{\mathbf k}$ gives
\begin{equation} 
{\cal G}_E({\mathbf x};\epsilon)=
\int{\epsilon^D d^D{\mathbf p}\over(2\pi)^D}
{e^{-i\mathbf{ p.x}}\over (1-2e^{-\mu(\epsilon)\epsilon}
\sum^D_{j=1}\cos p_j\varepsilon)} 
\label{capgcape}
\end{equation} 

This is an exact result in the lattice and we now have to
take the limit $\epsilon\to 0$ in a suitable manner to keep the limit finite. 
As $\epsilon\to 0$, the denominator of the integrand
becomes
\begin{eqnarray} 
 1-2e^{-\epsilon\mu(\epsilon)}
\left( D-{1\over 2}\epsilon^2|{\mathbf p}|^2\right)=
 \epsilon^2 e^{-\epsilon\mu(\epsilon)}
\left[|{\mathbf p}|^2+
{1-2De^{-\epsilon\mu(\epsilon)}\over \epsilon^2 e^{-\epsilon\mu(\epsilon)}}\right]
 \end{eqnarray}
so that we  get, for small $\epsilon$,
\begin{equation} 
{\cal G}_E({\mathbf x};\epsilon)\simeq
\int{d^D{\mathbf p}\over (2\pi)^D}
{A(\epsilon)e^{-i{\mathbf p.x}}\over |{\mathbf p}|^2+B(\epsilon)} 
\end{equation} 
where
$ 
 A(\epsilon)= \epsilon^{D-2}e^{\epsilon\mu(\epsilon)}
 $
 and
 $
B(\epsilon)=(1/ \epsilon^2)
[e^{\epsilon\mu(\epsilon)}-2D].
$
The continuum theory has to be defined in the limit of
$\epsilon\to 0$ with some measure $M(\epsilon)$; that is, we want to choose $M(\epsilon)$ such that the limit
\begin{equation} 
G({\mathbf x})=
\lim_{\epsilon\to 0}
\left\{M(\epsilon){\cal G}_E({\mathbf x};\epsilon)\right\} 
\end{equation} 
is finite. It is easy to see that 
we only need to demand
 near $\epsilon\approx 0$, the validity of the conditions:
\begin{equation}
 \mu(\epsilon)\approx {\ln 2D\over \epsilon}
+{m^2\over 2D}\epsilon
\approx {\ln 2D\over \epsilon}; \qquad M(\epsilon)={1\over 2D}{1\over \epsilon^{D-2}} 
\label{qmeas}
\end{equation} 
 With this choice, we get
\begin{equation} 
G_R(\mathbf{x})=\lim_{\epsilon\to 0}{\cal G}_E
({\mathbf x};\epsilon)M(\epsilon)=\int
{d^D{\mathbf p}\over(2\pi)^D}
{e^{-i\mathbf{ p.x}}\over |{\mathbf p}|^2+m^2} 
\label{eightfour}
\end{equation} 
which is the usual (Euclidean) Feynman propagator now obtained from a path integral using a lattice regularization. On analytic continuation to Lorentzian sector, it gives the expression in \eq{gfinft}. So we have succeeded in defining the relativistic path integral and evaluating it to give the Feynman propagator. We will now highlight several aspects of this approach.

\subsection{Comments on the lattice regularization approach}\label{sec:comlattice}

The scaling of $\mu(\epsilon)=\ln 2D/\epsilon$ might appear quite strange and I will provide two alternative routes to this scaling which might demystify it a little bit. First one proceeds as follows:  Let $\mathcal{N}(\ell)$ be the number of paths of length $\ell$ connecting the origin to the event $\mathbf{x}$ in the continuum limit. Then our propagator is given by
\begin{equation}
 G(\mathbf{x}) = \int_0^\infty d\ell\ \mathcal{N}(\ell; \mathbf{x})\ e^{-m\, \ell}
 \label{tpnotetwo}
\end{equation}
This expression, which is the continuum analogue of \eq{latticesum1}, is only a formal expression since $\mathcal{N}(\ell; \mathbf{x})$ is divergent in the continuum limit. 
To give meaning to this equation we have to define $\mathcal{N}(\ell)$ on a lattice with spacing $\epsilon$ and take the appropriate limit after the integral is performed. We also need to replace $m$ by the mass parameter $\mu(\epsilon)$ in the lattice. The Fourier transform of 
$\mathcal{N}_\epsilon(\ell)$ on the lattice is then given by \eq{sevenfive}. 
Switching to the continuum with the replacements $\mathbf{x} = \epsilon \mathbf{R}$ and $\mathbf{p} = \mathbf{k}/\epsilon$, it is easy to see that
\begin{equation}
\mathcal{N}_\epsilon(\ell; \mathbf{p}) \equiv \int d^D\mathbf{x}\ \mathcal{N}_\epsilon(\ell; \mathbf{x}) \, e^{i\mathbf{p\cdot x}} \simeq (2D - \epsilon^2 \mathbf{p}^2)^{\ell/\epsilon}
 \label{tpnotethree}
\end{equation} 
where we have set $N\approx \ell/\epsilon$. Taking the Fourier transform of \eq{tpnotetwo},  using \eq{tpnotethree} and performing the integral over $\ell$, we find that
\begin{equation}
 G(\mathbf{p}; \epsilon) = -\frac{2D}{\epsilon} \left[ \mathbf{p}^2 + \frac{2D}{\epsilon} \, \mu -  \frac{2D}{\epsilon^2} \, \ln 2D\right]^{-1}
 \label{tpnotefour}
\end{equation} 
If we now assume that $\mu(\epsilon)$ scales as in the first equation of \eq{qmeas}, the expression in the square bracket in \eq{tpnotefour} reduces to $\mathbf{p}^2 + m^2$. The overall factor in front can be taken care of by a suitable measure $M(\epsilon)$. You see that $(\log 2D)/\epsilon$ scaling of $\mu(\epsilon)$ arises\footnote{You might wonder whether one can give meaning to the Lagrangian path integral which led to the result in \eq{disaster1} by allowing $m$ to depend on $\epsilon$. Unfortunately this does not give the correct relativistic propagator. The best one can do is the following. If we  assume that $m\to\infty$ as $\epsilon\to0$, then \eq{disaster1} will reduce  to an expression proportional to 
$
\exp[-(p^2/2m^2)((n+1)t/\epsilon)]$
for a suitable choice of $C(N,\epsilon)$. The best we can do is to assume a scaling of the form 
$
m^2(\epsilon) = [m_0(n+1)/\epsilon]
$ 
where $m_0$ is a constant. This will lead to an expression proportional to 
$\exp[-(p^2/2m_0)t]$.
This expression is finite and, rather curiously, results in the propagator for NRQM but not the correct one we want.} due to the pre-factor $(2D)^{\ell/\epsilon}$ in \eq{tpnotethree}.

The second approach to understand the scaling $\mu\epsilon \approx \ln 2D$, which is of interest in its own sake, is to think of the propagator as a solution to the KG equation with a delta function source and compare the versions in the continuum and in the lattice.
 Let us consider a path of $N$ steps connecting the origin to a lattice site labeled by integer valued lattice points $\bm{n}$. Then the lattice propagator is given by the sum over all paths of the form
\begin{equation}
 \mathcal{G}_{\bm{n}} = \sum_{\rm paths} \exp(-m\epsilon \, N) \equiv \sum_{\rm paths} K^N
\end{equation} 
where $K = e^{-m\epsilon }$. We now interpret $K$ as the probability (amplitude) for the particle to hop between two nearby cells of the lattice. This immediately allows us to write the recurrence relation to reach a specific lattice point $\bm{n}$ as
\begin{equation}
 \mathcal{G}_{\bm{n}} = \delta_{\bm{0},\bm{n}} + K\sum_{j=1}^D  (\mathcal{G}_{\bm{n}+\bm{m}_j} + \mathcal{G}_{\bm{n}-\bm{m}_j})
 \label{latticerec}
\end{equation} 
where $\bm{m}_j$ is the unit vector in the $j$-th direction. This recurrence relation determines the lattice propagator. On the other hand, in the continuum limit the propagator satisfies the Klein-Gordon equation with a delta function source:  
The lattice version of this differential operator can be easily obtained by using the Taylor series relation 
\begin{equation}
 G(x+h) + G(x-h) - 2G(x) = h^2 G''(x)
\end{equation} 
for each direction.
Converting this relation into a lattice with lattice spacing $\epsilon$, the discretized Klein-Gordon equation for the propagator becomes
\begin{equation}
 \frac{1}{\epsilon^2} \sum_{j=1}^D  (\mathcal{G}_{\bm{n}+\bm{m}_j} + \mathcal{G}_{\bm{n}-\bm{m}_j} - 2 \mathcal{G}_{\bm{n}}) - m^2 \mathcal{G}_{\bm{n}} = \delta_{\bm{0},\bm{n}}
\end{equation} 
This equation can be re-written in the form
\begin{equation}
 \left( \frac{1}{m^2\epsilon^2 +2D}\right) \sum_{j=1}^D  (\mathcal{G}_{\bm{n}+\bm{m}_j} + \mathcal{G}_{\bm{n}-\bm{m}_j}) + \delta_{\bm{0},\bm{n}} = \mathcal{G}_{\bm{n}}
 \label{kglattice}
\end{equation} 
where we have rescaled the Dirac delta function by $\epsilon^2$ on the lattice. Comparing \eq{latticerec} with \eq{kglattice}, we see that $\exp(m_0 \epsilon)$ gets replaced by ($m^2 \epsilon^2 + 2D$) on the lattice. This is equivalent to the replacement of $m_0$ by $\epsilon^{-1}\ln 2D$ in the limit of $\epsilon \to 0$ which is precisely the mass renormalization we saw earlier.

How come the Lagrangian path integral, originally evaluated with the time slicing method led to a meaningless expression (viz. \eq{disaster1}) while the lattice regularization method leads to the Feynman propagator? The reason has to do with the different kinds of paths which are summed over in the two approaches. When you define the path integral by time slicing, you implicitly assume that the any path which is included in the sum cuts the intermediate time slices at only one point. That is, you only sum over paths which are always going forward (or always going backwards) in time. But when you sum over paths on the lattice, the paths can go back and forth in time. So the two sets of paths which are summed over are completely different and we have no reason to expect them to give the same answer.

This connection can be made more quantitative by examining a lattice regularization scheme for paths which go only forward in time in the Lorentzian sector. On analytic continuation to the Euclidean sector, they will go only forward in one of the axis, which we take to be the $x^0$ direction. 
(We will now label the $D=(n+1)$ axes as $(x^0, x^1, \ldots x^n)$, restoring the $x^0$ axis which is treated as special.)
Our aim is to see whether such a condition will lead to anything which resembles the non-relativistic propagator.

We know that a relativistic scalar field in the Euclidean sector will satisfy the Euclidean Klein-Gordon equation 
$
(-\Box_E + m^2)\phi =0
$, 
while its non-relativistic counterpart $f(x)$, related to $\phi(x)$ by 
$
\phi(x) \equiv e^{-mt}\, f(x)
$, 
will satisfy the Euclidean Schroedinger equation 
$
(\partial_t - (1/2m)\nabla^2)_E \, f=0
$.
 The latter is obtained from the former by \textit{approximating} the second time derivative $\ddot \phi$ by 
$
\ddot \phi \approx m^2 \phi - 2 me^{-mt} \, f
$.
In the momentum space, this involves replacement of
$
({p}^2 + m^2)_E \equiv \Omega^2 +\bm{p}^2 + m^2 
$ 
by
$
2mi\omega + \bm{p}^2
$ 
where
$
\Omega \equiv \omega + i m
$ 
and we have ignored the $\omega^2$ term in comparison with $m\omega$. This requires the  denominator $(\Omega^2 + \bm{p}^2+ m^2)$ in the Euclidean relativistic propagator (written as a Fourier transform with respect to Euclidean time), 
\begin{equation}
 G_R = \int \frac{d\Omega \, d^n \bm{p}}{(2\pi)^D} \frac{e^{i(\Omega t + \bm{p\cdot x})}}{(\Omega^2 + \bm{p}^2+ m^2)}
\end{equation} 
 given by \eq{eightfour}, to be replaced by $(2mi\omega + \bm{p}^2)$ to give the non-relativistic propagator:
\begin{equation}
 \bar{G}_{NR} = \int \frac{d\omega \,d^n \bm{p}}{(2\pi)^D}\ \frac{2me^{i(\omega t + \bm{p\cdot x})}}{(2mi\omega + \bm{p}^2)}
 =  \int \frac{d\omega \,d^n \bm{p}}{(2\pi)^D}\ \frac{(2m)e^{i(\omega t + \bm{p\cdot x})}}{ \bm{p}^2+2mi(\Omega -im)}
 \label{14oct9}
\end{equation} 
Let us see how this comes about when we restrict paths to go only forward along $x^0$.

Each of the $2\cos p_j \epsilon = e^{ip_j \epsilon} + e^{-ip_j\epsilon}$ in the denominator of \eq{capgcape} is contributed by paths going forward along the $j$-th direction (contributing $
e^{ip_j \epsilon}$) and paths going backward along the $j$-th direction (contributing $ e^{-ip_j\epsilon}$). 
So, when we restrict the paths to move only forward along $x^0$ axis and repeat the analysis,  along the $0$-th direction we only pick up a $e^{ip_0\epsilon_0}$ factor.
This modifies the denominator $\mathcal{D}$ of \eq{capgcape} to the expression:
\begin{equation}
 \mathcal{D} = 1 - 2\, e^{-\mu(\epsilon_1)\epsilon_1} \sum_1^n \cos p_j \epsilon_1 - e^{\mu(\epsilon_0)\epsilon_0} \, e^{i p_0 \epsilon_0}
 \label{14oct10} 
\end{equation} 
 We have taken the lattice spacing to be $\epsilon_0$ along the time direction and $\epsilon_1$ for all the space directions. This is  essential because the transition from Klein-Gordon equation to Schroedinger equation involves a transition from wave equation to a diffusion equation; the propagation in the Euclidean lattice will mimic a diffusion only if $t\propto x^2$ requiring $\epsilon_0\propto \epsilon_1^2$ when we take the continuum limit. (If you don't do this and assume the same lattice spacing along both direction you will \textit{not} reproduce the form of the propagator in \eq{14oct9}.) Straightforward computation now reduces \eq{14oct10}  to the form 
\begin{equation}
 \mathcal{D} = (A p_0 + \bm{p}^2 +B) C
\end{equation} 
 where 
 \begin{equation}
  C=\epsilon_1^2 \, e^{-\mu_1 \epsilon_1}
 \end{equation} 
 \begin{equation}
  B = \frac{1}{\epsilon_1^2} e^{\epsilon_1\mu_1} \left(1- 2 n e^{-\epsilon_1\mu_1} - e^{-\epsilon_0\mu_0}\right)
 \end{equation} 
 \begin{equation}
 A = \frac{1}{\epsilon_1^2} \, e^{\epsilon_1\mu_1} \left( -i\epsilon_0 \, e^{-\epsilon_0\mu_0}\right)
 \end{equation} 
 Ignoring the overall constant $C$ --- which merely defines the overall measure like $M(\epsilon)$ in the previous analysis --- and comparing $\mathcal{D}$ with the denominator in \eq{14oct9}, we find that  the following conditions need to be satisfied
 \begin{equation}
 A = 2 m_0 i; \qquad B=  2 m_0^2
 \end{equation} 
 Some more algebra now shows that this can indeed be achieved with the choices 
 \begin{equation}
 \epsilon_0 = \frac{2m}{(2n-1)}\, \epsilon_1^2 \propto \epsilon_1^2
  \label{14oct16}
 \end{equation} 
 and 
 \begin{equation}
  \mu_0 = - \frac{1}{\epsilon_0} \, \ln(2n-1); \qquad \mu_1 = 2m^2 \epsilon_1
  \label{14oct17}
 \end{equation} 
 Equation (\ref{14oct16}) shows that $\epsilon_0\propto \epsilon^2$ has to be expected in a diffusion process; \eq{14oct17} shows the scaling of $\mu_1$ and $\mu_0$ for this result to hold.

This feature can also be made more  transparent along the following lines. While the real space expressions for $G_{NR}(x)$ (given by \eq{nrqmg}) and $G_R(x)$ (given by \eq{gfinft} look very different, their spatial Fourier transforms are very   similar:
\begin{equation}
G_R(t,\bm{p}) \equiv \int d^3\bm{x}\ G(x_2;x_1) e^{-i\bm{p\cdot x}} = 
\begin{cases}
   e^{-i\omega_{\bm{p}}t} &\text{(non-relativistic)}\\
{ } \\
{\displaystyle{\frac{1}{2\omega_{\bm{p}}}}}\, e^{-i\omega_{\bm{p}}|t|} &\text{(relativistic)}
\end{cases}
\label{nrrcomp1}
\end{equation}
where $\omega_{\bm{p}}= \bm{p}^2/2m$ in the non-relativistic case, while $\omega_{\bm{p}}= (\bm{p}^2+m^2)^{1/2}$ in the relativistic case. 
Using the Fourier transform of $G_+(x)$ in \eq{spaceft}, it is easy to relate $G_R(x)$ and $G_+(x)$. We find that
$ G_R(x_2,x_1)=G_+(x_2;x_1)$ when $t_2>t_1$ and $G_R(x_2;x_1)=G_+^*(x_2;x_1)$ when $t_2<t_1$ where $G_-(x_2;x_1) \equiv  G_+^*(x_2;x_1)=G_+(x_1;x_2)$
is the complex conjugate of $G_+(x_2;x_1)$. That is:\footnote{The $\theta(t)$ is Lorentz invariant only when $x_2$ and $x_1$ are separated by a timelike interval in which case the $\theta(t)$ picks out one of the two terms. When $x_2$ and $x_1$ are separated by a spacelike interval, $\theta(t)$ is not Lorentz invariant and hence the expression could pick either of the the two terms depending on the Lorentz frame. But since we know --- from any of the explicit expressions like \eq{macd} --- that $G_R(x_2;x_1)$ is Lorentz invariant, it follows that $G_+(x_2;x_1)=G_-(x_2;x_1)$ when $x_2$ and $x_1$ are separated by a spacelike interval. This is indeed true and can be explicitly verified. (See Appendix \ref{sec:equalgs}.)}
\begin{eqnarray}
 G_R(x_2;x_1) &=& \theta(t) G_+(x_2;x_1) + \theta(-t)G_+^*(x_2;x_1)\nonumber\\
 &=& \theta(t) G_+(x_2;x_1) + \theta(-t)G_-(x_2;x_1)
\label{amppmnew}
\end{eqnarray}
Since we know that the $G_+$ uses only paths which go forward in time it is clear that
$G_R$ propagates particles with energy $\omega_p$ forward in time and propagates particles with energy $-\omega_p$ backward in time.
This feature arises from summing over paths which go back and forth in time direction.
So, $G_R(x_2,x_1)$ is actually two propagators rolled into one; we will come back to this aspect in Sec.\ref{sec:seamless}.

It is obvious that, while the relativistic propagator $G_R$ in \eq{eightfour} arises very naturally through the lattice regularization approach, we have to  make several artificial choices \textit{based on our hindsight} for obtaining non-relativistic propagator by lattice regularization. Once again there is no natural limiting process within the lattice regularization which allow us to obtain the non-relativistic propagator from the relativistic one. 

\subsection{Jacobi action and its path integral}\label{sec:jacobi}

A convenient expression for $G_R$ in the coordinate space is obtained using the Schwinger's proper time representation.\footnote{This representation, using a ``fifth time'' was introduced by Stueckelberg \cite{E1,E2} and developed further extensively by Schwinger.} 
 We write $(|{\bm p}|^2+m^2)^{-1}$ as a integral over $\lambda$ of $\exp[-\lambda(|{\bm p}|^2+m^2)]$ and do the $\bm{p}$ integration to obtain:
\begin{equation}
G_R = \int_0^\infty \frac{d\lambda}{(4\pi\lambda)^{D/2}}\, \exp\left({-\lambda m^2-\frac{|\mathbf{x}|^2}{4\lambda}}\right)
\Rightarrow\frac{1}{16\pi^2} \int_0^\infty\frac{d\lambda}{\lambda^2} \exp \left({-m^2\lambda-\frac{|\mathbf{x}|^2}{4\lambda}}\right)
\end{equation}
where the second expression is for 
$D=4$.
The analytic continuation from the Euclidean to the Lorentzian spacetime changes the sign of one of the coordinates in $|{\mathbf x}|^2$ to give $|{\bm x}|^2-t^2=-x^2$ and we set $\lambda = i s$. This gives the final result:
\begin{equation}
G_R 
=-\frac{i}{16\pi^2}\int_0^\infty\frac{ds}{s^2}\, \exp\left({-im^2 s- \frac{i}{4s}x^2}\right)=\frac{m}{4\pi^2 i \sqrt{x^2}} \, K_1(im\sqrt{x^2})
\label{macd}
\end{equation} 

This  proper time representation of the $G_R$ has an alternative interpretation. The integral expression in \eq{macd} can be expressed, after a rescaling of $s\to s/m$, as:
\begin{equation}
G_R(x_2;x_1) \propto \int_{0}^\infty ds\, e^{-ims} \amp{x_2,s}{x_1,0} = C_m\int_{0}^\infty ds\, e^{-ims}\sum_{x(\tau)} e^{iA[x(\tau)]}
\label{relpi}
\end{equation} 
where $C_m$ is an unimportant constant and
\begin{equation}
 \amp {x_2,s}{x_1,0} =\theta(s) i\left( \frac{m}{4\pi is}\right)^2 \exp\left( -\frac{i}{4} \frac{mx^2}{s}\right)
\label{leu}
\end{equation}
can be thought of as a propagator for a (fictitious) particle moving in the four dimensional Lorentzian spacetime from $x_1^i$ at $\tau=0$ to $x_2^i$ at $\tau =s$, where $\tau$ parameterizes the  path in spacetime $x^i(\tau)$. The relevant action for this  particle is a quadratic one given by
\begin{equation}
A[x(\tau)] = -\frac{1}{4} m \int_0^s d\tau\, \dot x_a \dot x^a
 \label{relact}
\end{equation} 
Classically, this action could \textit{also} be thought of as  representing the free relativistic particle (since it leads to the equation of motion $d^2x^i/d\tau^2 =0$). But --- unlike the action in \eq{sqrtact} --- it is (i) not reparametrisation invariant  and it (ii) does not have a geometrical interpretation. The path integral in \eq{relpi} gives the amplitude $\amp{x_2,s}{x_1,0}$ for the particle to propagate from $x_1$ to $x_2$ during the proper time interval $s$. The Fourier transform of this amplitude with respect to $s$  can be thought of as giving the amplitude for this propagation to occur with the energy $mc^2$ in the rest frame. This suggests that $G_R(x_2,x_1)$ gives an amplitude for propagation \textit{at a constant energy} rather than for a given \textit{time interval}. Such a path integral can be defined in a more general context using what is known as Jacobi action functional. We will now discuss this interpretation of the Feynman propagator.

The Jacobi action $A_J$ can be thought of as the integral of $\bm{p\cdot}d\bm{{x}}$ where $\bm{p}$ is expressed as a function of energy $E$ by solving the equation $H(\bm{p}) = E$. In our case, for a system with $H(\bm{p}) = H(|\bm{p}|)$, the $\dot{\bm{x}}$ and $\bm{p}$ will be in the same direction allowing us to write $\bm{p\cdot}d\bm{{x}}= \mathcal{P}(E) d\ell$ where $\ell$ is the arc-length of the path and $\mathcal{P}(E)$ is the magnitude of the momentum $|\bm p|$, expressed as a function of $E$. Since $E$ is constant, the Jacobi action in our case reduces to 
\begin{equation}
A_J = \mathcal{P}(E) \int d\ell = \mathcal{P}(E) \ \ell(\bm{x}_b, \bm{x}_a)
\end{equation} 
which has the geometrical meaning of the length of the path connecting the two events. This expression
is  manifestly re parameterization invariant with no reference to the time coordinate.

Since $A_J$ describes an  action principle for determining the path of a particle with energy $E$ classically, the sum over $\exp(iA_J)$ could be interpreted  as the amplitude for the particle to propagate from $x^\alpha_1$ to  $x^\alpha_2$ \textit{with energy} $E$.  Since $A_J$ is not quadratic 
in velocities,  even for a non-relativistic free particle, (because $d\ell  $ involves a square root) one has to  again do a lattice regularization to compute the result, just as we did for a relativistic particle. The path integral defined using the Jacobi action then reduces to the sum over paths of the kind considered in \eq{qlat} with $m$ replaced by $\mathcal{P}(E)$. So the propagator for the Jacobi action will be given by the expression obtained earlier in \eq{eightfour} with $m^2$ replaced by $\mathcal{P}^2(E)$. That is, the Jacobi action propagator will be 
\begin{equation}
 \mathcal{G}(\bm{x},E) = \sum\exp (-A_J)=\int \frac{d^D\bm{p}}{(2\pi)^D} \, \frac{ e^{-i\bm{p\cdot x}}}{p^2+\mathcal{P}^2(E)}
\end{equation} 
In the case of a non-relativistic free particle with $\mathcal{P}^2(E) = 2mE$,  this gives us the result:
\begin{equation}
 (2m) \ \mathcal{G}(\bm{x},E) = \int \frac{d^D\bm{p}}{(2\pi)^D}\ \frac{e^{-i\bm{p\cdot x}}}{E + (p^2/2m)}
\end{equation} 
which makes sense.\footnote{The Fourier transform of a Lorentzian propagator $G(t, \bm{x})$ with respect to Lorentzian time $t$ will be defined using an $\exp(-i\omega t)$ factor while the Fourier transform of a Euclidean propagator $G_E(t_E, \bm{x})$ with respect to Euclidean time $t_E$ will be usually defined using an $\exp(-i\omega t_E)$ factor. Sometimes it is more convenient to use the factor $\exp(-\omega t_E)$ (and restrict the time integration to the range $(0,\infty)$) and define the Laplace transform in the Euclidean sector. This is what we have done here; the usual Fourier transform can be obtained by the replacement $E\to -iE$.}

 But there is another way of determining $\mathcal{G}(\bm{x},E)$. Since we already have the standard  path integral defined for the  non-relativistic particle, we can use it \textit{to give meaning} to this sum over $\exp(-A_J)$. In the process, we would have obtained a procedure for \textit{defining} the sum over paths for any non-quadratic action that is proportional to the length of  the path. 
The idea is to write the sum over  all paths in the conventional Lagrangian action principle (with amplitude $\exp(iA_{\bm{x}})$) as a sum over paths with energy $E$ followed by a sum over all $E$. So we write, formally, 
\begin{equation}
 \sum_{0,\bm{x}_1}^{t,\bm{x}_2} \exp(iA_{\bm{x}}) = \sum_E \sum_{\bm{x}_1}^{\bm{x}_2} e^{-iEt} \, \exp iA_J[E,\bm{x}(\tau)] \propto \int_0^\infty dE\, e^{-iEt} \sum_{\bm{x}_1}^{\bm{x}_2} \exp(iA_J)
\end{equation} 
In the last step we have treated the sum over $E$ as an integral over $E>0$ (since, for any Hamiltonian which is bounded from below, we can always achieve this by adding a suitable constant to the Hamiltonian) but there could be an extra proportionality constant which will depend on the measure used to define the sum over $\exp(iA_J)$. Inverting the Fourier transform, we get the Jacobi propagator to be: 
\begin{equation}
\mathcal{G}(\bm{x}_2,\bm{x}_1;E) \equiv \sum_{\bm{x}_1}^{\bm{x}_2} \exp(iA_J) = C \int_0^\infty dt \, e^{iEt} \sum_{0,\bm{x}_1}^{t,\bm{x}_2} \exp(iA)
= C \int_0^\infty dt \, e^{iEt} G(x_2;x_1)
\label{pofe}
\end{equation} 
where we have denoted the proportionality constant by $C$. 
 This result shows that the sum over the  Jacobi action $A_J$ involving a \textit{square root of velocities} can be re-expressed in terms of the standard path integral; if the latter can be evaluated for a given system, then the sum over Jacobi action can be defined by this procedure.
 For the case of a free particle we get:
 \begin{equation}
 \sum_{\bm{x}_1}^{\bm{x}_2} \exp i\sqrt{2mE}\, \ell(\bm{x}_2,\bm{x}_1) = C \int_0^\infty dt \, e^{iEt}\sum_{0,\bm{x}_1}^{t,\bm{x}_2}\exp\frac{im}{2} \int_0^t d\tau \left( g_{\alpha\beta} \dot x^\alpha\dot  x^\beta\right)
\label{finres1}
\end{equation} 
where we have denoted the length of the path connecting $x^\alpha_1$ and  $x^\alpha_2$ by $\ell(\bm{x}_2,\bm{x}_1)$, 
Since the action for the relativistic particle in \eq{sqrtact}  has the same structure as the Jacobi action for a non-relativistic free particle, the propagator   $G_R(x_2;x_1)$, can be obtained directly from \eq{finres1}.
We first take the complex conjugate of \eq{finres1} (in order to get the overall minus sign in the action in \eq{sqrtact}) and generalize the result from space to spacetime, 
leading to 
\begin{equation}
 \sum_{\bm{x}_1}^{\bm{x}_2} \exp -i\sqrt{2mE}\, \ell(\bm{x}_2,\bm{x}_1) = C \int_0^\infty d\tau  e^{-iE\tau}\sum_{0,\bm{x}_1}^{t,\bm{x}_2}\exp-\frac{im}{2} \int_0^\tau d\lambda \left( g_{ab} \dot x^a\dot  x^b\right)
\label{finres2}
\end{equation} 
In order to get $-im\ell(\bm{x}_2,\bm{x}_1)$ on the left hand side we take $E=m/2$ and put $\tau=2s$ to get an $\exp(-ims)$ factor.
The path integral over the quadratic action trivial and in $D=4$, we get the expression in \eq{leu}.
Therefore the path integral propagator reduces to the expression in \eq{relpi}:
\begin{eqnarray}
 G(x_2;x_1) &=& -(2Cm) i\left(\frac{m}{16\pi^2}\right)\int_0^\infty \frac{ds}{s^2} \exp\left( -ims - \frac{i}{4} \frac{mx^2}{s}\right)\nonumber\\
 &=& - \frac{i}{16\pi^2}\int_0^\infty\frac{d\mu}{\mu^2}  \exp\left( -im^2\mu - \frac{i}{4} \frac{x^2}{\mu}\right)
\label{rpi1}
\end{eqnarray} 
where we have  rescaled the variable $s$ to $\mu$ by $s\equiv m\mu$ and made the choice $C=1/2m$ to  match with conventional result in \eq{macd}.

Once we introduce the idea of a fictitious particle propagating in spacetime, governed by a quadratic action in \eq{relact}, we can also introduce a complete set of (spacetime) position eigenkets $\ket{x}$ and momentum eigenkets $\ket{p}$. The Hamiltonian relevant for the action in  
\eq{relact} will be $H=-p^2$ (corresponding to the mass $m=1/2$) and the matrix element of the proper time evolution operator will be
\begin{equation}
\amp{x_2,s}{x_1,0}\equiv \bk{x_2}{e^{-isH}}{x_1}=\bk{x_2}{e^{isp^2}}{x_1}                                                  
\end{equation} 
So, the relativistic propagator $G_R$, treated as a function of $\mu =m^2$ can be expressed as the integral 
\begin{equation}
 G_R (x_2, x_1; \mu) \equiv \int_0^\infty ds\,  \bk{x_2}{e^{-is (H+\mu)}}{x_1} = - i \bk{x_2}{(\mu + H)^{-1}}{x_1}
 \label{why1}
\end{equation} 
This result will be useful later on.

Our propagator can also be obtained by using the quadratic action  \eq{relact} in the path integral and imposing the reparametrisation invariance through a Lagrange multiplier. (This is also equivalent to imposing the condition $H=-p^2=-m^2$ on the Hamiltonian.). The path integration over the Lagrange multiplier will reduce to integration over $\tau$ leading to the same final expression.
I will quickly run through this procedure \cite{C1,C2a,C2b} to connect up with our previous discussion. 
  We begin by recalling that, for the relativistic Lagrangian,
 $
 L_R = - m \left[ \eta_{mn} \dot x^m \dot x^n\right]^{1/2} = - m(\dot x^2)^{1/2}
 $ 
 the momenta $p_m = \partial L/\partial \dot x^m$ satisfy the constraint $\mathcal{H} \equiv p_m p^m - m^2 = 0$. While constructing the Hamiltonian form of the action, this constraint is incorporated through a Lagrange multiplier $N(\tau)$, leading to  
 \begin{equation}
 A_R = \int_{r_1}^{r_2} d\tau \left( p_m \dot x^m + N \mathcal{H}\right)
 \end{equation} 
 This action, in turn, retains the memory of re parameterization invariance of $L_R$ because it remains invariant under the gauge transformation generated by $\mathcal{H}$ given by 
 \begin{equation}
 \delta x = \epsilon(\tau) \{ x, \mathcal{H}\}, \qquad \delta p = \epsilon(\tau) \{ p , \mathcal{H}\}, \qquad \delta N = \dot \epsilon(\tau)
 \end{equation} 
 where $\epsilon(\tau)$ vanishes at the end points. The simplest gauge fixing condition \cite{3ofa} is to take $\dot N =0$ making $N$ a constant. The Hamiltonian path integral will now require an integration over the parameter $N$ which will lead to the correct propagator $G_R$ if the range of integration is restricted to $0\le N< \infty$. That is, 
 \begin{equation}
 G_R(x_2-x_1) = - i \int_0^\infty  dN\int \mathcal{D}p \mathcal{D}x \ \exp \left( i \int_{r_1}^{r_2} d\tau \left( p \dot x + N \mathcal{H}\right)\right)
 \end{equation} 
This is yet another popular route to the Feynman propagator discussed in the literature. 

To avoid possible misunderstanding, I stress the following fact. It is certainly possible to come up with schemes by which the path integral for a relativistic particle can be evaluated. We have already seen three such procedures which lead to the ``correct'' propagator, $G_F(x)$: (a) Lattice regularization, (b) Jacobi action and (c) Gauge fixing approach.  
(The approach based on lattice regularization or the one based on the interpretation of \eq{sqrtact} as a Jacobi action seems more transparent than the one in which gauge-fixing is used but this could be a matter of taste.) The key common feature is that all `successful' approaches --- which lead to the `correct' $G_F(x)$ --- allow for the paths to go backwards and forwards in Minkowski time coordinate $t$ which will not be allowed in the standard time-slicing approach to a path integral based on the Lagrangian $L_R=-m(1-\dot{\bm x}^2)^{1/2}$. In fact, the class of paths summed over in each of the three approaches are formally very different. For example, it is certainly true that the Lagrangian $L_R(\dot{\bm x})$ arises from the `gauge-invariant' Lagrangians, in a specific gauge. But the path integral involves the sum over totally different sets of paths ($x^a(\tau)$ versus $\bm x(t)$) in these two approaches; summing over $x^a(\tau)$ with time-slicing in $\tau$ allows for paths  $\bm x(t)$ which go backwards in Minkowski time coordinate $t$.

The existence of these three (and possibly many other) procedures does not provide the answer to the simple question: \textit{How come the most natural procedure, based on paths $\bm x(t)$ and time slicing in $t$, which works so well in the case of a NR particles, fails for a relativistic particle?} 
Given the action for a \textit{non-relativistic} particle I can construct NRQM by path integral, just with time-slicing, without knowing Schrodinger equation or Heisenberg operator algebra. But given the action for the \textit{relativistic} particle, I cannot do it in a natural fashion and, in fact, the corresponding single-particle  RQM does not exist!
Of course, if you think of relativistic particles as excitations of an underlying field and quantize the field --- rather than use the action principle for the particle --- you will get $G_F(x)$ as well as the antiparticles. You can then cook up several ways to get it from path integrals; that is hardly satisfactory if you want to do everything upfront from the path integral.

The technical reason, as we will see later in Sec.\ref{sec:seamless}, has to do with the fact that $G_F(x)$ actually propagates \textit{two} fields and \textit{two} kinds of particles, not one. The procedures which actually ``work'' ,for defining the path integrals, have this feature built into them one way or another --- usually by allowing paths to go backwards and forwards  in Minkowski time coordinate $t$ --- so that they can lead to the ``correct'' propagator, $G_F(x)$. This is hardly a satisfactory situation because we already need to know the answer (and  the existence of pairs of particles) from some other approach to define the suitable procedure for path integral. I will say more about this in Sec. \ref{sec:impli} around \eq{mystery1}.

I conclude this section with a technical comment related to the time slicing approach for determining the relativistic propagator which, as we saw earlier, does not work. In \eq{relpi}, the amplitude $\amp{x_2,s}{x_1,0}$ has a natural path integral expression with time slicing in the proper time $s$. If we divide the proper-time interval into $N$ slices, and write the usual time-sliced expression for $\amp{x_2,s}{x_1,0}$ in \eq{leu}, we can write  the relativistic propagator in \eq{relpi} in the form
 \begin{equation}
 G_R = \int_0^\infty ds  \int \prod^{N-1}_{n=1} id^4x_n\left(\frac{mN}{4\pi is}\right)^2 \exp\left( -i \sum_{n=0}^{N-1} \frac{m(x_{n+1} - x_n)^2}{4s/N} - i  m s\right)
  \label{20ofb}
 \end{equation} 
 In the absence of the integration over $s$, the propagator $\amp{x_2,s}{x_1,0}$ satisfies the non-relativistic composition law in \eq{complaw}. But once we introduce the integration over $s$, this composition law fails and --- as we will see later in \eq{tptwo} --- is replaced by a composition law involving Klein-Gordon inner product. The crucial point is that the expression in \eq{20ofb}, after integration of $s$, is not related to the exponential of the infinitesimal action. If we define the sum in the exponent as 
 \begin{equation}
  R^2 \equiv \frac{N}{4} \sum_{n=0}^{N-1} (x_{n+1} - x_n)^2
 \end{equation} 
 then the integration over $s$ will lead to a weight for each path given by (with $s=m\bar s$):
 \begin{equation}
 W = \int_0^\infty d\bar s\ \bar s^{2-2N} \exp \left(-\frac{iR^2}{\bar s} - i m^2\bar s \right) 
 = 2\left(\frac{-R^2}{m^2}\right)^{\nu/2}e^{-i\pi\nu/2}K_{-\nu}(2miR)
 \label{notaction}
 \end{equation} 
with $\nu=3-2N$. Obviously, this expression has no simple relation with the exponential of relativistic action.

\subsection{Non-relativistic limit of the Feynman propagator}\label{sec:nroff}

We have obtained the propagator $G_+$ using the Hamiltonian path integral and $G_R$ from the lattice regularization of the Lagrangian path integral. We already know that $G_+$ is related to $G_{NR}$ through the limit in \eq{plustonr} which is not very surprising because we could derive both $G_+$ and $G_{NR}$ at one go, in \eq{ggen}. But the derivation of $G_R$, using the lattice regularization, was quite different and there is no simple correspondence to $G_{NR}$. So the question arises as to whether one can get the non-relativistic propagator $G_{\rm NR}(x_b,x_a)$ from the relativistic propagator $G_{R}(x_b,x_a)$ in the limit of $c\to\infty$? 

This is \textit{not} possible in spite of occasional claims  to the contrary made in literature.
This should, in fact, be obvious from \eq{nrrcomp1}. When you take $c\to \infty $ limit of $G_R(t,\bm{p})$, the prefactor becomes $1/2m$ which is an inconsequential scaling.  In the phase $\omega_p$ can be approximated as $m+\bm{p}^2/2m$. The factor $\exp(-imt)$ could have been interpreted as due to the rest energy $mc^2$ contributing to the phase.
\textit{But the $|t|$ never becomes $t$ when we take this limit}.\footnote{The Feynman propagator is an even function of $t$  and it will remain an even function of $t$ even when you take the limit $c\to \infty$. On the other hand, $G_{NR}(t,\bm{x})$ is not an even function of $t$ since $G_{NR}(-t,\bm{x})= G_{NR}^*(t,\bm{x})$. So you can't get $G_{\rm NR}$ from $G_R$ for all $t$.}
So the factor $\exp(-m|t|)$ does not have a straightforward interpretation. Thus, while we can barely escape\footnote{\eq{plustonr} tells you that for $t>0$, we have $G_R=G_+$ and we already know the relation, given by \eq{plustonr},  between $G_+$ and $G_{NR}$. So this is not a big deal. The real issue is in the comparison of $G_R$ and $G_{NR}$ for $t<0$ where they do not match.}  in the case of $t>0$, the expressions are quite different for $t<0$. 

To see this more explicitly, we have only have to evaluate $G_R$ in the $c\to \infty$ limit using the saddle point approximation to the integral.
Rescaling $\lambda \to \lambda/m$ we can express the Euclidean $G_R$ in the form:
\begin{equation}
 G = \frac{1}{(4\pi)^2} \int_0^\infty \frac{d\lambda}{\lambda^2} \, e^{-\lambda m^2} \ e^{-(1/4\lambda) (t^2+x^2)}
 \to \frac{m}{(4\pi)^2} \int_0^\infty \frac{d\lambda}{\lambda^2} \, e^{-m\lambda - (m/4\lambda)(t^2+x^2)}
\end{equation}
We need the saddle point of the function $f(\lambda) = m\lambda + (mt^2/4\lambda)$ 
which occurs at $\lambda =\lambda_c= |t|/2$.
  The value of the function at the saddle point is $f_c = m|t|$
and the prefactor is given by $(2\pi/f'')^{1/2} = (\pi |t|/2m)^{1/2}$. 
  So we find that, in the limit $c\to \infty$, we get the propagator
\begin{equation}
 G = \frac{1}{2m}\frab{m}{2\pi |t|}^{3/2} \, e^{-m|t| - mx^2/2|t|}
\end{equation} 
The overall scaling by $2m$ is of no consequence and arises from  $2\omega_p$ in the limit $c\to \infty$. But you find  
 that the result has $|t|$ rather than $t$ in the expression. When you analytically continue to Lorentzian sector, this effect will persist and you will get
 \begin{equation}
 G = \frac{1}{2m}\frab{m}{2\pi i|t|}^{3/2} \, e^{-im|t| + imx^2/2|t|}
\end{equation} 
 One can understand the factor $\exp(-imc^2 |t|)$ as signaling the rest energy $mc^2$ of the particle, which has to be taken away from the phase of the wave function, to reach the non-relativistic limit when $t>0$. But, one cannot make sense of this phase for $t<0$; more generally we cannot interpret the occurrence of $|t|$ in NRQM. This issue is actually quite non-trivial and we will discuss it again in Sec.\ref{sec:seamless} from a different perspective. 
 
 The usual folklore that Feynman propagator has the correct NRQM limit originates (i) either from considering only $t>0$ case or (ii) from mixing up momentum space and real space descriptions. The momentum space argument goes along the following lines: In momentum space, the Feynman propagator is governed by a term in the denominator $(p^2-m^2-i\epsilon)$ where $p^a=(E,\bm p)$. If we write $E\equiv m+\epsilon$ removing the rest energy, then $p^2-m^2=\epsilon^2+2m(\epsilon-\bm{p}^2/2m)$ and when we study processes involving non-relativistic energies, one can ignore the $\epsilon^2$ term and use the approximate  expression proportional to $(\epsilon-\bm{p}^2/2m)$  in the \textit{momentum} space. This approximation completely changes the pole structure of propagator from two poles in complex plane to one. To get the real space propagator from the momentum space propagator, you need to integrate over all $\epsilon$ \textit{without} ignoring the $\epsilon^2$ term. Making the approximation $\epsilon\ll1$, obtaining an approximate momentum space propagator and \textit{then} integrating over \textit{all} $\epsilon$ to get the real space propagator is conceptually incorrect. 
  
  \subsection{Feynman propagator as a matrix element of time evolution operator}\label{sec:fmatrix} 
  
 The non-relativistic propagator $G_{NR}$ can be expressed as the matrix element  $G(x_b , x_a) = \bk{\bm{x}_b}{e^{-itH}}{\bm{x}_a}$ of the 
  time evolution operator in a straightforward manner. In the case of $G_{+}$, we could again do this but the the states $\ket{\bm{x}}$ did not have the interpretation as eigenstates of the position operator; instead we had to define them using a Fourier transform. 
  Let us now address the corresponding question for the relativistic propagator $G_R(x_b,x_a)$ obtained above from lattice regularization, viz, whether it can  be expressed in the form $G(x_b , x_a) = \bk{\bm{x}_b}{e^{-itH}}{\bm{x}_a}$ where $H=H(\bm{p}) =({\bm p}^2+m^2)^{1/2}$.  We already know that this is not going happen  with the  procedure we have adopted for defining the states $\ket{\bm x}$; it only leads (at best) to $G_{+}$ and $G_R\neq G_{+}$. Obviously we have to cheat a little bit 
  somewhere along the line if such a relation should hold. I will now describe how this can be achieved (with a bit of cheating) because the procedure highlights some key issues we have been discussing. 
  
To do this, we will first consider the case when $t>0$ and use the easily proved (operator) identity:
\begin{equation}
 2H\int_0^\infty d\mu \, \exp\left( - i\mu^2 H^2 - \frac{i t^2}{4\mu^2}\right) =  \frab{\pi}{i}^{1/2}\,  e^{-iHt}
 \label{oneoneseven}
\end{equation} 
which allows us to  write:
\begin{eqnarray}
 \bk{\bm{x}_b}{e^{-iHt}}{\bm{x}_a} &=&  \frab{i}{\pi}^{1/2}  \int_0^\infty d\mu \, e^{(-it^2/4\mu^2)}\ \bk{\bm{x}_b}{2H(\bm{p}) e^{-i\mu^2H^2(\bm{p})}}{\bm{x}_a}\nonumber\\
 &=&\frab{i}{\pi}^{1/2}  \int_0^\infty d\mu \, e^{(-it^2/4\mu^2)}\ e^{-i\mu^2m^2}\bk{\bm{x}_b}{2H(\bm{p}) e^{-i\mu^2\bm{p}^2}}{\bm{x}_a}
 \label{trick}
\end{eqnarray} 
The matrix element can be evaluated by introducing a complete basis of momentum eigenkets $\ket{\bm{p}}$ with integration measure $d\Omega_p=d^n\bm{p}/(2\pi)^n(1/2\omega_p)$ for the momentum integration. This will give us, in three dimensions with $\bm{\ell} \equiv \bm{x}_b - \bm{x}_a$:
\begin{equation}
\bk{\bm{x}_b}{2H(\bm{p})e^{-i\mu^2\bm{p}^2}}{\bm{x}_a} =
  \int \frac{d^3\bm{p}}{(2\pi)^3}\frac{1}{2\omega_p} \, e^{i\bm{p}\cdot  \bm{\ell}} \, [2\omega_p e^{-i \mu^2 p^2}] = \frab{\pi}{i\mu^2}^{3/2} \frac{1}{8\pi^3}\exp \left(\frac{i\bm{\ell}^2}{4\mu^2}\right)
  \label{onetwozero}
\end{equation} 
Note that the $2\omega_p$ arising from $2H$ in the left hand side of \eq{oneoneseven} cancels nicely with the $(1/2\omega_p)$ in the measure of integration in the momentum space, giving a simple result. Substituting \eq{onetwozero}
into \eq{trick} we get the final result, with $x^2 = x^ax_a=t^2  - \bm{\ell}^2$
\begin{eqnarray}
  \bk{\bm{x}_b}{e^{-iHt}}{\bm{x}_a} &=&  \frab{i}{\pi}^{1/2}  \frab{\pi}{i}^{3/2}\,\frac{1}{8\pi^3}\int_0^\infty \frac{ds}{2s^2} \, \exp\left( - \frac{ix^2}{4 s} - i m^2 s\right)\\
  &=& \frac{1}{i} \frac{1}{16\pi^2} \int_0^\infty\frac{ds}{s^2} \, \, \exp -i\left( \frac{x^2}{4 s} +  m^2 s\right)
  \label{oneoneight}
\end{eqnarray}
This is, of course, the standard expression for the Feynman propagator and we have obtained earlier in \eq{macd}; it as equal to the matrix element in the left hand side. So where did we cheat? 

The identity in \eq{oneoneseven} is actually valid when the right hand side has $\exp(-iH|t|)$. Note that, in our final expression given by \eq{oneoneight}, the right hand side is an even function of $t$. So the left hand side should also be an even function of $t$. This is ensured only because the result we have proved continues to be valid for $t<0$ as well, if we replace $\exp(-iHt)$ by $\exp(-iH|t|)$. In other words, the evolution operator we have sandwiched between the eigenkets is \textit{not} $\exp(-iHt)$ but 
\begin{equation}
 U(t) = e^{-iH|t|} = \theta(t) e^{-iHt} + \theta(-t) e^{iHt} 
\end{equation} 
So we are not computing the matrix element of the evolution operator $e^{-iHt}$ as per the standard rule but evaluating matrix element of the operator $U(t) = e^{-iH|t|}$. This modification of the evolution operator, in which propagation forward in time is dictated by $H$ and the propagation backward in time is dictated by $-H$, makes all the difference in the world. 

But the real surprise is the following: We have now shown that the propagator $G_R$ can be expressed as $\bk{\bm y}{\exp(-iH|t|)}{\bm x}$ where $\ket{\bm x}$ and $\ket{\bm y}$ are \textit{non-localized states!}. It is not obvious that, merely by using $U(t) = e^{-iH|t|}$ rather than $e^{-iHt}$ we can still express the correct propagator without solving the problem of localized particle state. This is the real surprise I want to highlight about the discussion in this section.

\subsection{Aside: Composition law for propagators}\label{sec:compo}

 In NRQM, the propagator $G_{\rm NR}(x_b,x_a)$ actually propagates the wave function from the event $\mathcal{A}$ to the event $\mathcal{B}$. Such an interpretation relies crucially on the propagator satisfying the composition law in \eq{complaw}. This composition law, in turn,  is a \textit{trivial} consequence of two facts: (i) $G_{\rm NR}$ can be expressed as the matrix element $\bk{\bm{x}_b}{\exp[-iH(t_b - t_a)]}{\bm{x_a}}$ and (ii) the set $\ket{\bm{x}}$ forms a complete set of orthonormal basis. So multiplying two propagators $G_{\rm NR}(x_b,x_c)$ and $G_{\rm NR}(x_c,x_a)$  and integrating over the variable occurring in $\ket{\bm{x}_c}$  reduces the composition law to an identity:
 \begin{equation}
  \int d\bm{x}_c\ \bk{\bm{x}_b}{e^{-iH(t_b - t_c)}}{\bm{x}_c} \bk{\bm{x}_c}{e^{-iH(t_c - t_a)}}{\bm{x}_a}
  = \bk{\bm{x}_b}{e^{-iH(t_b - t_a)}}{\bm{x}_a}
 \end{equation} 
 Obviously, this will not hold for the relativistic propagator because  the condition  (ii) is violated. 
 
 It is, however, straightforward to derive the corresponding composition law \textit{with integration over spacetime} rather  than just space, for the relativistic propagator. 
 From the integral representation of the propagator in \eq{why1}, we immediately see that\footnote{You can also obtain the same result by multiplying the two propagators and using their explicit form in Fourier space. For a discussion of composition laws for the propagators, see ref. \cite{B,A1} and references therein.} with $\mu=m^2$:
\begin{equation}
 \int d^Dx \, G(\mu; x_2,x) G(\mu; x,x_1) = - \bk{x_2}{(\mu +H)^{-2}}{x_1} = i \frac{\partial}{\partial \mu} G(\mu; x_2,x_1)
 \label{twentysix}
\end{equation} 
But the integration now is over, say, $d^Dx$ at the intermediate event, rather than over $d^n\bm{x}$; so the physical meaning of this composition law is unclear; you certainly cannot use it to propagate a wave function. (It does not help to restrict the integration over spatial coordinates in \eq{twentysix}; see Appendix \ref{sec:appencompo}.) It is also obvious from the derivation of \eq{twentysix} that
  it is  the integration over $ds$ in \eq{why1}  which makes the relativistic case very different from the non-relativistic one. 
 
 Incidentally, this composition law can be  iterated $N$ times to give the result:
\begin{equation}
\int d^Dx_1 \cdots d^Dx_N\ G(\mu; x_b,x_N) \cdots G(\mu; x_1,x_a) = (i)^N  \frac{\partial}{\partial \mu^N} G(\mu; x_b,x_a)
\end{equation}
This result suggests a  curious way of reconstructing $G(\mu; x_b, x_a)$. We first note that the Euclidean version of \eq{twentysix} (in which the  $i$ factor on the right hand side is replaced by $-1$) can be rewritten, after integrating over $\mu$ in the range $m^2<\mu<\infty$, in the form 
 \begin{equation}
 \int_{m^2}^\infty d\mu\, d^Dx_1\ G(\mu; x_b, x_1) G(\mu; x_1,x_a)\equiv
 \int d\mathcal{M}_1\  G(\mu; x_b,x_1) G(\mu; x_1,x_a)   = G(m^2; x_b, x_a)
 \end{equation} 
 where we have treated the propagator as a function of the variable $\mu$ and defined the measure of integration as $d\mathcal{M} \equiv d\mu\, d^Dx$ and used the fact that the Euclidean propagator vanishes when $m^2\to\infty$. This equation can be iterated infinite number of times by keeping two events in $G(\mu; x_j,x_{j-1})$ infinitesimally
  close to each other.  Iterating $N$ times will give the result
  \begin{equation}
  \int d\mathcal{M}_1 \cdots d\mathcal{M}_N \ G(\mu_N; x_b,x_N) \cdots G(\mu_1; x_1,x_a) = G(m^2; x_b,x_a)
   \label{thirtysix}
  \end{equation} 
 This is very similar in structure to the non-relativistic composition law in \eq{complaw}. Therefore, one can, in principle, convert \eq{thirtysix} to some kind of sliced up path integral prescription. Unfortunately, the form of $G(\mu; x,y)$, when $x$ and $y$ are infinitesimally separated, is not the  exponential of the action for the relativistic particle and, in fact, has no simple interpretation.

The expression for the  relativistic propagator in terms of the Jacobi action offers some further insight into the composition law and demystifies it. Even in NRQM, the \textit{energy propagator} $\mathcal{G}(\bm{x}_2,\bm{x}_1;E)$, obtained from the path integral sum over the  Jacobi action, does not obey the composition law in \eq{complaw}; instead it satisfies an analogue of the composition law in 
\eq{twentysix}.
This is, again, obvious from the structure of 
$\mathcal{G}(\bm{x}_2,\bm{x}_1;E)$, defined in \eq{pofe}. Expressing the propagator $G(x_2,x_1)$ in \eq{pofe} as the matrix element of the time evolution operator of NRQM, we get: 
\begin{equation}
 \mathcal{G}(\b x_2, \b x_1;E) = \int_0^\infty dt \, e^{it(E+i\epsilon)}\bk{\b x_2}{e^{-it\hat H}}{\b x_1} 
= i\bk{\b x_2}{(E - \hat H +i\epsilon)^{-1}}{\b x_1}
\label{xxx}
\end{equation} 
where we have introduced an $i\epsilon$ factor, with an infinitesimal $\epsilon$, to ensure convergence.
From  \eq{xxx}, it immediately follows that:
\begin{equation}
 \int d^D\b y \, \mathcal{G}(\b x_2, \b y; E) \mathcal{G}(\b y, \b x_1;E) =  -i\left[\frac{\partial  \mathcal{G}(\b x_2, \b x_1;E)}{\partial E}\right]
\end{equation} 
which has the same form as the result in \eq{twentysix}, for pretty much the same algebraic reasons. So this composition law in \eq{twentysix} has \textit{nothing to do} with relativity; it arises because the $G_R$ can be interpreted as arising from a Jacobi action. (One can also write down an iterated relation, identical in form to \eq{thirtysix} in this case as well; unfortunately its physical meaning is not clear.)
 
The composition law in \eq{twentysix} induces corresponding composition laws in the Fourier transform of the propagators. Consider first $G_{\bm{p}}(t_b-t_a)$ which is the spatial Fourier transform of the propagator in \eq{nrrcomp1}. This function satisfies, in the Euclidean sector, the composition law: 
\begin{equation}
 \int_{-\infty}^\infty dt\ G_p(t_2 -t) G_p(t- t_1) = - \frac{\partial}{\partial \mu} G_p(t_2 - t_1)
 \label{twentynine}
\end{equation}
It is straightforward to verify that the integrals on both the left hand side and right hand side can be expressed in 
 the form
 \begin{equation}
 I_{\rm RHS}  = - \frac{G}{2\omega_p} \frac{\partial \ln G}{\partial \omega_p}
 =\frac{G}{2\omega_p} \left\{ (t_2-t_1) + \frac{1}{\omega_p}\right\}= I_{\rm LHS}
\end{equation}
 It would be interesting to ask whether  one can recover the non-relativistic composition law in \eq{complaw} from this result --- which looks quite different ---  in the appropriate limit. This \textit{cannot} be done with the expressions in \eq{nrrcomp1} but if we change the propagator for the non-relativistic case by multiplying it by a $\theta(t)$ [that is, we take the non-relativistic propagator in the Fourier space to be $G_{\rm NR}(t, \bm{p}) = \theta(t) \exp(-i\omega_p t)$] then one can  obtain the non-relativistic limit correctly. This is based on the fact that in the non-relativistic limit we have the approximate form:
 \begin{equation}
 -\frac{\partial \ln G}{\partial \mu} \bigg|_{\rm NR} = + \frac{1}{2\omega}  \left\{ (t_2-t_1) + \frac{1}{\omega}\right\}
 \approx \frac{t_2 - t_1}{(2m)}
\end{equation}
 Then, as long as $t_1<t<t_2$ then the composition law in \eq{twentynine} reduces to the composition law of NRQM in \eq{nrrcomp1}. (Some of the details of these computations are given in Appendix \ref{sec:appencompo}.)

 One can also consider the Fourier transform of the Euclidean propagator with respect to time obtaining $G_E (\bm{x})$. A simple calculation shows that 
\begin{equation}
 G_E(\bm{x}) \equiv \int_{-\infty}^\infty G \, e^{iEt} dt = \int \frac{d^n\bm{p}}{(2\pi)^3} \, \frac{e^{i\bm{p\cdot x}}}{E^2+ \bm{p}^2 + m^2}
\end{equation} 
 This function satisfies  the composition law 
\begin{equation}
 \int d^n \bm{x} \, G_E (\bm{x}_2, \bm{x}) \, G_E(\bm{x}, \bm{x}_1) = - \frac{\partial}{\partial \mu} G_E(x_2, x_1)
\end{equation} 
which is easy to verify. 

Finally, let us consider the composition law which \textit{does} lead to the propagator in terms of two other propagators in the relativistic case. Since the scalar product for the relativistic Klein-Gordon equation is defined as 
\begin{equation}
(\phi_1,\phi_2)\equiv i\int d\sigma^a [\phi_1^*\partial_a\phi_2-\phi_2\partial_a\phi_1^*] = i \int d\sigma^a \ \phi_1^*\overleftrightarrow{\partial_a}\phi_2
\end{equation} 
It is straightforward to show that the propagator, treated as a function of $x$, satisfies the composition law:
\begin{equation}
 (G^*(x_2,x),G(x,x_1))=G(x_2,x_1)
 \label{tptwo}
\end{equation}
This result holds \textit{only} as long as $x_2^0>x^0>x_1^0$. On the other hand, if $x^0>x_2^0 >x_1^0$, say,  the integral on the left hand side vanishes. One simple way to prove this result is to 
Fourier transform \eq{tptwo} with respect to spatial coordinates and write the corresponding condition involving  $G_R(t,\bm{p}) $ and $\partial_t G_R(t,\bm{p}) $. We next note from \eq{nrrcomp1} that $G_R(t,\bm{p}) $ and its time derivative can be expressed as 
\begin{equation}
 G = \frac{1}{2\omega} \left[ \theta(t) e^{-i \omega t} + \theta(-t) e^{+i\omega t}\right]; \qquad 
 \partial_t G = + \frac{i}{2} \left[-\theta(t) e^{-i \omega t} + \theta(-t) e^{i\omega t}\right]
\end{equation} 
leading to:
$
 \partial_t G = - i \omega G[{\rm Sg}(t)]
$ where ${\rm Sg}(t)=t/|t|$ is the sign function. With this result, it is easy to show that the combination occurring on the left hand side of \eq{tptwo} in Fourier space  is proportional to
$G_R(t_2-t,\bm{p})G_R(t-t_1,\bm{p})[{\rm Sg}(t_2-t)+{\rm Sg}(t-t_1)]$
which is non-zero only in the interval 
 $t_1<t<t_2$.
  In this interval the relation \eq{tptwo} is identically satisfied.\footnote{If you do the integrals in \eq{tptwo} in the real space, this result arises, after a bit of tedious algebra, because a factor in the numerator cancels a pole in the denominator in a rather subtle manner. 
 It is also related to the orthogonality of $G_+(x,y)$ and $G_-(x,y) \equiv G_+^*(x,y)$.
 I do not know of any simple way to ``guess'' this result.} So, even though the Feynman propagator can propagate backwards in time, it does not work in the composition laws.

\section{A seamless route  from QFT to NRQM}\label{sec:seamless}

In Sec. \ref{sec:ffrpr} we found that one is led to a notion of a field operator $A(x)$ fairly naturally from the propagator both in RQM and in NRQM. This was done by introducing the  
``creation'' and ``annihilation'' operators in the Fourier space, $A_{\bm{p}}$ and $A^\dagger_{\bm{p}}$, and defining $A(x)$ by \eq{fop1}. This approach, therefore, holds promise for a seamless transition from QFT to NRQM. 

There was, however, one serious difficulty.  We found that, in QFT, the commutator $[A(x_2), A^\dagger(x_1)]=\amp{x_2}{x_1}$ (where the state $\ket{x}$ is defined by \eq{nintynine}) does not reduce to a Dirac delta function on a spacelike hypersurface. This is a reflection of the non-localisability of the particle position. 
 So if you build observables from $A$ and $A^\dagger$, they will not commute for events separated by a space-like interval. A sensible way of incorporating causality into quantum theory will be to arrange matters such that commutator between observables vanish for space-like separated events. So we cannot treat $A(x)$ as the basic building block in the theory and  need to do a little bit more work. 

To tackle this issue, we will introduce another field $B(x)$ whose commutator will lead to $G_-(x_2, x_1)=G_+^*(x_2,x_1)$ just as the commutator in \eq{hundredthree} lead to $G_+(x_2,x_1)$. This is achieved through the definition  
\begin{equation}
B(x) \equiv \int d\, \Omega_{\bm{p}} B_{\bm{p}}e^{-ipx}
\end{equation} 
with the assumption that $B(x)$ commutes with $A(x)$. It is straightforward to verify that 
 $
 [B(x_2), B^\dagger(x_1)]  \equiv G_-(x_2;x_1)
$.
  Let us now  define the combination:
$
 \phi(x) = A(x) + B^\dagger(x)
$. This field $\phi$ will also satisfy the Klein-Gordon equation since $A$ and $B$ do. But $\phi$ has better behaviour as regards causality. It is straightforward to show that 
\begin{eqnarray}
[\phi(x_2), \phi^\dagger(x_1)]&=& [A(x_2) + B^\dagger(x_2), A^\dagger(x_1) + B(x_1) ] \\
&=& [ A(x_2), A^\dagger(x_1)] - [B(x_1), B^\dagger(x_2)]
=  G_+(x_2;x_1) - G_+(x_1;x_2)\nonumber 
\end{eqnarray} 
This commutator vanishes at spacelike separation because $G_+(x_2;x_1) = G_+(x_1;x_2)$ in that case. (See Appendix \ref{sec:equalgs}; for a nice discussion of the role of causality in QFT, see \cite{NO}.)

). 
So we find that to maintain relativistic causality we need to work with two fields $A$ and $B$ and define the physical field as $
 \phi(x) = A(x) + B^\dagger(x)
$.
We also have the relations giving the propagator directly in terms of $\phi$ and $\phi^\dagger$:
  \begin{eqnarray}
 &&\bk{0} {\phi(x_2) \phi^\dagger(x_1)}{0} =\bk{0} {A(x_2) A^\dagger(x_1)}{0}
 = \int d\,\Omega_{\bm{p}} e^{-ipx} = G_+(x_2;x_1) \nonumber\\
&&\bk{0} {\phi^\dagger(x_1) \phi(x_2)}{0} = \bk{0} {B(x_1) B^\dagger(x_2)}{0}
=\int d\,\Omega_{\bm{p}} e^{+ipx} = G_-(x_2;x_1)
\label{fieldexps}
\end{eqnarray} 
  These relations, in turn, allows us to express our relativistic propagator  entirely in terms of $\phi$ through the relation 
\begin{eqnarray}
 &&G(x_2,x_1) 
 =\theta(t_2-t_1) \bk{0}{A(x_2)A^\dagger(x_1)}{0} + \theta(t_1-t_2) \bk{0}{B(x_1)B^\dagger(x_2)}{0}\nonumber\\
 &&\hskip 5cm =  \bk{0}{T(\phi(x_2)\phi^\dagger(x_1))}{0}
 \label{conc1}
\end{eqnarray} 
 
 To summarize, we first introduced a primitive field $A(x)$  based on the relationship between 
  $\ket{x}$ and $\ket{\bm{p}}$. This field satisfies the Klein-Gordon equation but not our notion of causality. Looking at the structure of the commutator of $A$ field, we introduced another field $B(x)$ which also satisfies the Klein-Gordon equation and, finally, a physical field  $\phi(x)$ which obeyed the Klein-Gordon equation and the notion of causality. Obviously, the notion of causality introduced here will disappear in the non-relativistic limit and the two primitive fields $A$ and $B$ will --- so to speak --- be liberated. They will have appropriate non-relativistic limits which will allow us to construct NRQM in a proper manner. 

  To see how this comes about, we first introduce two  ``non-relativistic''  fields $a(x)$ and $b(x)$ in place of $A(x)$ and $B(x)$ by
  \begin{equation}
 A(x) \equiv \frac{e^{-imt}}{\sqrt{2m}} \, a(x); \qquad B(x)\equiv \frac{e^{-imt}}{\sqrt{2m}}\, b(x)
\end{equation} 
 This rescaling does two things: (i) It separates out a rapidly oscillating phase $\exp(-i mc^2t)$ from the fields; this phase arises from the relativistic rest energy of the particle. (ii) It factors out $(1/\sqrt{2m})$ which is a vestige of the relativistic momentum measure $(1/2\omega_p)$ that goes over to $(1/2m)$ in the non-relativistic limit. Thus we have eliminated two key relativistic factors (one due to rest energy, $mc^2$, and the other due to the change of measure in momentum integration) from the fields $A$ and $B$ to define $a$ and $b$. 
 
 We next express the  Lagrangian $L=\partial_a \phi \, \partial^a \phi^\dagger-m^2\phi\phi^\dagger$ for the physical field $\phi$ in terms of $a$ and $b$ fields. The kinetic energy part is:
\begin{eqnarray}
 \partial_a \phi \, \partial^a \phi^\dagger &=& \left(\partial_a A+\partial_a  B^\dagger \right) \, \left(\partial^a A^\dagger + \partial^a B\right) \\
 &=& \left(\partial_a A \partial^a A^\dagger\right) +  \left(\partial_a B^\dagger \partial^a  B\right) + \partial_a A\partial^a B + \partial_a A^\dagger \partial^a  B^\dagger\nonumber\\ 
 &=& \left(\partial_a A \partial_a A^\dagger\right) +  \left( A\Rightarrow B \right) + \cdots
\end{eqnarray} 
Here and in what follows, the $\cdots$ represent terms with factors $\exp(\pm 2imt)$ which can be ignored
since they rapidly oscillate and average out to zero in the non-relativistic limit.\footnote{One can do this more formally in an RG-type analysis by integrating out the high frequency modes and defining a low-energy effective Lagrangian. But since the modes are decoupled in the free field theory, this is equivalent to just dropping the rapidly oscillating terms.}
In terms of the ``non-relativistic'' fields, the first term is given by
\begin{equation}
 2m \partial_a A \, \partial^a A^\dagger= \left[ \left( - i m a + \dot a\right) \left( i m a^\dagger + \dot a^\dagger\right) - \partial_\mu a\, \partial^\mu a^\dagger\right]
 \label{aadag1}
\end{equation} 
and corresponding terms for $b$.
Similarly
\begin{equation}
 m^2 \phi^\dagger \phi  = m^2 AA^\dagger +  \left( A\Rightarrow B \right) + \cdots
 = m^2 aa^\dagger +  \left( a\Rightarrow b \right) + \cdots
\end{equation} 
Using these results we can express the  Lagrangian in terms of $a(x)$ and $b(x)$ as:
\begin{equation}
 2 L = a^\dagger \left( i \partial_t - H\right)\, a + b^\dagger \left( i \partial_t - H\right) \, b + \text{h.c} + \cdots
 \label{onetwoeight}
\end{equation} 
where $H=-(1/2m)\nabla^2$ is the non-relativistic Hamiltonian for free particle --- obtained by writing $\partial_\mu a\, \partial^\mu a^\dagger=
\partial_\mu (a\, \partial^\mu a^\dagger)-a\,\partial_\mu\partial^\mu a^\dagger$ in \eq{aadag1} and ignoring the total divergence ---
and
the dots indicate terms which can be ignored   in the non-relativistic  limit. These are terms of the kind:
\begin{equation}
 Q = |\dot a|^2  + |\dot b|^2 + e^{-2imt} (\ ) +  e^{2imt} (\ )
\end{equation} 
 The terms $|\dot a|^2$ and $|\dot b|^2$ are ignorable because the leading time variation, viz. the $e^{-imt}$ factor has been pulled out  while defining the non-relativistic fields $a$ and $b$; therefore, it is justifiable to retain only up to first time derivative while working with $a$ and $b$. We can also ignore terms multiplied by factors $\exp(\pm 2imt)$ since they rapidly average out to zero in the non-relativistic limit.
  
  From the structure of \eq{onetwoeight} we see that, in the non-relativistic limit, our system is described by \textit{two} fields $a$ and $b$ which actually represent the particle and anti-particle of the original system. Both of them satisfy the non-relativistic Schroedinger equation in operator form. \textit{So, anti-particles do not go away when you take the non-relativistic limit if you do it correctly. }
  
  We worked with the primitive fields $A$ and $B$ (which actually corresponds to the particle and anti-particle respectively) in order to show that the non-relativistic limit leads to a pair of fields $a$ and $b$ both obeying the Schroedinger equation. It is however possible to work entirely with the physical field $\phi$ and obtain the appropriate limit. To do this, we start with the definition of $\phi$, viz.: 
\begin{equation}
 \phi= A + B^\dagger = \left( a e^{-imt} + b^\dagger e^{imt}\right) \, \frac{1}{\sqrt{2m}}
\end{equation}
  The canonical momentum associated with $\phi$ is:
\begin{equation}
 \Pi = \dot \phi \approx i \sqrt{\frac{m}{2}} \left( -a e^{-imt} + b^\dagger e^{imt}\right) 
\end{equation} 
 where we have ignored the time derivatives of $a$ and $b$ in comparison with the time derivatives coming from $\exp(\pm imt)$ factor. This allows us to write 
\begin{equation}
 a =  e^{imt} \left(\sqrt{\frac{m}{2}}\, \phi  + \frac{i}{\sqrt{2m}} \, \Pi\right); \qquad
  b^\dagger =  e^{-imt} \left(\sqrt{\frac{m}{2}}\, \phi  - \frac{i}{\sqrt{2m}} \, \Pi\right)
\end{equation} 
This procedure works even for a real scalar field for which the anti-particle is identical to the particle. So, even real scalar fields have a natural non-relativistic limit contrary to what is sometimes claimed in the literature.

 The most important feature which has come about in the non-relativistic limit is the transition from second time derivatives to first time derivatives in the equation obeyed by the operators. That is, the relevant operator changes from $(\partial_t^2-\nabla^2+m^2)$ to
 $(i\partial_t+(1/2m)\nabla^2)$ or, equivalently
 $(\nabla^2 -m^2-\partial_t^2)$ goes over to $ \nabla^2+(2mi\partial_t)$. So the net effect is the replacement
 \begin{equation}
  (\partial_t^2 +m^2)\Longrightarrow (-2mi\partial_t)
  \label{opchange}
 \end{equation} 
 Almost all the key differences between QFT and NRQM are directly or indirectly connected with this change. In view of its importance, it is worth going over the algebraic reasons which led to this reduction. 
 
Since the spatial dependence is governed by the \textit{same} operator $\nabla^2$ \textit{both }in the relativistic and non-relativistic field equations, we can work in the Fourier space --- with modes labeled by the magnitude of a wave vector $k$ --- in both cases.  In the relativistic case the Fourier mode will satisfy a harmonic oscillator equation with frequency $\Omega_k^2\equiv k^2+m^2$. All  we need to do is to look at appropriate features of harmonic oscillators to understand what is going on. So consider a dynamical degree of freedom $f(t)$ which satisfies the harmonic oscillator equation of the form:
\begin{equation}
 \left( \frac{d^2}{dt^2} + \Omega_k^2\right)\, f = \left( \frac{d^2}{dt^2} + k^2 + m^2 \right)\, f
 \label{shoqft}
\end{equation} 
 To study NRQM, we want to look at the limit $k^2 \ll m^2$ when the frequency of oscillation of $f$ will be dominated by a factor like $\exp(\pm imt)$. It makes sense to pull this factor out of $f$ and redefine another dynamical variable $F$ by the relation $f= e^{-imt}\, F$.  It is now straightforward to show that the Lagrangian that leads to \eq{shoqft} can be re-expressed in terms of $F$ as:
\begin{equation}
 L= f^\dagger\, \left( \frac{d^2}{dt^2} + \Omega_k^2\right)\, f = F^\dagger\, \left( \frac{k^2}{2m} - i \partial_t\right) \, F + \frac{F^\dagger\ddot F}{2m}
\end{equation} 
The first term on the right hand side involves only the first time derivative. The second term contains $\ddot F$ which can be ignored compared to $\dot F$ in the limit we are interested. This is how the reduction of time derivatives occur when we proceed from QFT to NRQM. The culprit is the rest energy which introduces rapid time oscillations through the factor $\exp(-imt)$. 

 The idea of the primitive fields $a$ and $b$ and the physical field $\phi$ can also be understood  without worrying about the spatial dependence. and working with Fourier modes which behave like oscillators. To do this, let us consider a dynamical variable $q(t)$ described by a Lagrangian
\begin{equation}
 L = \dot q^\dagger \dot q - \Omega^2 q^\dagger q + \text{h.c}
\end{equation} 
 We now introduce two primitive fields $a(t)$ and $b(t)$, such that $ q = a+b^\dagger$,
  and re-express $L$ in terms of $a$ and $b$.  You will find that the Lagrangian separates into two parts as $L=L_1+L_2$ where
  \begin{equation}
   L_1=|\dot a|^2 - \Omega^2 |a|^2+|\dot b|^2 - \Omega^2 |b|^2
  \end{equation} 
  and 
  \begin{equation}
   L_2=- a (\ddot b - \Omega^2 b) - a^\dagger (\ddot b - \Omega^2 b)^\dagger
  \end{equation} 
  The second part of the Lagrangian $L_2$ actually leads to the identical field equations as $L_1$. For example, if you vary $a$ in $L_2$ you get $\ddot b = -\Omega^2 b$ which is identical to the field equation you get from the second pair of terms in $L_1$. Therefore we can ignore $L_2$ and think of the dynamics as being dictated by $L_1$ itself. The $L_1$ describes two independent oscillators $a$, $b$ with frequency $\Omega$. By an analysis similar to the one done before, we can reduce this system to one which involves only first time derivative. This is exactly analogous to what we have done earlier in the case of the field.

One feature which emerges out of this analysis is the sharp distinction between (i) any direct approach to quantum theory of relativistic particle  and (ii) relativistic particles emerging as excitations of a quantized field. Conceptually these constructions are completely different. To describe a relativistic particle, we can start with an eigenstate $\ket{\bm{p}}$ of its 3-momentum (with its energy $\omega_{\bm{p}}$  determined by $\omega_{\bm{p}} = + (\bm{p}^2 + m^2)^{1/2}$. One can build further states like, for example, $\ket{\bm{x}}$ and other useful operators like, for example, $A(x)$ etc. and build a theory in a suitable Hilbert space. But such a field $A(x)$ will not obey a sensible notion of causality. To remedy this situation \textit{we have to double up the number of particles by associating with each particle another particle} with (an unfortunate) nomenclature anti-particle. This is roughly what the introduction of the field $B(x)$ does. Then, the combination $\phi(x) = A(x) + B^\dagger(x)$ obeys a natural notion of micro-causality.  So the answer to the question ``why do anti-particles exist'' is simply ``to ensure causality in a Lorentz invariant theory''. It has nothing to do with square roots in Hamiltonians  or some funny notion of negative energy states; there are no negative energy states in the one-particle sector of the Fock space. Both the fields $A(x)$ and $B(x)$ have to be treated at equal footing and both have a  right to exist in NRQM. In short, a pair of fields in NRQM gets mapped to a single field in QFT.    

\section{Propagators as correlators}\label{sec:prcor}

We have seen that when you take the non-relativistic limit properly, an operator remains an operator. All that happens is that the Lagrangian and the field equation describing the field operator $\hat{\phi}(x)$ is different from the one describing the field operator $\hat{a}(x)$ and $\hat{b}(x)$. The $\phi(x)$ satisfies a field equation which is second order in time while $a(x)$ and $b(x)$ satisfy field equations which are first order in time. These non-relativistic fields are what are usually called --- in a confusing and incorrect nomenclature --- the ``second quantized'' version of Schroedinger wave functions. 
By using the language of field operators both in QFT and NRQM we can make a seamless transition from QFT to NRQM. (This is a familiar aspect of condensed matter physics but is not usually explored in detail in the context of non-relativistic limit of QFT; some earlier work is cited in \cite{G}.) The last issue which remains to be answered is the role of the propagators: How do we obtain the non-relativistic propagator in the appropriate limit since we are no longer talking about particle positions and trajectories \textit{even }in the NRQM limit? 
 
 To answer this question, let us start by examining an action which is quadratic in the fields and can be expressed in the form:
\begin{equation}
 A = \int d^D x\ \Phi^* \, \hat{D}(i \partial_a) \Phi = \int d^D x\, \Phi^*(\hat Q + \mu) \Phi
\end{equation} 
 The first equation defines an operator $\hat{D}$ which is built from the time and space derivatives $\partial_a$; for convenience we have introduced a parameter $\mu$ and written this operator as $\hat D \equiv \hat Q +\mu$. (In the case of Klein-Gordon field, for example, $\mu$ could be identified with $m^2$.) We will now define the propagator for the field as the correlator averaged using $e^{iA}$ through 
\begin{equation}
 G(x,y) \equiv\langle \Phi(x)\Phi^*(y)\rangle \equiv  \frac{1}{Z} \int \mathcal{D}\Phi\, \mathcal{D}\Phi^*\ \Phi(x) \Phi^*(y) \, e^{iA}; \quad Z\equiv \int \mathcal{D}\Phi\, \mathcal{D}\Phi^* e^{iA}
\end{equation} 
Since the action is quadratic, it is straightforward to evaluate this correlator, which is the matrix element of $D^{-1}$  in Fourier space. We get:
\begin{equation}
 G(x,y) = -i \bk{x}{D^{-1}}{y} = \int\frac{d^D p}{(2\pi)^D} \, \frac{e^{-ip(x-y)}}{iD(p)}
\end{equation} 
  As an example,  consider the standard Klein-Gordon field. In this case, we have 
$
 \hat D = (\Box + \mu - i \epsilon) = (-p^2 + \mu - i \epsilon)
$ so that 
$
 - i D^{-1} = +i (p^2  - \mu + i \epsilon)^{-1}
$. This will lead to the standard Feynman propagator $G_R$.  

One can give a nicer interpretation to \textit{any} such propagator by using the integral representation for $D^{-1}$ and writing:
\begin{equation}
 G(x,y)= \int_0^\infty ds \, \bk{x}{e^{-is\hat D}}{y} = \int_0^\infty ds \int e^{-ip\cdot (x-y)} \, e^{-isD(p)} \frac{d^D p}{(2\pi)^D}
 \label{onefourfive}
\end{equation} 
The second expression is obtained by introducing a complete set of momentum eigenstates and using $\amp{x}{p}=e^{-ipx}$ etc. The matrix element $\bk{x}{e^{-isD}}{y}$ can be thought of as a quantum mechanical propagator for a particle to go from $y$ to $x$ under the action of a Hamiltonian $D$ in ``time'' interval $s$. The structure of this expression immediately leads to the composition law for the propagator. Since  
\begin{equation}
 \int dx\ \bk{x_2}{(iD)^{-1}}{x} \bk{x}{(iD)^{-1}}{x_1} =  \bk{x_2}{(iD)^{-2}}{x_1} = i \frac{\partial}{\partial \mu } \bk{x_2}{(iD)^{-1}}{x_1}
\end{equation} 
we obtain the result 
\begin{equation}
 \int d^D x\ G(x_2,x) \, G(x, x_1) = i \frac{\partial}{\partial \mu }\,  G(x_2,x_1)
\end{equation} 

The discussion so far has been completely general. Let us now consider the question recovering NRQM from this approach. We start with the Fourier transform of the propagator with respect to spatial coordinates which can be expressed as:
\begin{equation}
 G_{\bm{k}} (t) \equiv \int d^D \bm{x}\ G(t,\bm{x})\, e^{-i\bm{k\cdot x}} = \int_{-\infty}^\infty \frac{d\omega}{(2\pi)} \, \frac{e^{-i\omega t}}{iD(\omega, \bm{k})} 
 \label{onefoureight}
\end{equation}
Obviously, the form of the propagator depends on the pole structure of $D(\omega, \bm{k})$ in the complex plane. We saw in the last section that the essential difference between QFT and NRQM is in the reduction of second time derivatives to first time derivatives, indicated by \eq{opchange}. This, in turn, suggests that in the Fourier domain, a second order pole is replaced by a first order pole in $\omega$. In fact,  
this is indeed the case.
You will get the standard form of NRQM if the pole structure of $D(\omega, \bm{k})$ has the form:
\begin{equation}
 D(\omega, \bm{k} ) - i \epsilon = \left[ - \omega + F(\bm{k}) - i \epsilon\right] (2\Omega_{\bm{k}})
\end{equation}
Then, a simple contour integration of the integral in \eq{onefoureight} will give the momentum space propagator to be:
\begin{equation}
G_{\bm{k}} (t) = \frac{\theta(t)}{2\Omega(\bm{k})} \, \exp\left[ - i t F(\bm{k})\right]; \qquad F\equiv H + \mu 
\label{onefivezero}
\end{equation} 
This will lead to standard NRQM if $2 \Omega(\bm{k})=$ constant. This propagator obeys  the composition law in Fourier space given by:
\begin{equation}
 \int_{-\infty}^\infty dt\, (2\Omega_k) G_k(t_2,t) \, (2\Omega_k) G_k(t,t_1) = i \frac{\partial}{\partial \mu} (2\Omega_k) G_k(t_2,t_1)
 \label{comp1}
\end{equation} 
From the explicit form of the propagator in \eq{onefivezero}, we see that the right hand side of \eq{comp1} is given by:
\begin{equation}
  i \frac{\partial}{\partial \mu} (2\Omega_k) G_k(t_2,t_1) = (t_2-t_1)\, G_k(t_2,t_1)
\end{equation} 
The left hand side of \eq{comp1} will also reduce to this expression because of the theta functions in time and we will recover the standard result in NRQM.  
Thus it is clear that NRQM is recovered when $D(\omega, \bm{k})$ has a single pole in the lower half plane. 

We can also construct the propagator directly from \eq{onefourfive} along the following lines. Introducing a complete set of momentum eigenkets $\ket{p}$ in the matrix element, this expression can be reduced to
\begin{eqnarray}
 G(x) &=& \int_0^\infty ds \, \int \frac{d^n\bm{p}}{(2\pi)^{n}} \, e^{i\bm{p\cdot x}}\, \int_{-\infty}^\infty \frac{d\omega}{2\pi} \, e^{-i\omega t}\, e^{-is(2\Omega) (-\omega + F - i\epsilon)}\\
 &=& \int \frac{d^n\bm{p}}{(2\pi)^{n}} \int_0^\infty ds\ \delta\left((2\Omega_p)\,s - t\right) \ e^{-isF(\bm{p}) + i \bm{p\cdot x}} \\
 &=&\theta(t)  \int \frac{d^n\bm{p}}{(2\pi)^{n}}\frac{1}{2\Omega_{\bm{p}}} \ e^{-it F(\bm{p}) +i \bm{p\cdot x}}
\end{eqnarray} 
Notice that, when there is only one pole for $D(\omega, \bm{k})$ making it a linear function of $\omega$, the $\omega$ integration in the first line leads to a Dirac delta function in time. This allows us to identify the ``internal time'' $s $ with the physical time $t$, leading to the final result. The final expression also has a direct interpretation in terms of the Hamiltonian form of the action principle. Thus the definition of propagators as correlators work consistently both in QFT and in NRQM. The key difference between the two is in the pole structure of the operator $\hat D$, which, in turn, is related to the conversion of second time derivatives to first time derivatives as explained in the previous section.

\section{Discussion}
\label{sec:impli}

This has been a rather long journey and --- for the sake of clarity --- let me briefly describe the path we have followed and the landmarks on the way. (The reader is invited to revisit the summary of the results given in Sec. \ref{sec:presum} at this stage, for more details.) I will then conclude by highlighting two important results we have obtained.

\subsection{Brief Overview}

One main conclusion --- which we have reached from several different perspectives --- is that, to make the seamless transition from QFT to NRQM, you need to describe NRQM in a language
which is closer to that of QFT and  not the other way around. This conclusion by itself may not be surprising but it was necessary to demonstrate it from different perspectives, which was one of the main objectives achieved in the paper. 

The NRQM limit can be obtained for a free particle by working with relativistic particle \textit{and} antiparticle field operators
$A^\dagger(x)$ and $B^\dagger(x)$. (The antiparticles do \textit{not} ``go away'' in the NRQM limit.) These operators are, in turn, defined in terms of operators which create fixed 3-momentum states from the no-particle state. The 3-momentum continues to be a ``good" operator in QFT while the 3-position is not. I have commented on this aspect extensively, contrasting  the 
the non-relativistic and relativistic cases, where the Hamiltonian takes the forms $H(\bm{p}) = \bm{p}^2 /2m$, or $H(\bm{p}) = \bm{p}^2 + m^2$, respectively. 

A closely related question is whether the non-relativistic wave function can be recovered through some limiting
procedure from a relativistic field operator. I addressed this  by focusing
on the propagator, an object that is well-defined in both NRQM and QFT. The technical issue, which makes all the difference between the two cases, is the fact that the measure of integration in momentum space has to be different in the two cases, which --- in turn --- arises from the requirement of Lorentz invariance.  This difference features  throughout the discussion, and makes it 
 impossible to perform a Fourier transform in the relativistic
case that will yield Lorentz covariant coordinate wave functions  representing spatially localized
particles.  

After discussing these aspects, I turned to the   issue of obtaining NRQM from QFT using the path integral formalism. Once again, the simplest route is to try and define the respective propagators from the path integrals. You then find that  the Lagrangian path integral cannot be defined through time slicing  in the relativistic
case for any sensible choice of measure. The Hamiltonian approach does work in both cases but does \textit{not} lead to the correct Feynman propagator.  

The best route seems to be the one based on  Euclidean lattice regularization scheme which \textit{does} lead to the Feynman propagator. In this approach we sum over the paths, parametrized
 by proper time,  including implicitly those that proceed both forward and backward in \textit{coordinate}
time. Exploiting the mathematical similarity of this method to  the approach based on the Jacobi action principle, one can again understand the origin of the difficulties in obtaining a single particle wave function. The Jacobi action approach tells us that, in the non-relativistic case, we need to 
construct  a propagator for fixed energy and then  sum over all energies while, in  the relativistic case, we need to sum over 
paths  for a fixed proper time followed by an integration over the proper time. It is
this last integration (over energy in NRQM and over propertime in RQM) that ruins the composition property of the propagator in either situation. 

To conclude this summary, I will comment briefly on two issues which are indirectly related to the discussion in this paper. The first comment  has to do with the  philosophical interpretation of  the  wave function in NRQM, which is still strongly debated. But note that (i)
 the QFT is more fundamental than NRQM and (ii) we do not have a sensible notion of single-particle wave function in RQM. Therefore, the debate over the ontological versus epistemological
status of the wave function \textit{within the context of NRQM} --- in which it is often attempted ---
seems irrelevant and misplaced. At the least one should escalate the debate to full QFT (say in Schrodinger functional formalism) for it to be meaningful; but then we will face several new serious, nontrivial, issues which might take precedence and change the nature of the debate.

The second comment is more technical. We saw that the consistent description of the NRQM limit of QFT requires us working with a \textit{pair} of fields, corresponding to a particle and its antiparticle.  
In the case of charged particles, these two will carry equal and opposite charges and hence there is a natural notion of charge conjugation with an associated operator in QFT. From our discussion it is clear that this is a purely relativistic feature and one does not have natural notion of charge conjugation operation in the NRQM, within the single particle sector. There are attempts in literature to introduce the notion of charge conjugation in NRQM but these attempts lack the naturalness with which one can introduce this notion in QFT.

\subsection{Two intriguing results}

The investigations of the path integral leads to some remarkable results, definitely worthy of further study. The first one is the expression for the 
the Feynman propagator, $G_R(x_2 , x_1 )=\bk{x_2}{e^{−iH|t|}}{x_1}$,  with the appearance of the absolute value of the time difference in the evolution operator. The second one is an intriguing relation between the path integral and the existence of antiparticles. I will now discuss these two results, starting from the second one.

The key result I want to highlight is contained in the beautiful --- and not adequately appreciated --- equation, which allows us to describe  relativistic particles as excitations of a Lorentz invariant, causal, quantum field:
\begin{eqnarray}
 \label{mystery1}
 &&\sum_{\rm paths} \exp\left( - \frac{im}{\hbar} \int_1^2 dt\, \sqrt{1-\bm{v}^2} \right)\\
 &&\hskip5em= \theta(t_2-t_1) \bk{0}{A(x_2)A^\dagger(x_1)}{0} + \theta(t_1-t_2) \bk{0}{B(x_1)B^\dagger(x_2)}{0}\nonumber
\end{eqnarray} 
The equality of the left hand side with the relativistic propagator $G_R(x_2,x_1)$ was demonstrated by lattice regularization in the Euclidean sector in Sec. \ref{sec:lattice}; the equality of the right hand side with the relativistic propagator 
$G_R(x_2,x_1)$ is provided by \eq{conc1}. 

\textit{The remarkable fact about \eq{mystery1} is that nobody understands it!}. That is to say,  no one has found a simple, physical argument suggesting why the left and right hand sides of  \eq{mystery1} should be equal without doing fairly elaborate calculations. This means we do not quite understand the conceptual basis of QFT --- and the structural implications of combining the principles of quantum theory and special relativity --- in spite of its remarkable success as a working tool. 

To see why `explaining' \eq{mystery1} is hard, consider the two sides separately. On the left hand side we have the action for a \textit{single} relativistic particle summed over all paths in spacetime connecting two events. So the left hand side combines the principles of quantum theory and special relativity in the most straightforward manner. The right hand side, on the other hand, describes \textit{two} kinds of particles propagating between the two events in spacetime. If $t_2 > t_1$, then $A$-type particle propagates forward in time, while if $t_2<t_1$, the $B$-type particle again propagates forward in time. (There is no propagation of particles backward in time which textbooks are fond of invoking.) It is a mystery how the path integral for  a \textit{single} relativistic particle gets an equivalent description in terms of \textit{two} kinds of particles --- both propagating forward in time with the choice of particles determined by the time ordering. It would be nice if a prescription for sum over paths can be devised which  nicely separates the contributions from $A$ and $B$ type particles on the right hand side. (I have some ideas on how to do this but --- as you could have guessed --- none of them works properly.) 
\footnote{Usually textbooks combine the two terms in the right hand side of \eq{mystery1} to a single time ordered product of the field $\phi = A + B^\dagger$ which does not help in resolving the mystery.}

The second result I want to highlight is the one we found in Sec. \ref{sec:fmatrix}. We found that the relativistic propagator can be expressed in the form
\begin{equation}
 G_R (x_2,x_1)  = \bk{\bm{x}_2}{e^{-iH|t|}}{\bm{x}_1}
  \label{mystery2}
\end{equation} 
We are working throughout with $H=\sqrt{\bm{p}^2+m^2} $ which is a positive definite operator.
But for $t=-|t|<0$, we have $\exp(-iH|t|) = \exp[-i(-H)t]$ and thus the minus sign in $t$ can be transfered to $H$ giving the illusion of a negative energy Hamiltonian. Since the operator $U(t) \equiv e^{-iH|t|}$ separates into two distinct evolution operators for $t>0$ and $<0$, it is obvious that two types of propagations are again incorporated in \eq{mystery2} just as in the case of \eq{mystery1}. This is understandable \textit{but the real surprise has to do with the quantum states between which the time evolution occurs} in \eq{mystery2}. We saw repeatedly that there are no localized particle states in QFT and hence we necessarily have to interpret $\ket{{\bm x}_1}$ and $\ket{{\bm x}_2}$ as some kind of smeared particle position states. Then \eq{mystery2} tells you that the relativistic propagator is obtained by the \textit{standard} time evolution operator used with either $H$ or $-H$ between such \textit{smeared} states. Clearly, there is some subtle interplay is going on between non-localisability of particle states and the existence of two kinds of propagation. As in the case of \eq{mystery1} I do not know of any simple way of explaining  \eq{mystery2} vis-a-vis the occurrence of smeared states.

Finally, let me comment on some broader implications of these and other results highlighted in the paper. 
I believe we can learn lessons regarding combining General Relativity (GR) with QM from carefully exploring the new features which arise when we combine Special Relativity (SR) with QM, which is the motivation for these comments.

It often happens in physics that certain well-defined notions  become approximate or, sometimes, even lose their utility when we proceed from an approximate description of Nature to a more exact description. It is possible that the  spatial location of an event is one of such concepts. In classical physics, both relativistic and non-relativistic, the notion of a spatial location $\bm{x}$ is operationally identified either with the position of the particle $\bm{x}(t)$ at some time $t$ or through the intersection of the world lines of two particles. Both these notions assume the existence of particles with arbitrarily small dimensions. 

In the conventional formulation of NRQM this idea is retained except for elevating $\bm{x}(t)$ to a Heisenberg operator $\hat{\bm{x}} (t)$ while retaining the purely parametric  (non-operator) status for time $t$. In NRQM you can still work with sharply localized one-particle states $\ket{t,\bm{x}}$ which are eigenstates of the operator $\hat{\bm{x}}(t)$, long as you don't care about the momentum of the particle. 
But, as we have seen,  the introduction of special relativity into QM makes this notion ill-defined. We no longer have localized particle states in RQM which, of course, is well known in the literature. But if you do not have localized particle states, can you still use the notion of spatial coordinates as though they are well-defined? The usual belief is that one can. For example, combining  the uncertainty principle of QM with the mass energy equivalence of SR, we immediately reach the conclusion that the notion of a single particle position becomes ill-defined, for a particle of mass $m$, at length scales below $\lambda_C \equiv \hbar/mc$.
So by considering hypothetical particles of arbitrarily high mass you can define spatial location with arbitrarily high accuracy. 

This idea, of course, breaks down when you approach Planck length, $L_P$. It is well known that one cannot (see, for e.g., \cite{tplp2,tplp1} and references therein) operationally define spatial locations with an accuracy better than a few Planck lengths, say. This in turn brings in an \textit{extra} non-localization  in the states $\ket{\bm{x}}$. In the absence of gravity, $\amp{\bm{y}}{\bm{x}}$ differs from a Dirac delta function and has significant support over a region of the size $|\bm{x}-\bm{y}|^2 \approx \lambda_C^2$. When we introduce Planck length into the consideration, we probably need to modify the form of  $\amp{\bm{y}}{\bm{x}}$ so that it has support in a region, say, $|\bm{x}-\bm{y}|^2 \approx \lambda_C^2+ L_P^2$ or something like that.\footnote{The standard smearing over Compton wavelength arises with the Klein-Gordan equation of the form $(\Box + \lambda_C^{-2})\phi =0$, while one can get the above modification if we use the equation $(\Box + \ell^{-2})\phi =0$ where $\ell^2 = \lambda_C^2 + L_P^2$. One can also introduce such a zero-point-length into spacetime in Lorentz invariant manner by modifying the path integral; see \cite{pid1,pid2}.} 

To incorporate any such modification at a fundamental scale, we may have to abandon the notion of  precise spatial location $\bm{x}$. 
Instead, one may want to consider creation and annihilation operators for spatial locations themselves; the action of these operators on a pre-geometric quantum state should produce the standard geometrical notion of a space-like hypersurface as a collection of spatial coordinates along with other geometrical notions. This is a coordinate based notion of a more abstract idea in which a creation operator $A^\dagger({}^3\mathcal{G})$ creates a 3-geometry ${}^3\mathcal{G}$ out of a pre-geometric state. Such an approach may be necessary to incorporate the breakdown of operational notion of spatial location at Planck scale. The de-localization of position by an amount $\lambda_C$ which arises when we combine SR with QM, suggests that some such structure is required to describe the spacetime when we combine GR with QM.

\section*{Acknowledgement}
The research work of the author is partially supported by the J.C. Bose research grant of DST, India.

\appendix

\section{Appendix A}
\label{sec:appencompo}

The left hand side of \eq{twentynine} has the form:
\begin{equation}
I_{\rm LHS}    = \int_{-\infty}^\infty dt\  \frac{1}{(2\omega_p)^2} e^{-\omega_p(|t_2 -t| + |t-t_1|)}
\end{equation}
which, on integration, gives:
\begin{equation}
 4\omega_p^2 \, I_{\rm LHS} = \left[ (t_2 - t_1) + \frac{1}{\omega_p} \right] \, e^{-\omega_p(t_2 -t_1)}
\end{equation} 
Consider now the right hand side of \eq{twentynine} which also evaluates to:
\begin{equation}
 I_{RHS} = -\frac{\partial}{\partial \mu} G_p(t_2 - t_1) = -  \frac{1}{2\omega_p}\frac{\partial}{\partial \mu} G = - \frac{G}{2\omega_p} \frac{\partial \ln G}{\partial \omega_p}
 =\frac{G}{2\omega_p} \left\{ (t_2-t_1) + \frac{1}{\omega_p}\right\}= I_{\rm LHS}
\end{equation}
Either side can also be expressed as:
\begin{equation}
 I_{\rm LHS} =  I_{RHS}=\frac{G(t_2-t_1)}{2\omega_p} \left\{ -\frac{\partial \ln G}{\partial \omega_p}\right\} = - \frac{1}{2\omega_p}\frac{\partial G}{\partial \omega_p} = - \frac{\partial G}{\partial \mu}
\end{equation} 
The non-relativistic limit corresponds to $\omega\approx m$ and $\omega(t_2-t_1)\approx m(t_2-t_1)\gg 1$. Then we find that
\begin{equation}
 -\frac{\partial \ln G}{\partial \mu} \bigg|_{\rm NR} = + \frac{1}{2\omega}  \left\{ (t_2-t_1) + \frac{1}{\omega}\right\}
 \approx \frac{t_2 - t_1}{(2m)}
\end{equation} 
The $(t_2-t_1)$ dependency will arise in the left hand of the standard composition law because of the appropriate $\theta(t)$ functions.

If one restricts the integration in \eq{twentysix} to just spatial coordinates of the intermediate point, we get the result:
\begin{equation}
 \int d\bm x \, G (x_2,x) G (x,x_1) = \int\frac{d\bm p}{(2\pi)^3} 
 \frac{1}{(2\omega_p)^2} e^{-\omega_p(|t_2 -t| + |t-t_1|)}e^{i\bm p\cdot(\bm x_2-\bm x_1)}
 \end{equation} 
This expression, of course, is not Lorentz invariant and will depend on the intermediate time $t$ except when $t_1<t<t_2$. Further it is not related in any simple manner to $G(x_2,x_1)$. It is also clear from the form of the expression how it reduces to the correct non-relativistic composition law when the modulus signs are omitted in $|t_2 -t|$ and $|t-t_1|$ and we set $\omega_p=m$.

\section{Appendix B}
\label{sec:equalgs}
This is most easily seen by manipulating this expression to the form:
\begin{eqnarray}
\label{delta}
G_+(x_2;x_1) - G_-(x_2;x_1) &=&
 G_+(x_2;x_1) - G_+(x_1;x_2)\\
 &=&\int d\,\Omega_{\bm{p}} [e^{-ipx} - e^{+ipx}]\\
&=&\int d\,\Omega_{\bm{p}} [ e^{-i\omega_{\bm{p}} t+ i\bm{p\cdot x}} -  e^{i\omega_{\bm{p}} t+ i\bm{p\cdot x}}]\nonumber\\
&=&\int d\,\Omega_{\bm{p}} e^{i\bm{p\cdot x}}\, [e^{-i\omega_{\bm{p}} t} - e^{i\omega_{\bm{p}} t}]\nonumber
\end{eqnarray} 
To arrive at the third line, we have flipped the sign of $\bm p$ in the second term.
Since the expression is  Lorentz invariant, we can always evaluate it in the frame with $t_2-t_1=t=0$, when the events are separated by a spacelike interval. It vanishes showing that $G_+(x_2;x_1) = G_-(x_2;x_1)$ when the events are separated by a spacelike interval.

\end{document}